\documentclass[aps,prd,preprint,superscriptaddress,,nofootinbib,amsmath,amssymb]{revtex4}
\usepackage{amssymb}
\usepackage{mathrsfs}
\usepackage{slashed}
\usepackage{graphicx,color}
\usepackage{amsmath}
\usepackage{subeqnarray}
\usepackage{multirow}
\allowdisplaybreaks

\usepackage{epsf}
\def\be{\begin{equation}}
\def\ee{\end{equation}}
\def\bea{\begin{eqnarray}}
\def\eea{\end{eqnarray}}
\def\nn{\nonumber}
\def\gsim{\ \rlap{\raise 2pt\hbox{$>$}}{\lower 2pt \hbox{$\sim$}}\ }
\def\lsim{\ \rlap{\raise 2pt\hbox{$<$}}{\lower 2pt \hbox{$\sim$}}\ }
\def\dslash{\kern-4pt \not{\hbox{\kern-2pt $\partial$}}}
\def\pslash{\not{\hbox{\kern-2pt p}}}

\begin{document}
\author{Monojit Ghosh\footnote{monojit@prl.res.in}}
\affiliation{
Physical Research Laboratory, Navrangpura,
Ahmedabad 380 009, India}

\author{Srubabati Goswami\footnote{sruba@prl.res.in}}
\affiliation{
Physical Research Laboratory, Navrangpura,
Ahmedabad 380 009, India}

\author{Shivani Gupta\footnote{shivani@cskim.yonsei.ac.kr}}
\affiliation{
Department of Physics and IPAP, Yonsei University, Seoul 120-479, Korea}

\author{C. S. Kim\footnote{cskim@yonsei.ac.kr,~~ Corresponding Author}}

\affiliation{
Department of Physics and IPAP, Yonsei University, Seoul 120-479, Korea}

\title{Implication of a vanishing element in 3+1 Scenario}


\begin{abstract}
 In this paper we study the phenomenological implications of the one zero textures of low energy
neutrino mass matrices in presence of a sterile neutrino.
We consider the 3+1 scheme and use the results from global fit for
short baseline neutrino oscillation data which provides the bounds on the
three additional mixing angles.
We find that the mass matrix elements
$m_{\alpha \beta}$ ($\alpha, \beta = e, \mu, \tau$)  involving
only the active states can assume vanishing values
in the allowed parameter space
for all the mass spectrum.
 Among the mass matrix elements
connecting the active and sterile states, $m_{es}$ and $m_{\mu s}$
can become small only for the quasi-degenerate neutrinos.
The element $m_{\tau s}$ on the other hand  can
vanish
even for lower values of masses
since the 3-4 mixing angle only has an upper bound
from current data.
The mass matrix element ($m_{ss}$)
involving only the  sterile state stays $\sim$ $\mathcal{O}$(1) eV in the
whole parameter region.
We study the possible correlations between the sterile mixing angles
and the Majorana phases to give a zero element in the mass
matrix.
\end{abstract}


\maketitle

\section{Introduction}

Light sterile neutrinos were invoked to explain the
results of the LSND experiment which reported
oscillation events in the $\bar{\nu}_\mu - {\bar\nu}_e$
mode corresponding to a mass squared difference $\sim$
eV$^2$ \cite{lsnd}.
Adding one sterile neutrino to the standard 3 generation framework
gives rise to two possible mass spectra
-- the 2+2 in which
two pairs of mass states are separated by a difference $\sim$ eV$^2$ and
3+1 in which a single predominantly sterile state differs by
$\sim$ eV$^2$ from the three active states \cite{earlier-papers}.
Subsequently the 2+2 schemes were found to be incompatible
with the solar and atmospheric neutrino data \cite{valle}.
The MiniBoone experiment was designed to test this
and its
antineutrino data confirmed the LSND anomaly \cite{miniboone}.
Both 3+1 and 3+2 sterile neutrino schemes have been considered
to explain these results \cite{3+1_fits,3+2_fits,giunti123}.
Such global fits aim to explain the non-observance of
eV$^2$ oscillations in the disappearance channel in other
short baseline experiments as well as the
reported evidence in LSND/MiniBoone experiments.
The relevant probabilities for 3+1 case is governed by
a single mass squared difference and hence is independent of
the CP phase.
In the 3+2 scheme, dependence on
CP phase comes into play and one gets a slightly better fit.

Other evidences in support of sterile neutrinos
include -- the reactor and the Ga anomaly. The first one
refers to the deficit in the
measured electron antineutrino flux in several experiments when
the theoretical predictions of  reactor neutrino fluxes were
reevaluated \cite{reactor}.
The second one implies shortfall of  electron neutrinos
observed in the solar neutrino detectors GALLEX and SAGE
using radioactive sources \cite{ga}.
Both these can be explained by adding light sub-eV sterile neutrinos
in the three generation framework.

There has also been some hint in favour of
sterile neutrinos
from cosmological observations of a "dark radiation" which is
weakly interacting and relativistic.
Attributing this to sterile neutrinos
one gets the bound on the number of neutrinos as
$N_{eff} = 4.34 \pm 0.87$ at 68\% C.L \cite{cosmo4}.
The Plank satellite experiment which has very recently
declared its first results \cite{Planck}, on the
other hand, gives $N_{eff} = 3.30 \pm 0.27$ at 68\% C.L.
which allows for an extra sterile neutrino at 95\% C.L.,
although its mixing with active species can be very tightly
constrained \cite{mirrizi} within the framework
of standard cosmology. Thus the sterile neutrinos continue
to be intriguing and
many new experiments are planned
proposed to test this \cite{sterile-future}.

Theoretically, sterile neutrinos are naturally included in
Type-I seesaw model
\cite{seesaw1}. But their mass scale is usually very high to
account for the small mass of the neutrinos.
Light sub-eV sterile neutrinos as suggested by the data
can arise in many models
\cite{sterile-future}.

Irrespective of  the mechanism for generation of neutrino masses
the low energy Majorana mass matrix  in presence of an extra sterile
neutrino will be of dimension $4\times4$
with  ten independent entries and is given as,
\begin{equation}
M_{\nu}=V^*M_{\nu}^{diag}V^{\dagger}
\label{mnu}
\end{equation} where, $M_{\nu}^{diag} = Diag({m_1,m_2,m_3,m_4})$
and $V$ denotes the leptonic mixing matrix in a basis
where the charged lepton mass matrix is diagonal.
One of the important aspects in the study of neutrino physics
is to explore the structure of the  neutrino mass matrices.
At the fundamental level
these are governed by Yukawa couplings which
are essentially free parameters in most models.
These motivated the study of texture zeros
which means one or more elements are relatively small compared to the others.
Texture zeros in the low energy mass matrices in the context of three
generations have been extensively explored both in the
quark and lepton sector \cite{frampton}, \cite{threegentexture}.
Such studies help in  understanding  the underlying parameter space
and the nature of the mass spectrum involved  and often predict
correlations between various parameters which can be
experimentally tested.
For three generation scenario it is well known that the number of
maximum texture zeros in low energy mass matrix is two \cite{frampton}.
In the context of the 4-neutrino case however  more than
two zeros  can be allowed \cite{ggg}. Two zero textures of sterile
neutrinos have been studied recently in \cite{ggg}
and three zero cases have been considered in \cite{chinachor}.
In this paper
we concentrate on the textures where one of the mass matrix elements
is vanishing.
For the $4\times4$ symmetric mass matrix it gives
total 10 different cases which needs to be investigated.
We study the implications of one zero  textures and the
possible correlations between the parameters.
We also compare our results with the 1 zero textures for three active neutrinos
\cite{werner-merle,lashin}.

The plan of the paper goes as follows. In section II we discuss the
possible mass spectra and the mixing matrix in the 3+1 scenario.
In the next section we present our study regarding the implications
of one vanishing entry in the low energy neutrino mass matrix.
We conclude in section IV.

\section{Masses and Mixing in the 3+1 scheme}

There are two ways in which one can add a
predominantly sterile state separated by $\sim$ eV$^2$ from the
standard 3 neutrino mass states.
In the first case the additional
neutrino  can be of higher mass than the other three
while in the second case the
the fourth neutrino is  the lightest state.
The later turns out to be incompatible with cosmology
since  in this case
three active neutrinos, each with mass $\sim$  eV results in an
enhanced cosmological energy density.
Thus it suffices to consider only the first case which admits
two possibilities displayed in Fig. 1.

\begin{figure}
 \begin{center}
 \includegraphics[scale=0.5,angle=0]{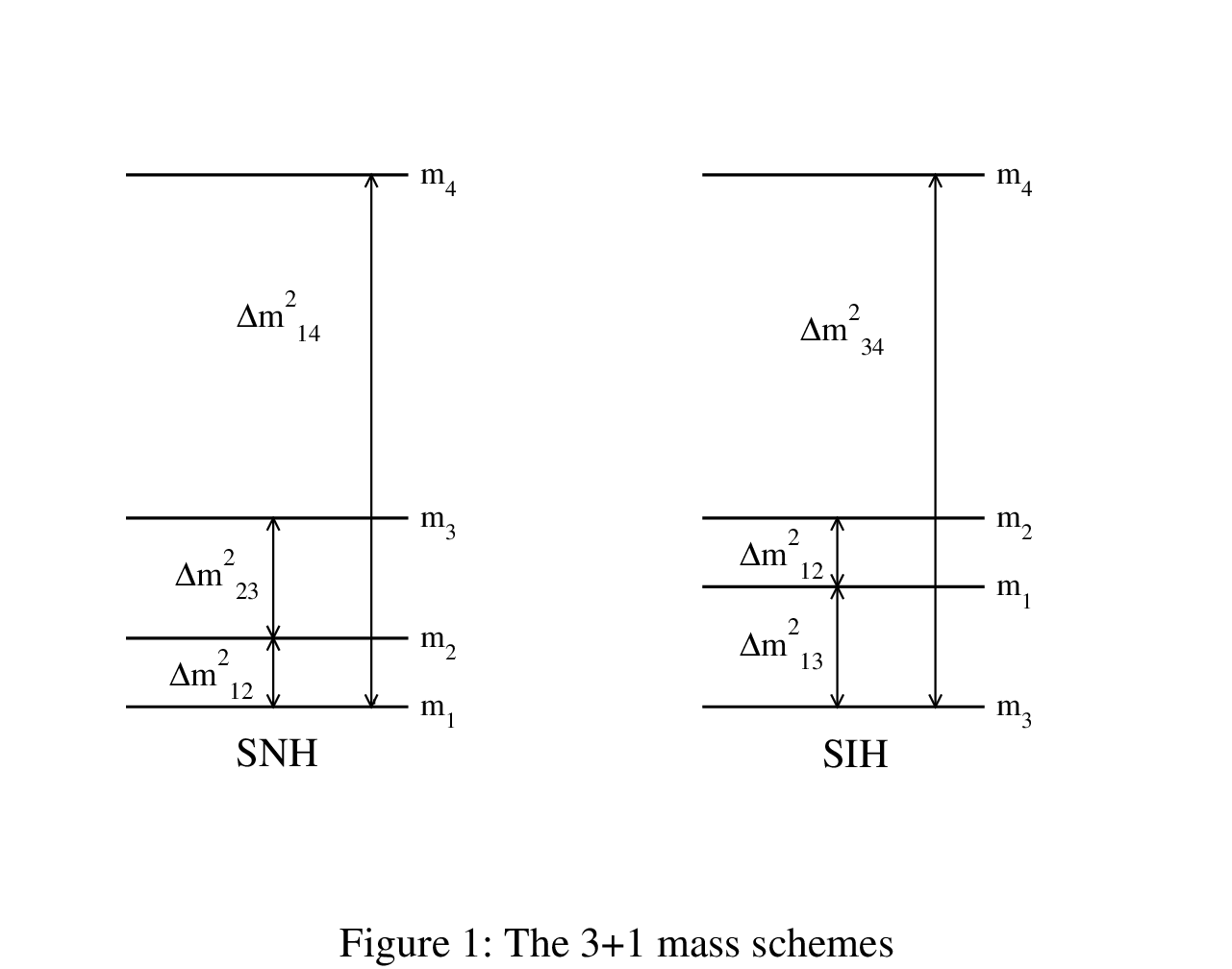}
 \end{center}
\label{fig1}
 \end{figure}

\begin{itemize}
\item[(i)] SNH: in this
$m_1 \approx m_2 < m_3 < m_4$  corresponding to a normal hierarchy (NH)
among the active neutrinos which implies,
\\
$m_2=\sqrt{m_1^2+\Delta m_{12}^2}~~ ,
m_3= \sqrt{m_1^2+\Delta m_{12}^2+\Delta m_{23}^2}~~ ,
m_4=\sqrt{m_1^2+\Delta m_{14}^2}.
$

\item[(ii)] SIH : this corresponds to
$m_3 < m_2 \approx m_1 < m_4$ implying an inverted ordering
among the active neutrinos with masses expressed as,
\\
$m_1= \sqrt{m_3^2+\Delta m_{13}^2}~~,
m_2=\sqrt{m_3^2+\Delta m_{13}^2+\Delta m_{12}^2}~~,
m_4=\sqrt{m_3^2+\Delta m_{34}^2}.
$
\end{itemize}
Here, $\Delta m_{ij}^2 = m_j^2 - m_i^2$.
 We define the ratio of the mass squared differences $\xi$ and $\zeta$ as
\begin{eqnarray}
 \xi= \frac{\Delta m_{14}^2}{\Delta m_{23}^2}~~{\mathrm(NH)} ~~ or ~~ \frac{\Delta m_{34}^2}{\Delta m_{13}^2}~~{\mathrm(IH)}, ~~
\label{xi}
\end{eqnarray}
\begin{eqnarray}
 \zeta =\frac{\Delta m_{12}^2}{\Delta m_{23}^2}~{\mathrm(NH)} ~~ or ~~
\frac{\Delta m_{12}^2}{\Delta m_{13}^2}~\mathrm{(IH)}.
\label{zeta}
\end{eqnarray}

In the extreme cases and using $\zeta \ll 1$, these masses can be written in terms of $\xi$ and $\zeta$ as \\
\be
SNH: |m_4| \approx \sqrt{ \Delta m_{23}^2 \xi} \gg |m_3|\approx \sqrt{(1 + \zeta)\Delta m_{23}^2} \approx \sqrt{\Delta m_{23}^2} \gg |m_2| \approx \sqrt{\Delta m_{23}^2 \zeta} \gg|m_1|
\label{xnh}
\ee

\be
SIH: |m_4|\approx \sqrt{\Delta m_{13}^2 \xi} \gg|m_2|\approx \sqrt{(1 + \zeta)\Delta m_{13}^2} \approx \sqrt{\Delta m_{13}^2} \approx |m_1| \gg|m_3|
\label{xih}
\ee

\be
SQD:~~|m_4|\gg|m_1|\approx|m_2|\approx|m_3|\approx m_0.
\label{qd}
\ee

The first two cases correspond to complete hierarchy among the active neutrinos
while the last one is the quasi-degenerate (QD) regime  where the three active
neutrinos have approximately equal masses.
\begin{table}
\centering
\begin{tabular}{lccc}
\hline
\hline
Parameter &  Best Fit values & $3\sigma$ range \\
\hline
$\Delta m^2_{12}/10^{-5}~\mathrm{eV}^2 $ (NH or IH)  & 7.54 & 6.99 -- 8.18 \\
\hline
$\sin^2 \theta_{12}/10^{-1}$ (NH or IH)  & 3.07  & 2.59 -- 3.59 \\
\hline
$\Delta m^2_{23}/10^{-3}~\mathrm{eV}^2 $ (NH) & 2.43 & 2.19 -- 2.62 \\
$\Delta m^2_{13}/10^{-3}~\mathrm{eV}^2 $ (IH) & 2.42  & 2.17 -- 2.61 \\
\hline
$\sin^2 \theta_{13}/10^{-2}$ (NH) & 2.41 & 1.69 -- 3.13 \\
$\sin^2 \theta_{13}/10^{-2}$ (IH) & 2.44 & 1.71 -- 3.15 \\
\hline
$\sin^2 \theta_{23}/10^{-1}$ (NH)  & 3.86  &  3.31 -- 6.37 \\
$\sin^2 \theta_{23}/10^{-1}$ (IH) & 3.92  & 3.35 -- 6.63 \\
\hline
$ \Delta m_{LSND}^2~\mathrm{eV}^2$  & 1.62 & 0.7 -- 2.5 \\
\hline
$ \sin^2\theta_{14} $ & 0.03 & 0.01 -- 0.06 \\
\hline
$ \sin^2\theta_{24} $  & 0.01 & 0.002 -- 0.04 \\
\hline
$ \sin^2\theta_{34} $ &  --  & $ <$ 0.18  \\
\hline
$\zeta /10^{-2}$ (NH) & --   & 2.7 -- 3.7 \\
$\zeta /10^{-2}$ (IH) & --   & 2.7 -- 3.8 \\
\hline
$\xi /10^3$ (NH) & --   &  0.27--1.14 \\
$\xi /10^3$ (IH) & --   &  0.27-- 1.15 \\
\hline
\label{parameters}
\end{tabular}

\begin{center}
\caption{$3 \sigma$ ranges of neutrino oscillation parameters \cite{fogli}.
The current constraints on
sterile neutrino parameters are from \cite{giunti12}, \cite{thomas-talk},
 where $\Delta m_{LSND}^2 = \Delta m_{14}^2(NH)$ or $\Delta m_{34}^2(IH)$.
Also given are the $3\sigma$ ranges of the mass ratios
$\zeta $ and  $\xi$.
}
\end{center}
\end{table}

In the 3+1 scenario,
the neutrino mixing matrix, $V$ in the flavor basis
will be a $4 \times 4$ unitary matrix. In general a N $\times$ N unitary mixing matrix contains $\frac{N(N-1)}{2}$ mixing angles and $\frac{1}{2}(N-1)(N-2)$ Dirac type CP violating phases.
It will also have
(N-1) number of additional Majorana phases if neutrinos are Majorana particles.
So in our case V can be parametrized in
terms of sixteen parameters. In addition to the three mixing angles between the active flavors,
($\theta_{13}$, $\theta_{12}$, $\theta_{23}$) we now have three more mixing angles from sterile
and active mixing, ($\theta_{14}$, $\theta_{24}$, $\theta_{34}$). There are six CP violating phases,
three Dirac ($\delta_{13}$, $\delta_{14}$, $\delta_{24}$) and three additional Majorana phases as
($\alpha$, $\beta$, $\gamma$) as neutrinos here are considered to be Majorana particles. Then,
there are four masses of neutrino $m_1$, $m_2$, $m_3$ corresponding to three active states and $m_4$
which is predominantly the mass of heavy sterile neutrino.

The mixing matrix $V$ can be expressed as
$V=U.P$ \cite{gr1} where
\begin{equation}
U={R_{34}}\tilde R_{24}\tilde R_{14}R_{23}\tilde R_{13}R_{12}
\end{equation}
where $R_{ij}$ denotes rotation matrices in the \textit{ij} generation space
and is expressed as,
\begin{center}
$R_{34}$=$\left(
\begin{array}{cccc}
1~ &~0 & 0 & 0 \\  0~ &~ 1 & 0 & 0 \\ 0~ & ~0 & c_{34}& s_{34} \\0 ~& ~0 & -s_{34} & c_{34}
\end{array}
\right)$ , $\tilde R_{14}$=$\left(
\begin{array}{cccc}
c_{14}~ & ~0 &~ ~0 &~s_{14}e^{-i \delta_{14}} \\ 0 ~ & ~ 1&~~ 0 & 0 \\ 0 ~& ~0 &~~ 1 & 0 \\-s_{14}e^{i \delta_{14}}  & ~ 0& ~~0 &c_{14}
\end{array}
\right)$ \\
\end{center}
Here we use the abbreviations $s_{ij}=\sin\theta_{ij}$ and $c_{ij}=\cos\theta_{ij}$. The  phase matrix is diagonal and is expressed as,
\begin{center}
$P=Diag(1,e^{i \alpha}, e^{i {(\beta+\delta_{13})}},e^{i {(\gamma+\delta_{14})}})$.
\end{center}
The best-fit values and the 3$\sigma$ ranges of the oscillation parameters
in the 3+1 scenario are given in Table I
where in addition to the masses and mixing angles we also
present the mass ratios $\zeta$ and  $\xi$
which would be useful in our analysis.
Note that the constraints on the three-neutrino
parameters may change slightly once a full four-neutrino fit combining
all global data is done. However since the sterile mixing
angles are small the change is not expected to be significant.
Therefore in absence of a full four-neutrino global fit we use
three-neutrino parameter values as obtained from three generation
analyses \cite{fogli,GonzalezGarcia:2012sz,Tortola:2012te}.

\begin{figure} \label{sigmam}
\begin{center}
\includegraphics[width=0.33\textwidth,angle=270]{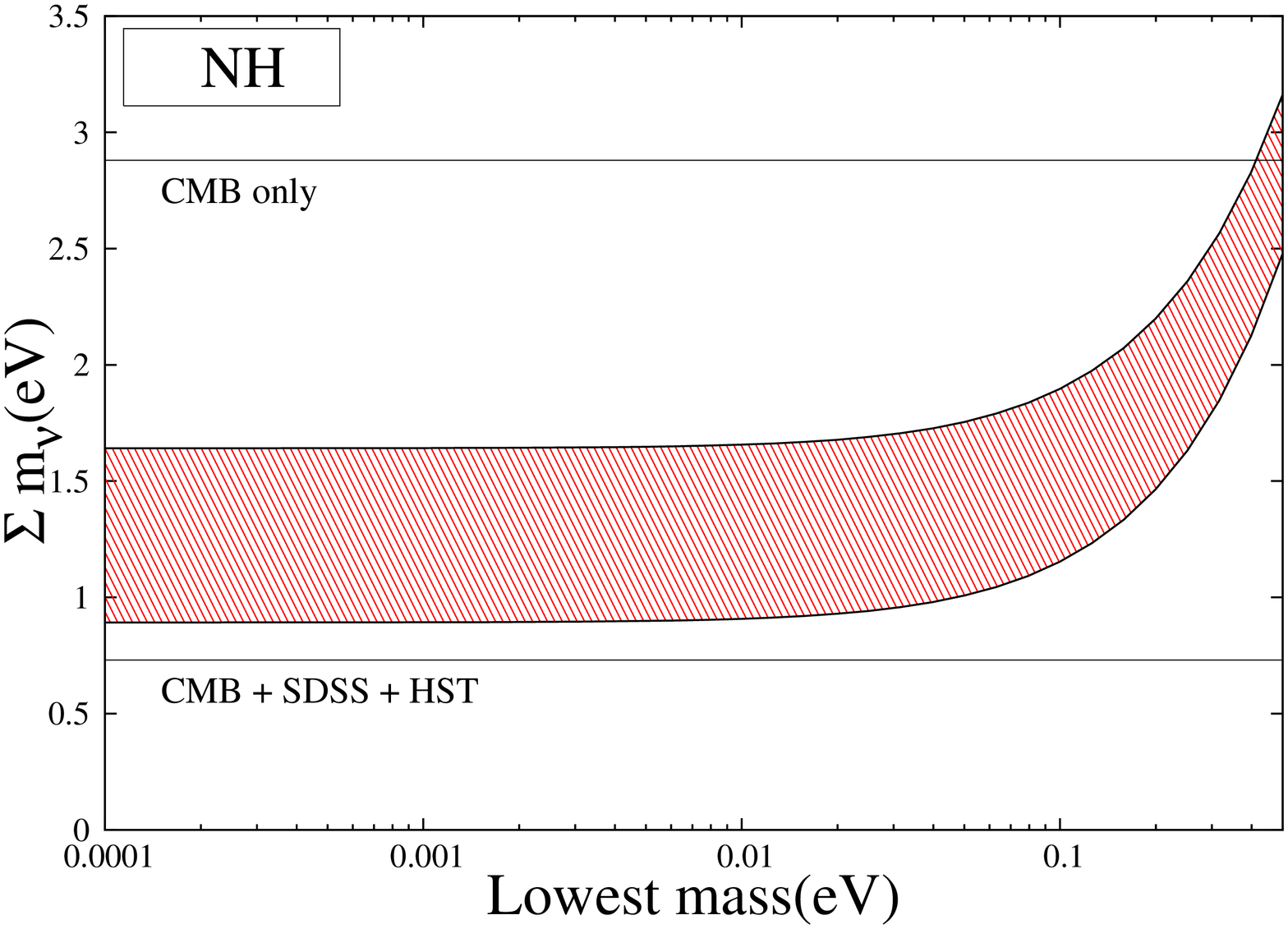}
\includegraphics[width=0.33\textwidth,angle=270]{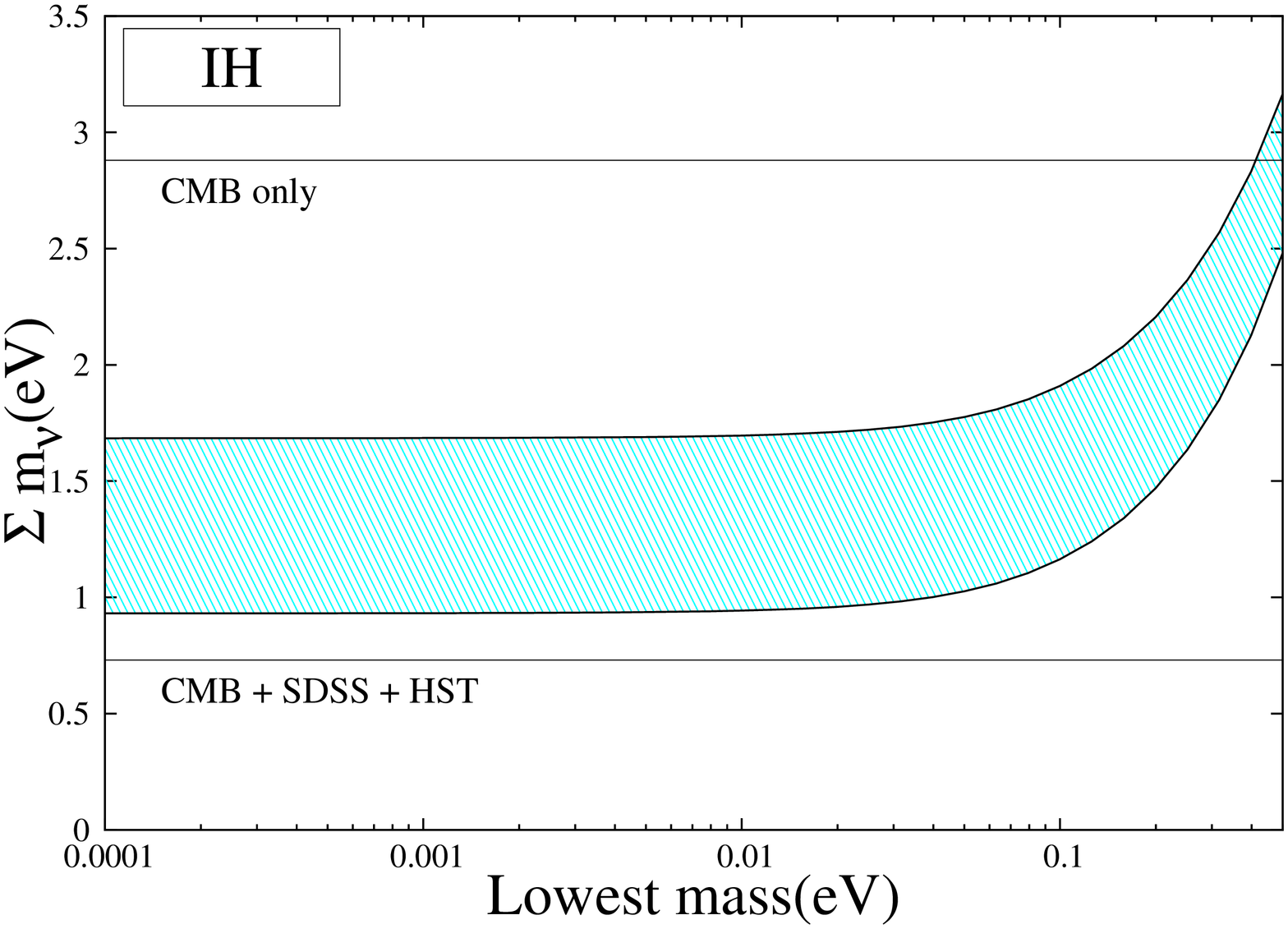}\\
\end{center}
Figure 2:  Plots of sum of light neutrino masses ($\Sigma m_\nu$)
vs the lowest mass in the 3+1 scenario.
Also shown are the
the cosmological upper bound on neutrino mass from the analysis of CMB data
plus matter
power spectrum information (SDSS) and a prior on $H_0$(HST) and from the
analysis of CMB data only from reference \cite{giunti12} for 3+1 scheme.
\end{figure}

In Fig. 2  we have plotted the sum of neutrino masses against the
lowest neutrino mass for both NH and IH.
The band corresponds to variation of the mass squared differences in their
current 3$\sigma$ range. We also show the cosmological upper bound on neutrino masses
in 3+1 scenario from \cite{giunti12}.
The combined analysis
of CMB + SDSS + HST  seems to rule out the mass spectrum of 3+1
scenario
in the framework of standard cosmology.
However, if only CMB data is taken then region for the lowest mass $< 0.4$ eV
gets allowed for both
the hierarchies.  Note that the analysis in \cite{giunti12} does not
incorporate the Planck results \cite{Planck} which can constrain the
sum of masses further. In our analysis
we have varied the lowest mass up to 0.5 eV.

\section{Neutrino Mass Matrix Elements}

In this section we study the implication of the condition of vanishing
$m_{\alpha \beta}$ for the 3+1 scenario,
where $\alpha, \beta = e, \mu,\tau, s$.
Since $m_{\alpha \beta}$ is complex the above condition implies
both real and imaginary parts are zero.
Therefore to study the 1-zero textures we consider $|m_{\alpha \beta}|=0$.
In our analysis we have varied the three Dirac phases in the range 0 to $2 \pi$ and the three Majorana phases from 0 to $\pi$.

\subsection{The Mass Matrix element $m_{ee}$ }

The matrix element $m_{ee}$ in the 3+1 scenario is given as,
\be
m_{ee}=m_1 c_{14}^2 c_{13}^2c_{12}^2+m_2 s_{12}^2c_{14}^2c_{13}^2e^{2i\alpha}+m_3 s_{13}^2 c_{14}^2
e^{2i\beta}+m_4s_{14}^2 e^{2i\gamma}.
\label{mee}
\ee
This is of the form
\be \label{meenh}
m_{ee}= c_{14}^2 (m_{ee})_{3\nu}+e^{2i\gamma}s_{14}^2 m_4,
\ee
where $(m_{ee})_{3\nu}$  corresponds to the matrix element in the
3 active neutrino case.
The  contribution of the sterile neutrino to the element $m_{ee}$
depends on the mass $m_4$ and  the
active-sterile mixing angle $\theta_{14}$.
Of all the mass matrix element $m_{ee}$
has the simplest form because of the chosen parametrization and
can be understood quite well.
Using approximation in Eq. (\ref{xnh}) for the case of extreme hierarchy
one can write this for NH as,
\be \label{meenh}
m_{ee} \approx c_{14}^2 (m_{ee})_{3\nu}+e^{2i\gamma}s_{14}^2\sqrt{\Delta m_{14}^2},
\ee
where $(m_{ee})_{3\nu}\approx \sqrt{\Delta m_{23}^2}(e^{2i\alpha}c_{13}^2s_{12}^2 \sqrt{\zeta}+s_{13}^2e^{2i\beta})$ and $\zeta$ is defined in
Eq. \ref{zeta}.
\begin{figure} \label{meefig}
\begin{center}
\includegraphics[width=0.33\textwidth,angle=270]{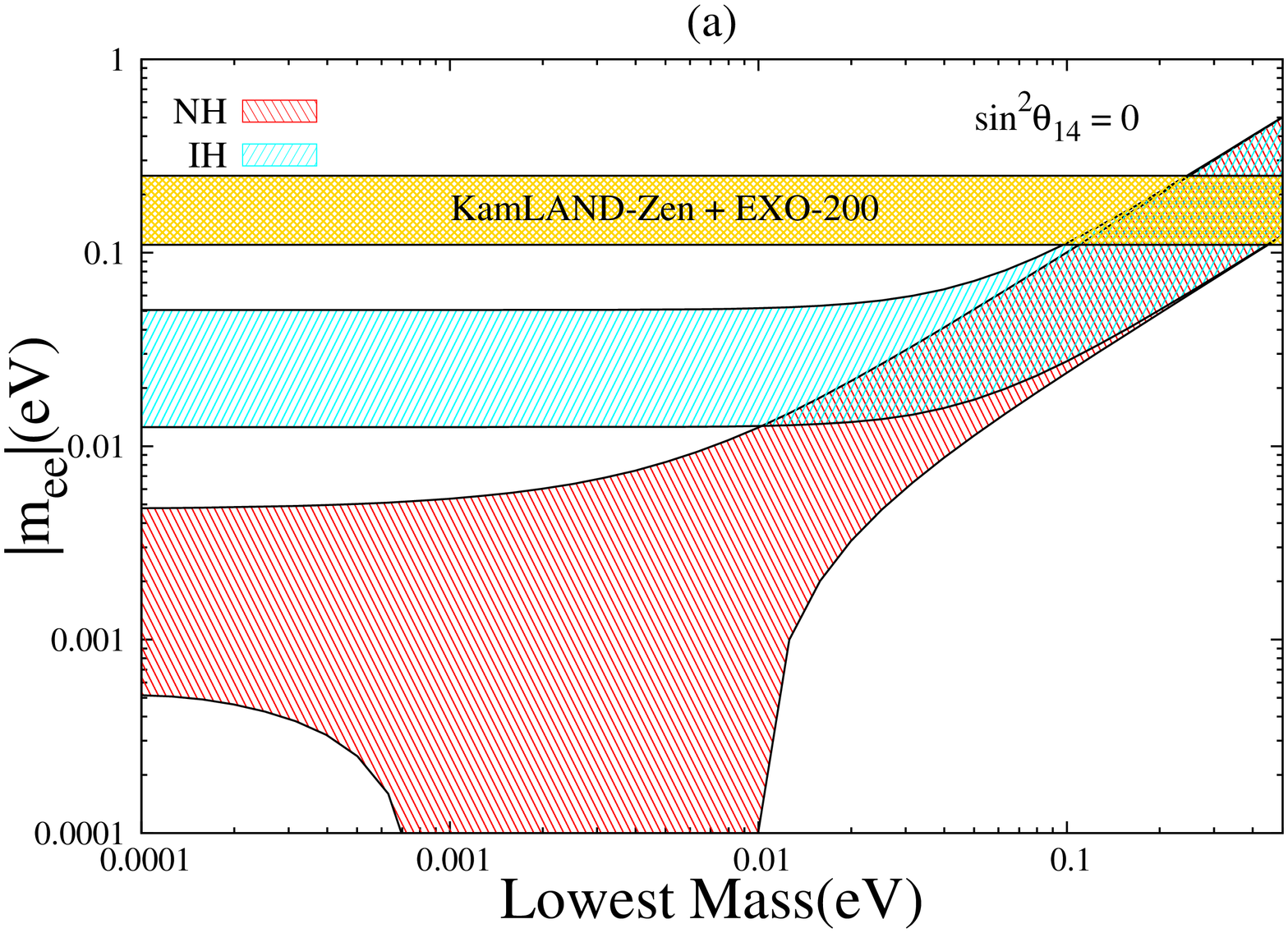}
\includegraphics[width=0.33\textwidth,angle=270]{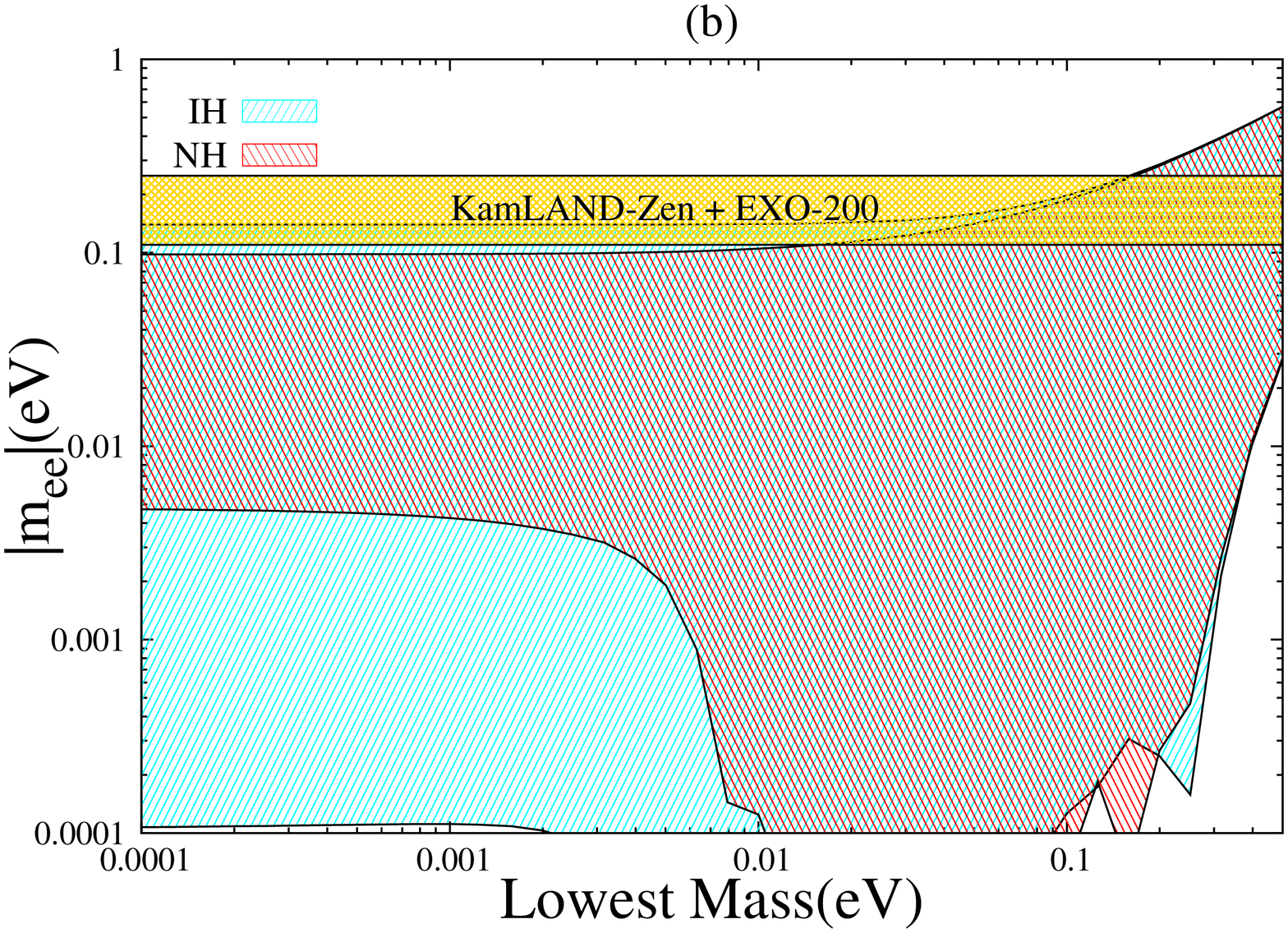} \\
\includegraphics[width=0.33\textwidth,angle=270]{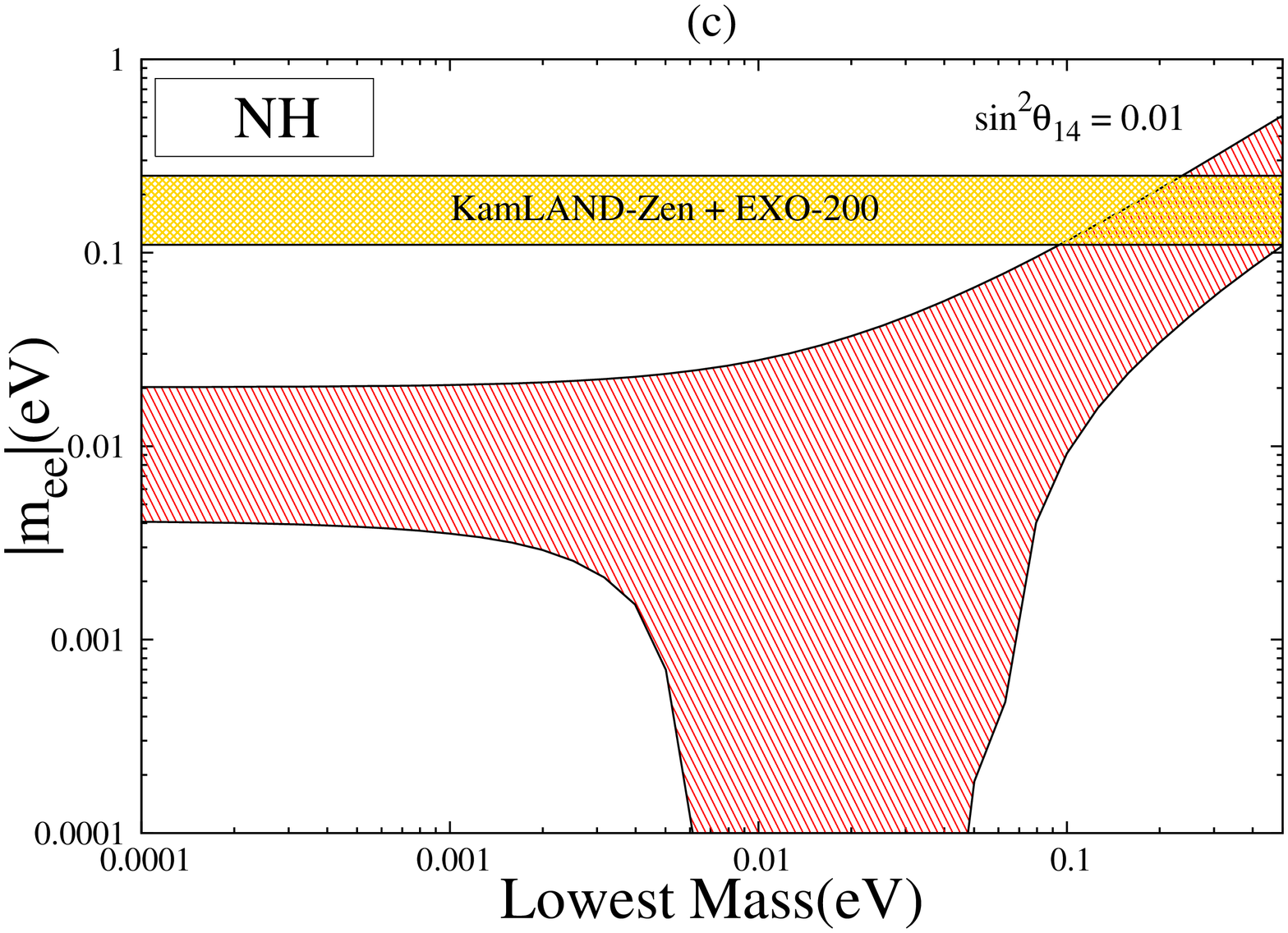}
\includegraphics[width=0.33\textwidth,angle=270]{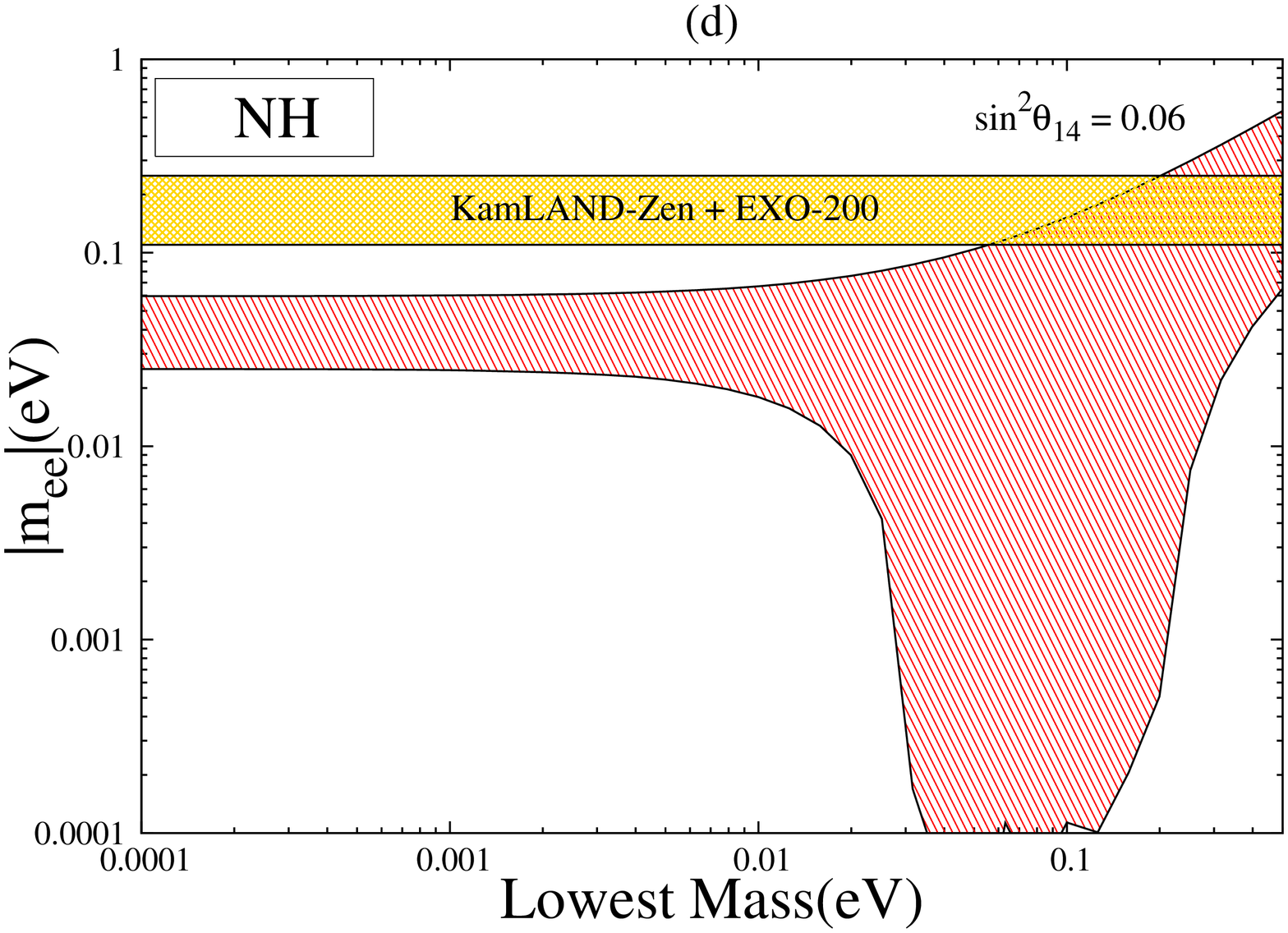}
\end{center}
Figure 3: Plot of $|m_{ee}|$ versus the lowest mass. The panel (a)
corresponds to the three generation case while panel (b) is for 3+1 case. In panel (b)
all the mixing angles are varied in their 3$\sigma$ range and the Majorana CP violating phases are varied in their full range (0-$\pi$).
The panel (c) and (d) are for specific values of
$\theta_{14}$ with all other parameters covering their full range.
\end{figure}
The modulus of $m_{ee}$ is the effective mass that can be extracted from
half life measurements in neutrinoless double beta decay.
In Fig. 3 we plot the effective mass  as a function of the smallest
mass by varying $\theta_{14}$ in its
complete 3$\sigma$ range from Table I as well as
for specific values of the mixing
angle $\theta_{14}$.
The Majorana phases are varied randomly
in the range  0 to $\pi$ in all the plots.
The first panel is for $\theta_{14}=0$ i.e the three generation case.
It is seen that for present values of the oscillation parameters
the cancellation condition is not satisfied for $m_1 \rightarrow 0$ for NH.
However, as one increases $m_1$, complete cancellation can be achieved.
For IH the complete cancellation is never possible.
These results change when we include the sterile contribution as is
evident from the panel (b) in  Fig. 3 which shows the effective mass
for NH and IH by varying all the parameters in their full 3$\sigma$
allowed range.
The behaviour can be understood from the expressions of $|m_{ee}|$ in
various limiting cases.
For NH, in the hierarchical limit of  $m_1 \rightarrow 0$
the major contributor will be the additional term due to
the sterile neutrinos because of higher value of $m_4$.
Complete cancellation is only possible for
smaller values of $\theta_{14}$ so that this contribution
is suppressed.
The typical value of $\theta_{14}$ required for cancellation can be
obtained by putting
$\alpha =\beta=0$ (which would maximize the three neutrino contribution)
and $\gamma = \pi/2$, as
\be
\tan^2\theta_{14} \approx \frac{(\sqrt{\zeta} c_{13}^2s_{12}^2+s_{13}^2)}{\sqrt{\xi}} \approx 10^{-3},
\ee
which lies outside the allowed range of $\theta_{14}$ given in Table I.
As we increase $m_1$, $(m_{ee})_{3\nu}$
increases and can be of the same order of magnitude of the
sterile term.  Hence one can get cancellation regions.
The cancellation is mainly controlled by the value of $\theta_{14}$. For higher values of $s_{14}^2$
one needs a higher value of $m_1$ for cancellation to occur. This correlation between $m_1$ and $\theta_{14}$ is brought
out by the panels (c) and (d) in Fig. 3.

For IH case, in the limit of vanishing $m_3$ using approximation in Eq. (\ref{xih}),
$m_{ee}$ in a 3+1 scenario can be written as
\be
|m_{ee}| \approx |c_{14}^2 c_{13}^2 \sqrt{\Delta m_{13}^2}(c_{12}^2+s_{12}^2 e^{2i\alpha})+\sqrt{\Delta m_{34}^2}s_{14}^2 e^{2i\gamma}|.
\ee
The maximum value of this  is achieved for $\alpha=\gamma=0$
which is slightly lower than that of NH in this limit.
The element vanishes in the limit $m_3\approx$ 0 eV when
$\alpha=0$  and $\gamma = \pi/2$ provided
\be \label{meeih}
\tan^2\theta_{14}  \approx \frac{c_{13}^2}{\sqrt{\xi}} \approx 0.05
\ee
This
is well within the allowed range.
This behaviour is in stark contrast to that in the 3 neutrino
case \cite{Werner-barry} .
There is no significant change in this behaviour as the
smallest mass $m_3$ is increased since this contribution
is suppressed by the $s_{13}^2$ term and the dominant
contribution to  $(m_{ee})_{3\nu}$ comes from the first two
terms in Eq. (\ref{mee}). Therefore in this case we do not
observe any correlation between $m_3$ and $s_{14}^2$.

While moving towards the quasi-degenerate regime of
$m_1 \approx m_2 \approx m_3$ we find that effective mass can still be zero.
However, when the
lightest mass approaches a larger value \textasciitilde~ 0.3 eV we need very large values of active sterile
mixing angle $\theta_{14}$, outside the allowed range, for cancellation.
Hence the effective mass cannot vanish for such
values of masses.

Also shown is the current limit on effective mass from combined
KamLAND-Zen and  EXO 200
results on the half-life of $0\nu\beta\beta$ in
$^{136}$Xe \cite{Gando:2012zm,Auger:2012ar}.
When translated in terms of effective mass this corresponds to the bound
$|m_{ee}| < 0.11 - 0.24$ eV including nuclear matrix element uncertainties.
For the three generation case, the hierarchical neutrinos
cannot saturate this bound.
But in the 3+1 scenario this bound can be reached even for
very small values of $m_3$ for IH and for some parameter
values it can even exceed the current limit. Thus from the present limits
on neutrinoless double beta decay searches a part of the parameter space
for smaller values of $m_3$ can be disfavoured for IH.
For NH,  the KamLAND-Zen + EXO 200
combined bound is reached for $m_1 = 0.02$ eV
and again  some part of the parameter space can be disfavoured by this bound.

\subsection{The Mass Matrix element $m_{e\mu}$ }
The mass matrix element $m_{e\mu}$ in the presence of extra sterile neutrino is given as
\bea
m_{e \mu}&=&c_{14}(e^{i (\delta _{14}-\delta _{24}+2 \gamma)}m_4s_{14}s_{24}+e^{i(\delta_{13}+2 \beta)} m_3s_{13}(c_{13}c_{24}s_{23}-e^{i
(\delta_{14}-\delta_{13}-\delta_{24})}\\ \nonumber && s_{13}s_{14}s_{24})
+c_{12}c_{13}m_1(-c_{23}c_{24}s_{12}+c_{12}(-e^{i \delta_{13}}c_{24}s_{13}s_{23}-e^{i (\delta_{14}-\delta_{24})}c_{13}s_{14}s_{24})) \\
\nonumber &+& e^{2i \alpha}m_2c_{13}s_{12} (c_{12}c_{23}c_{24}+s_{12}(-e^{i \delta_{13}}c_{24}s_{13}s_{23}-e^{i (\delta_{14}-\delta_{24})}c_{13}s_{14}s_{24}))).
\eea
Unlike $m_{ee}$ here the expression is complicated and an analytic understanding is difficult from the full expression.
The expression for $m_{e \mu}$ in the limit of vanishing active sterile mixing angle $\theta_{24}$ becomes
\bea
m_{e \mu}= c_{14}(m_{e \mu})_{3 \nu}. \nonumber
\eea
Since the active sterile mixing is small, in order to simplify
these expressions we introduce a quantity $\lambda\equiv$0.2
and define these small angles to be of the form $a\lambda$. Thus a
systematic expansion in terms of $\lambda$ can be done.
For sterile mixing angle
\bea \nonumber \label{chi1}
{\mathrm{sin}} \theta_{14} \approx \theta_{14} \equiv \chi_{14}\lambda, \\
{\mathrm{sin}} \theta_{24} \approx \theta_{24} \equiv \chi_{24}\lambda,
\eea
and the reactor mixing angle as
\be \label{chi2}
\sin \theta_{13} \approx \theta_{13} \equiv \chi_{13}\lambda.
\ee
Here $\chi_{ij}$ are parameters of $\mathcal{O}$(1) and their $3 \sigma$ range
from the current constraint on the mixing angles is given by
\begin{eqnarray}
\chi_{13} &=& 0.65 - 0.9, \\ \nonumber
\chi_{14} &=& 0.5 - 1.2, \\ \nonumber
\chi_{24} &=& 0.25 - 1.
\end{eqnarray}
 Note that for the sterile mixing angle $\theta_{34}$ we do not adopt the above approximation because this angle can be large compared to other two sterile mixing angles
and hence the small parameter approximation will not be valid.

Using the approximation in Eqs. (\ref{xnh}), (\ref{chi1}) and (\ref{chi2}) we get the expression for $|m_{e\mu}|$ for normal hierarchy as
\bea
|m_{e\mu}| &\approx& |\sqrt{\Delta m_{23}^2}\{\sqrt{\zeta}s_{12}c_{12}c_{23}e^{2i\alpha}+e^{i\delta_{13}}(e^{2i \beta}-e^{2i \alpha}\sqrt{\zeta}s_{12}^2)s_{23} \lambda \chi_{13}\\ \nonumber
&+&\lambda^2
e^{i(\delta_{14}-\delta_{24})}(e^{2i\gamma}\sqrt{\xi} -e^{2i\alpha}\sqrt{\zeta} s_{12}^2)\chi_{14}\chi_{24}\}|.
\eea
To see the order of magnitude of the different
terms we choose vanishing Majorana phases while Dirac CP phases are taken
as $\pi$. The mass matrix element  $m_{e \mu}$ vanishes when

\be \label{mem}
 \sqrt{\zeta}s_{12}c_{12}c_{23}-(1-\sqrt{\zeta}s_{12}^2)s_{23}\lambda \chi_{13}+ \lambda^2(\sqrt{\xi}-\sqrt{\zeta}s_{12}^2)\chi_{14}\chi_{24}=0.\\
\ee
The three generation limit is recovered for $s_{24}^2 = 0$ and in panel (a) of Fig. 4 we show $|m_{e\mu}|$ as a function of  $m_1$ of this case, for NH.
Panel (b) (red/light region) of Fig. 4 shows  $|m_{e\mu}|$ for the 3+1 case, with all parameters
varied randomly within their $3 \sigma$ range.
The figures show that $|m_{e\mu}|=0$ can be achieved over the
whole range of the smallest mass for both 3 and 3+1 cases. However, we find that in the
hierarchical limit cancellation is not achieved for
large values of $\theta_{24}$, since
in that case the third term of Eq. (\ref{mem}) will be of the $\mathcal{O}$ (10$^{-1}$) compared to the leading order term which is of the $\mathcal{O}$ (10$^{-2}$)
and hence there will be no cancellation of these terms.
This can be seen from panel (b) (green/dark region) of Fig. 4 for  $s_{24}^2=0.04$.
In the QD limit the contribution from the active terms are large enough to
cancel the sterile contribution and thus $|m_{e\mu}| = 0$ can be achieved.


\begin{figure}
\label{fig4}
\begin{center}
\includegraphics[width=0.33\textwidth,angle=270]{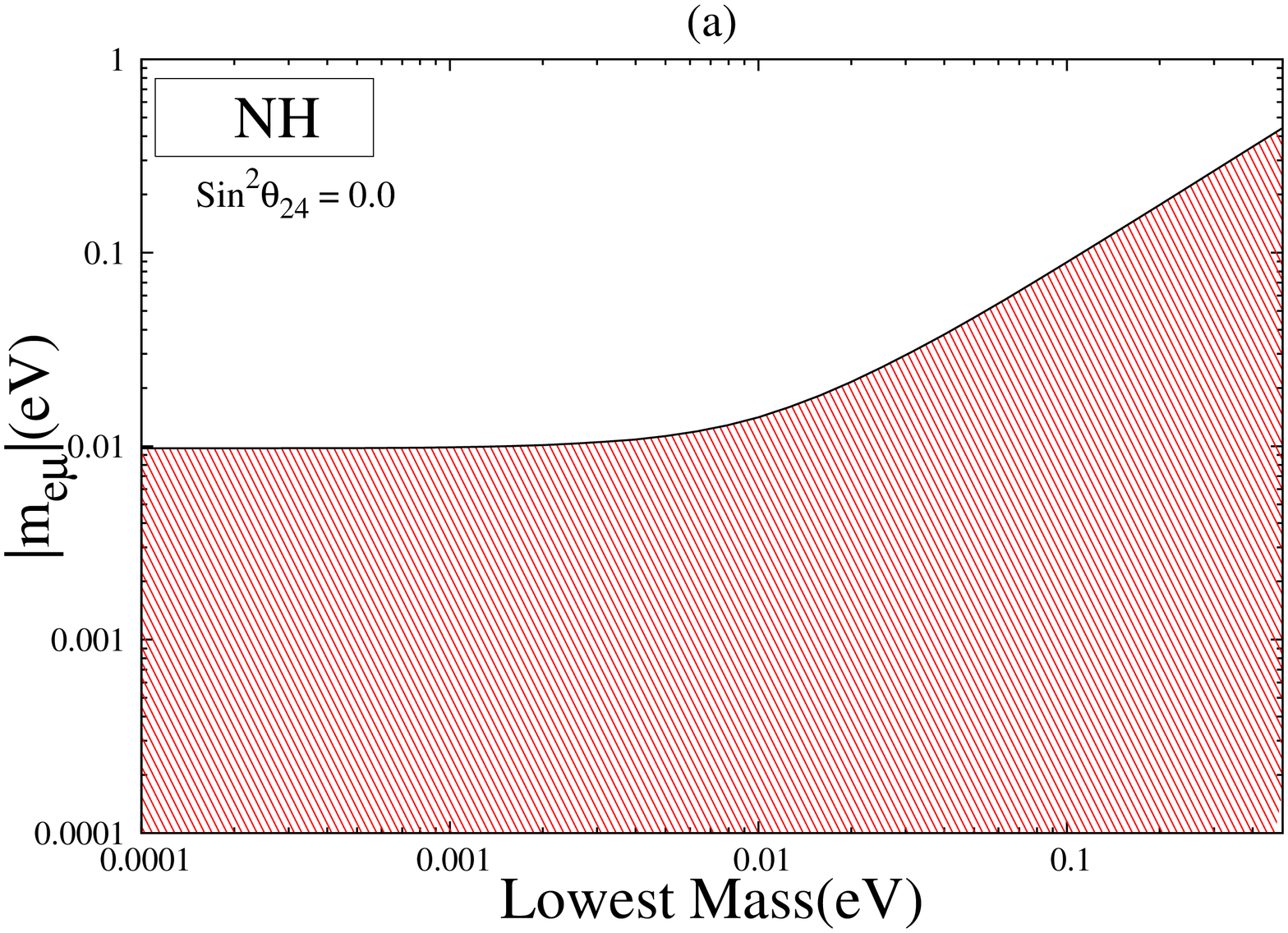}
\includegraphics[width=0.33\textwidth,angle=270]{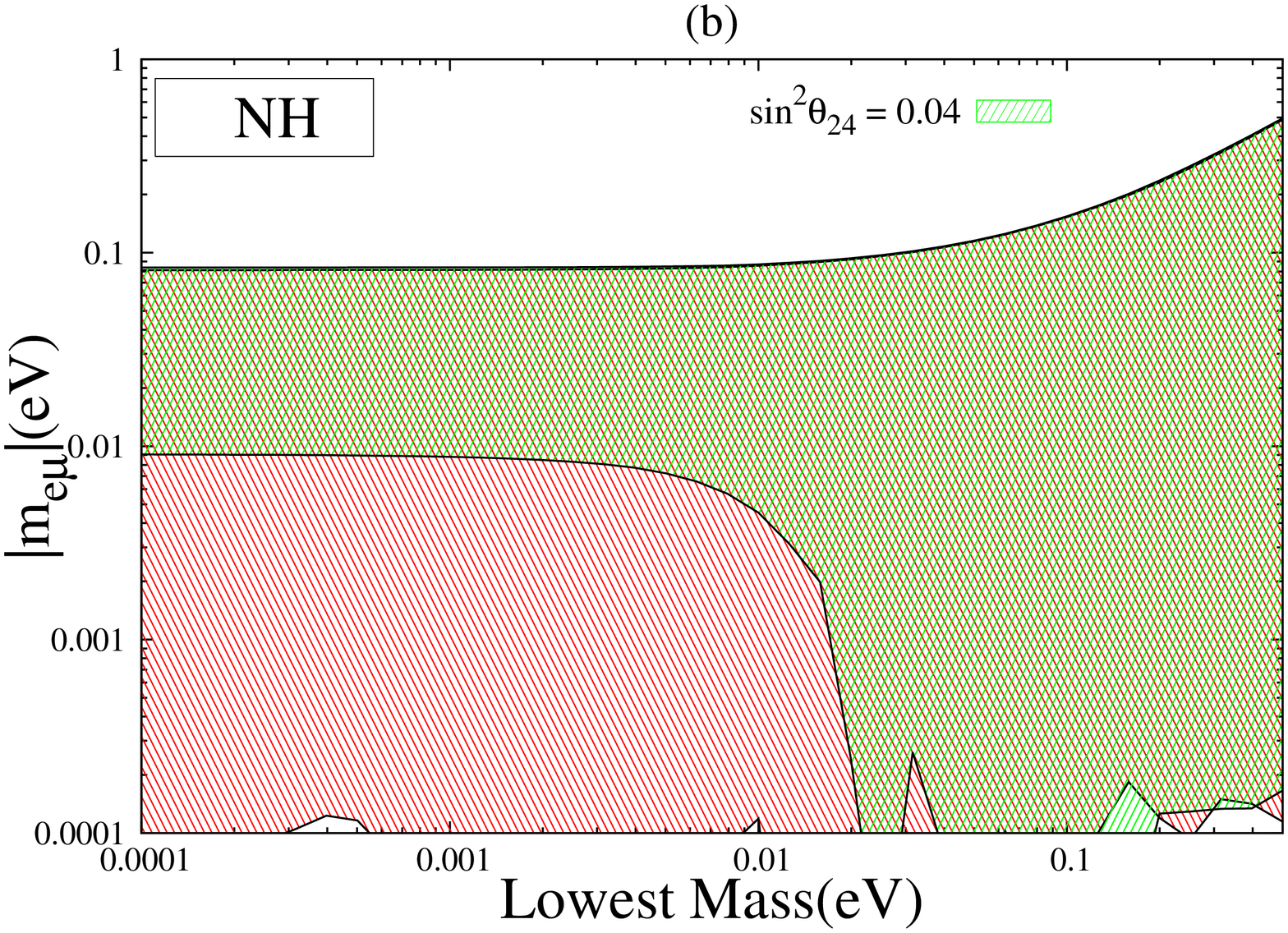}
\end{center}
Figure 4: Plots of $|m_{e\mu}|$ as a function of
the lowest mass $m_1$ for NH. Panel (a) correspond to the three generation case while (b) (red/light region) is for 3+1 case
and also for $s_{24}^2 = 0.04$ (green/dark region). All the parameters
are varied in their full 3$\sigma$ allowed range, the CP violating Dirac phases are varied from 0 to $2\pi$ and the Majorana phases are varied from 0 to $\pi$ unless
otherwise stated.
\end{figure}

\begin{figure}
\begin{center}
\includegraphics[width=0.33\textwidth,angle=270]{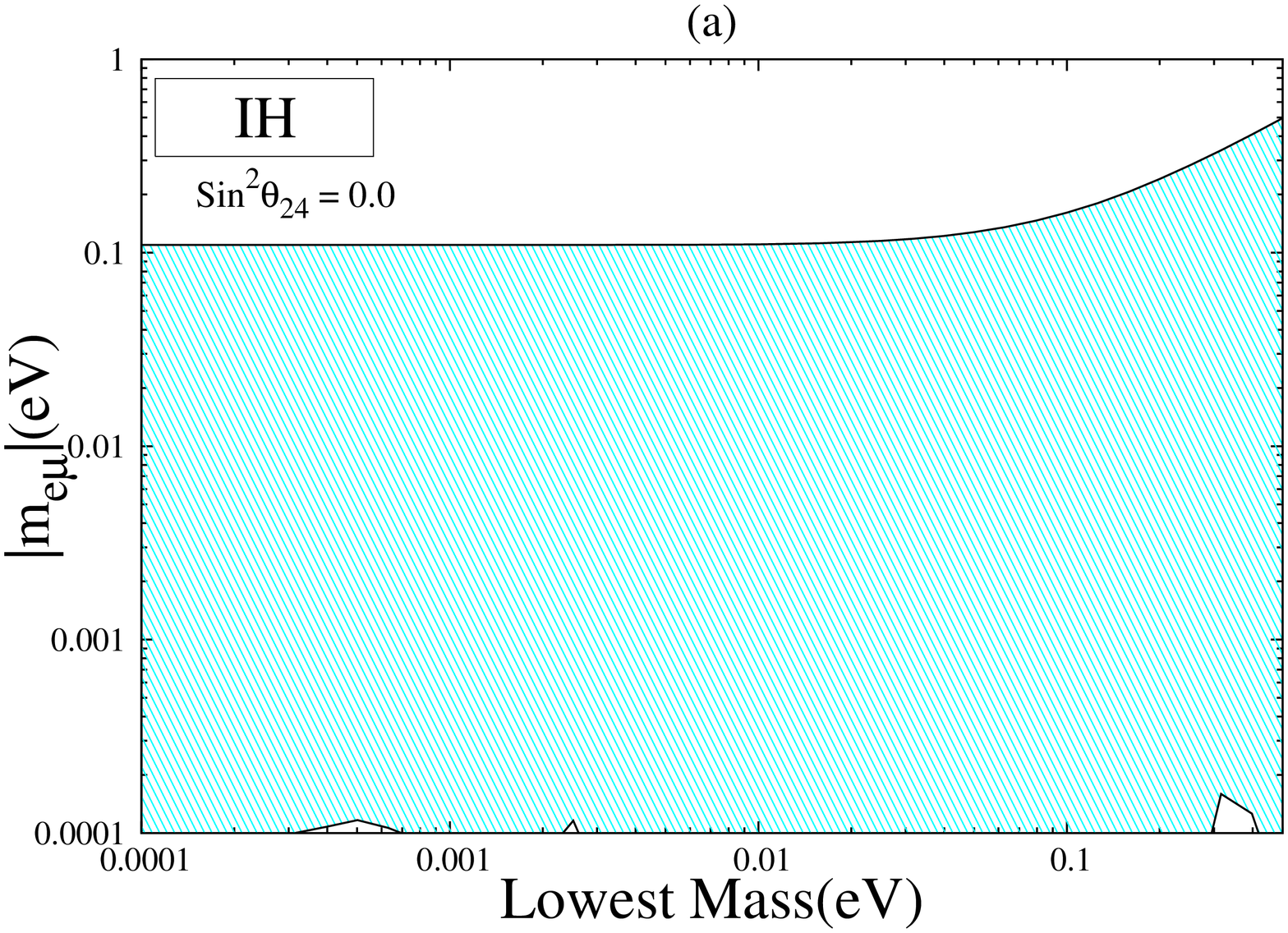}
\includegraphics[width=0.33\textwidth,angle=270]{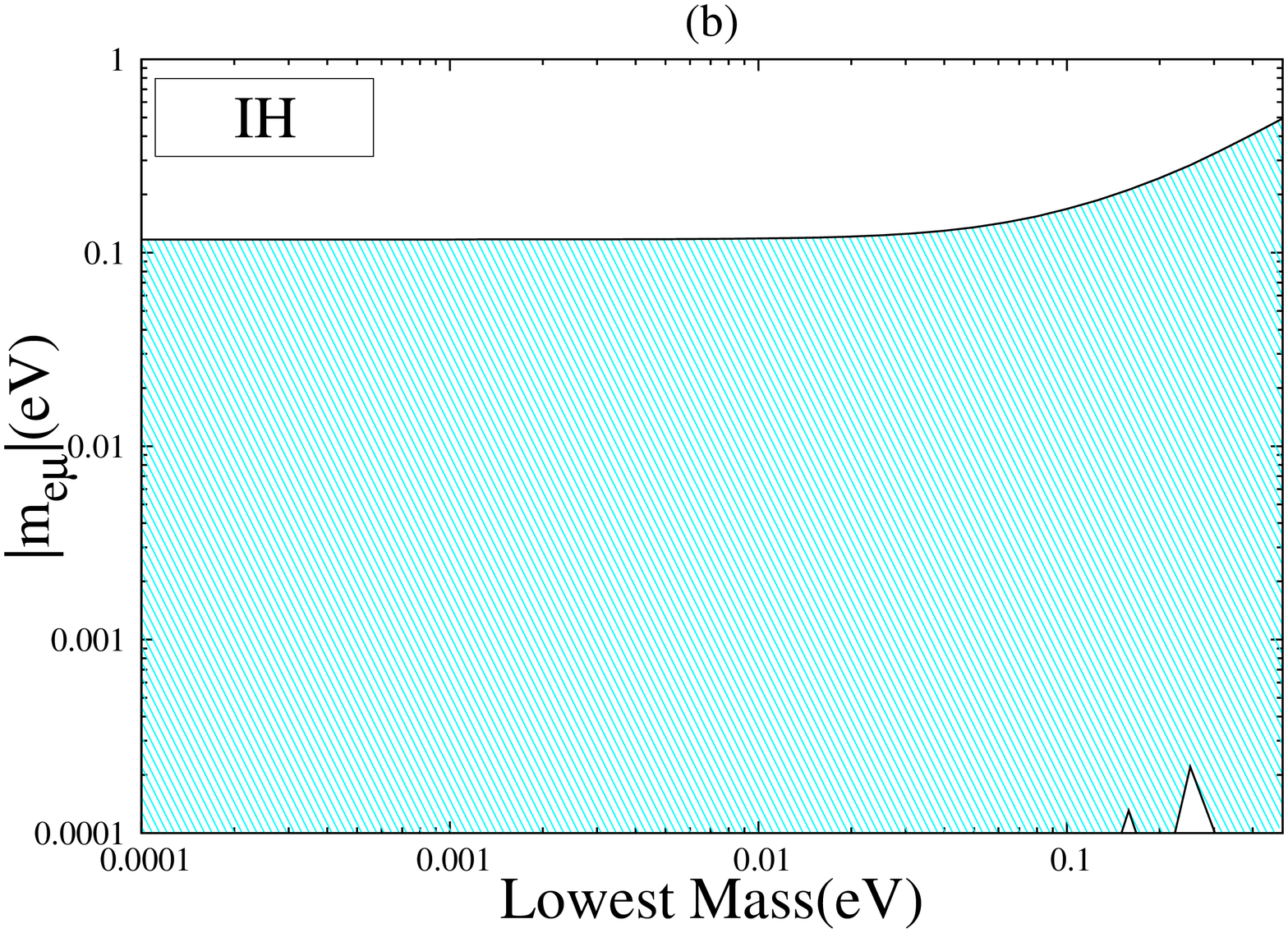}\\
\includegraphics[width=0.33\textwidth,angle=270]{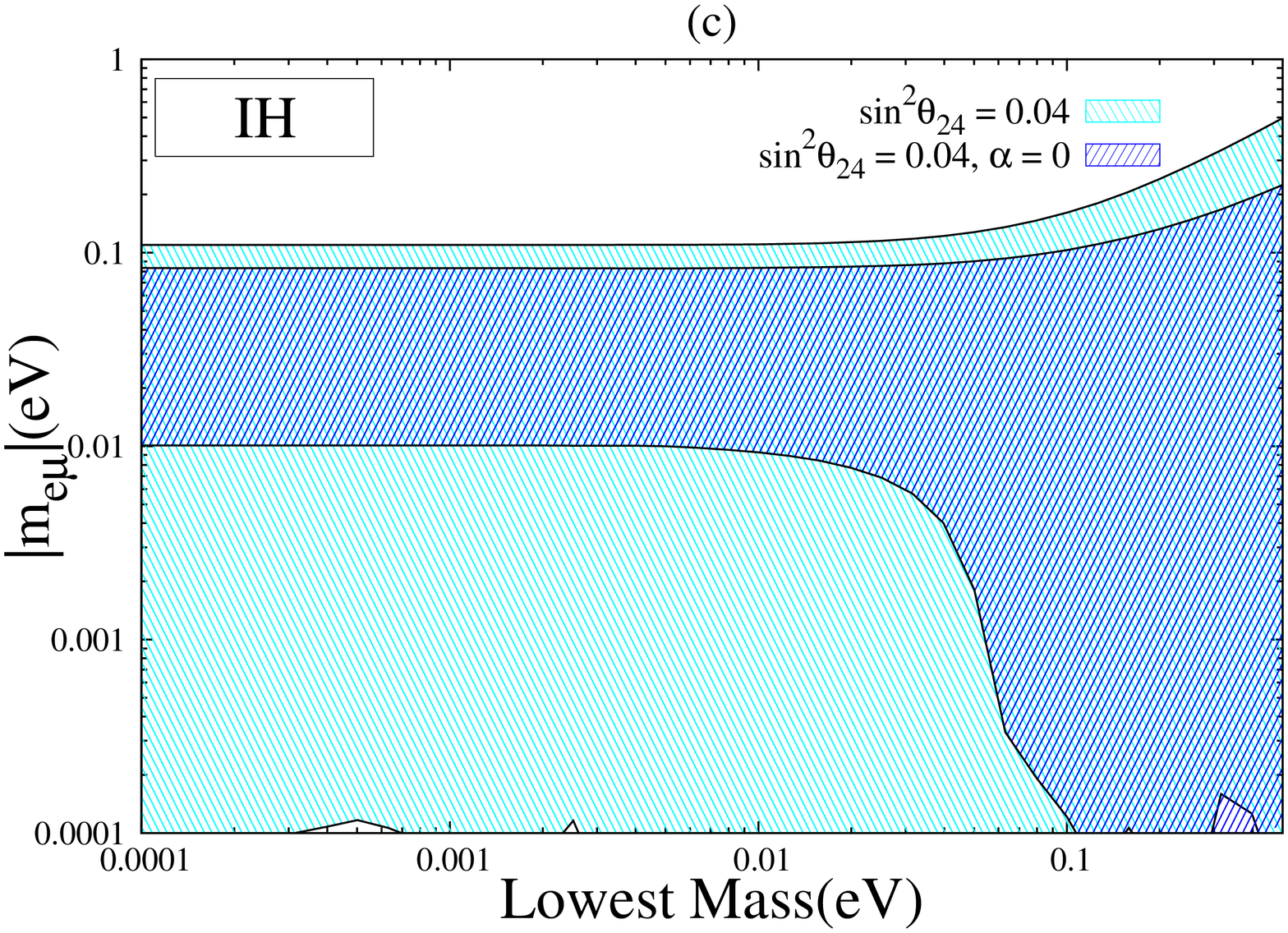}
\includegraphics[width=0.33\textwidth,angle=270]{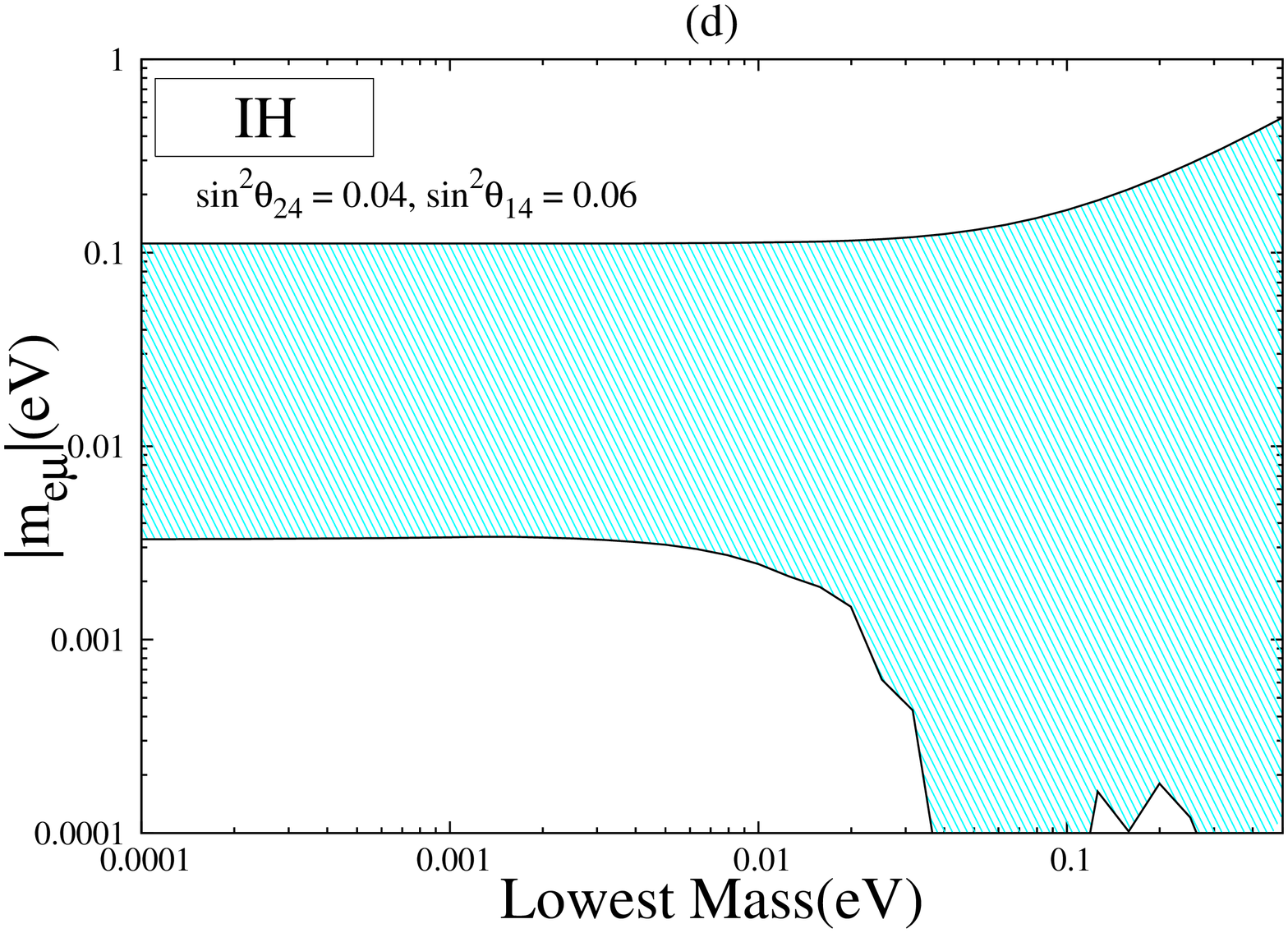}\\
\includegraphics[width=0.33\textwidth,angle=270]{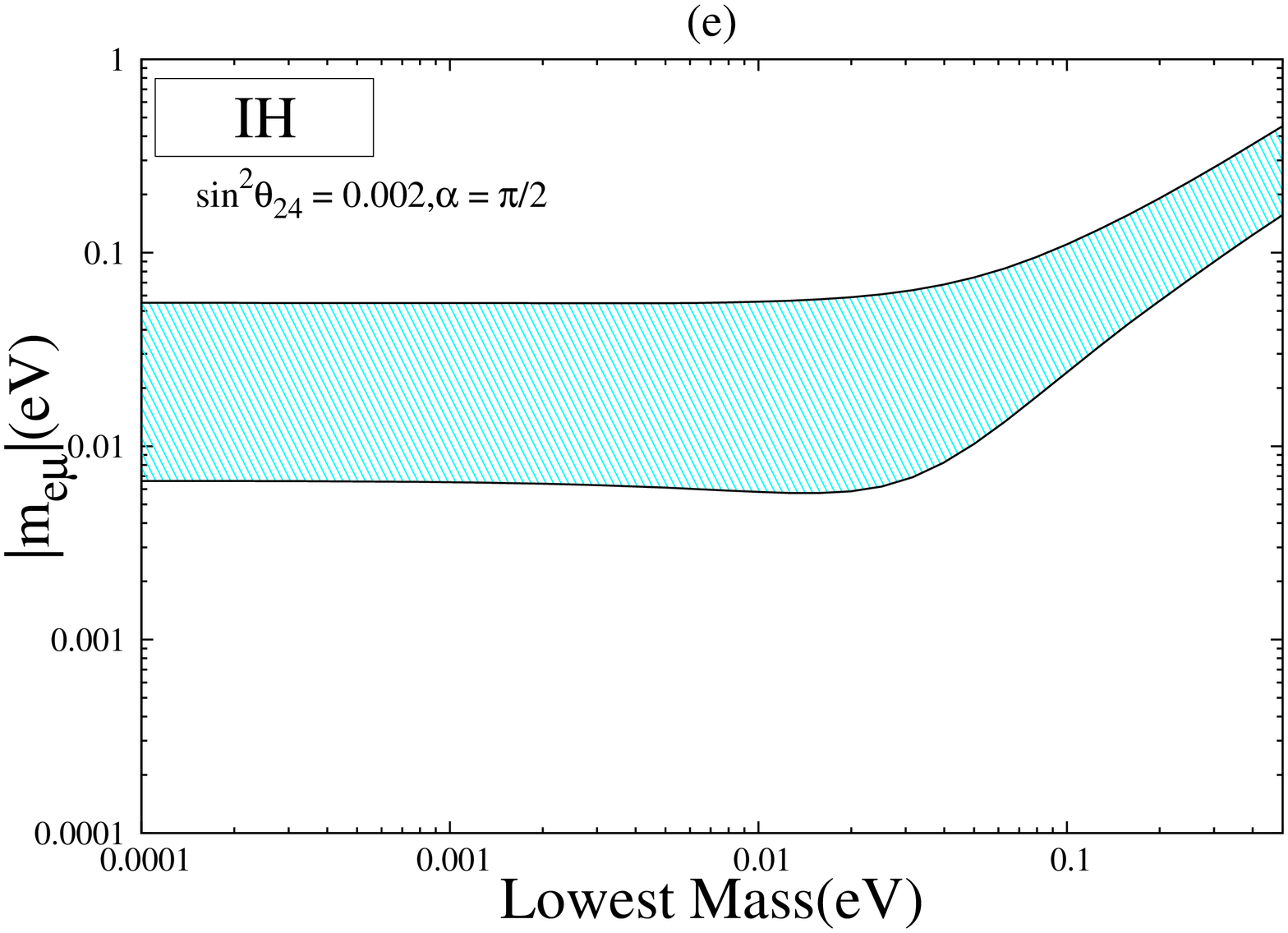}
\includegraphics[width=0.33\textwidth,angle=270]{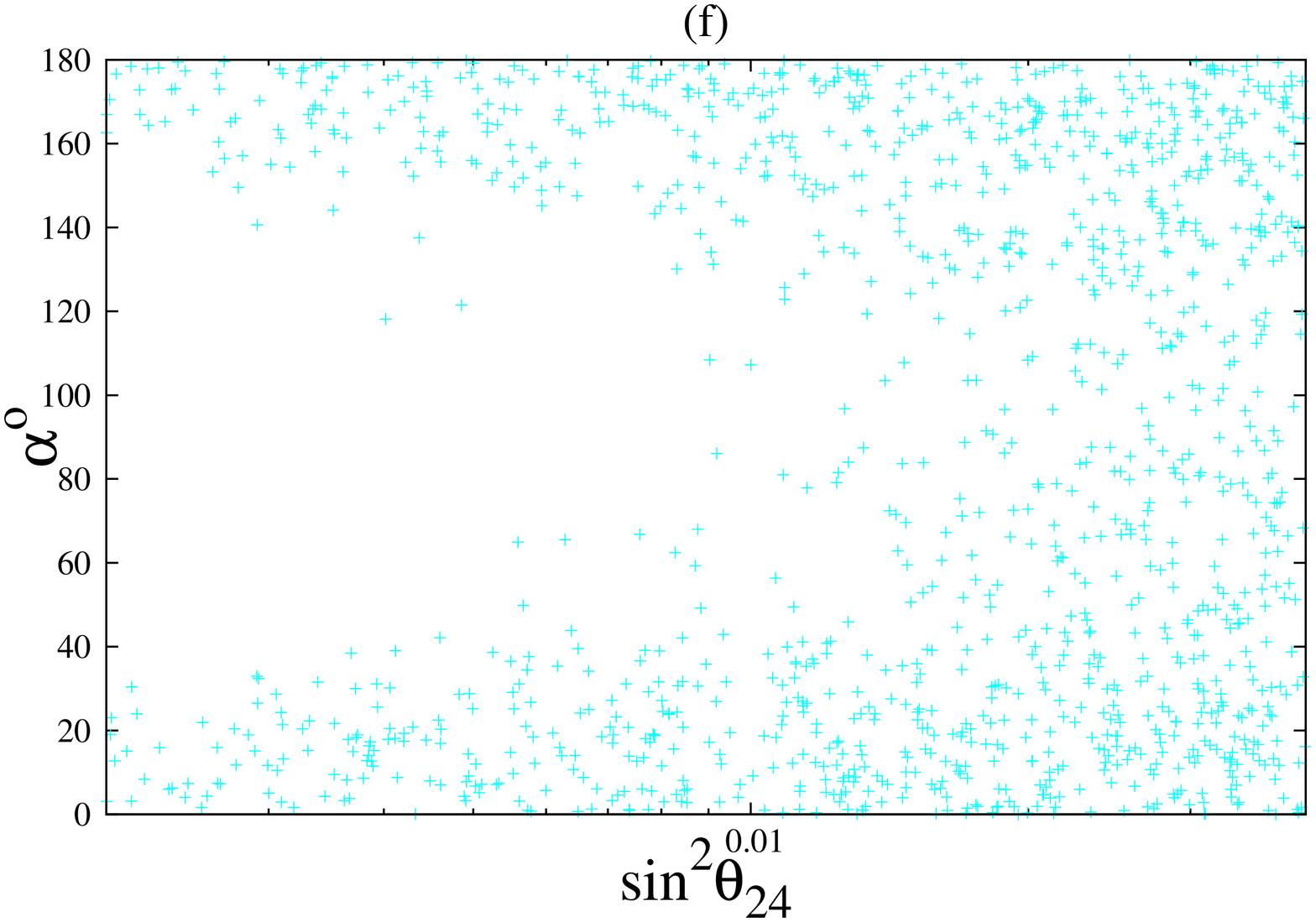}
\end{center}
Figure 5: Plots of $|m_{e\mu}|$ vs $m_3$ for inverted hierarchy for
(a) three generation case (b) 3+1 case with all parameters varied
randomly in their full range. Panel (c), (d) and (e) are for specific values of $s_{24}^2$ and
$\alpha$. The panel (f) shows the correlation between $\alpha$ and $s_{24}^2$ when all other parameters are randomly varied.
\label{fig5}
\end{figure}

For IH using the approximation Eq. (\ref{xih}) for the hierarchical
limit  we get the expression
\bea  \label{memih}
|m_{e \mu}| &\approx& |\sqrt{\Delta m_{13}^2}\{c_{12}s_{12}c_{23}(e^{2i\alpha}-1)-e^{i\delta_{13}}(c_{12}^2+s_{12}^2e^{2i\alpha})s_{23}\chi_{13}\lambda \\ \nonumber
&-& e^{i(\delta_{14}-\delta_{24})}\lambda^2
\chi_{14}\chi_{24}(c_{12}^2 -e^{2i\gamma}\sqrt{\xi}+e^{2i\alpha}s_{12}^2)\}|.
\eea
To see the order of magnitude of the
various terms we consider the case when Majorana phases vanish and the
Dirac phases assume  the value $\pi$. Then we get for vanishing $m_{e\mu}$,
\bea
 s_{23}\lambda\chi_{13}-\lambda^2(1-\sqrt{\xi})\chi_{14}\chi_{24}=0.
\eea
In panel (a) of Fig. 5 we display the plot of $|m_{e\mu}|$ with $m_3$ for the
3 generation scenario i.e for $\sin^2\theta_{24} =0$ for IH.
In panel (b) we consider the 3+1 case with all the parameters varying
in their allowed range. Note that in the small
$m_3$ limit (cf. Eq. 3.13) for $\alpha =0$ the leading order term vanishes.
For this case, for large active sterile mixing angle $\theta_{24}$, the
$\lambda^2$ term becomes large $\mathcal{O}$ (10$^{-1}$) and the cancellation with $\lambda$ term is not be possible.
When CP violating phase $\alpha$ is non zero, the leading order
term  can cancel the $\lambda^2$ term
even for large values of $s_{24}^2$.
These features are reflected in
panel (c) where we plot $|m_{e\mu}|$ for
$s_{24}^2 = 0.04$ and $ \alpha = 0$ (blue/dark region)
and by varying $\alpha$ in its full range (cyan/light region). As expected,
for $\alpha=0$, cancellation is not achieved for
smaller values of $m_3$.
Thus the condition $|m_{e\mu}|=0$ implies some correlation between
$m_3$ and $\alpha$ for IH.
Even if $\alpha$ is varied in its full range,
the absolute value of the
matrix element $|m_{e\mu}|$ can vanish only if the product $\chi_{14}\chi_{24}$ is small, i.e. $s_{14}^2$ and $s_{24}^2$ are simultaneously small .
This is because if they are large the $\lambda^2$ term becomes of the $\mathcal{O}$ (10$^{-1}$) and hence cancellation will not be possible.
This is seen from panel (d) where for
$s_{14}^2 = 0.06$ and $s_{24}^2 = 0.04$ the region where $m_3$ is small
gets disallowed.
Taking CP violating phase $\alpha=\pi/2$ makes the magnitude of
leading order term ($s_{12} c_{12}c_{23}\sqrt{\zeta}$) quite large and
smaller values of $\theta_{24}$ cannot give cancellation even for
large values of $m_3$ which can be seen from panel (e) of Fig. 5.
For the occurrence of cancellation $s_{24}^2$ has to be $\geq$ 0.01 for $\alpha=\pi/2$ as can be seen from panel (f) where we have
plotted the correlation between $\alpha$ and $s_{24}^2$ for $|m_{e\mu}| = 0$.


\subsection{The Mass Matrix element $m_{e\tau}$}
The mass matrix element $m_{e\tau}$ in the presence of an extra sterile neutrino is given by
\bea
m_{e\tau}&=&c_{14}c_{24}e^{i(2\gamma + \delta_{14})}m_4s_{14}s_{34}+m_3c_{14}s_{13}e^{i(2\beta+\delta_{13})}(-c_{24}s_{13}s_{14}s_{34}
e^{i(\delta_{14}-\delta_{13})}\\ \nonumber
&+&c_{13}(c_{23}c_{34}-e^{i\delta_{24}}s_{23}s_{24}s_{34}))+m_2s_{12}c_{13}c_{14}e^{2i\alpha}(c_{12}(-c_{34}s_{23}-c_{23}s_{24}s_{34}e^{i\delta_{24}}) \\ \nonumber
&+&s_{12}(-c_{13}c_{24}s_{14}s_{34}e^{i\delta_{14}}-e^{i\delta_{13}}s_{13}(c_{23}c_{34}-e^{i\delta_{24}}s_{23}s_{24}s_{34}))) \\ \nonumber
&+&m_1c_{12}c_{13}c_{14}(-s_{12}(-c_{34}s_{23}-c_{23}s_{24}s_{34}e^{i\delta_{24}})+c_{12}(-c_{13}c_{24}s_{14}s_{34}e^{i\delta_{14}}
-e^{i\delta_{13}}s_{13}\\ \nonumber
&&(c_{23}c_{34}-e^{i\delta_{24}}s_{23}s_{24}s_{34}))).
\eea
The elements $m_{e\tau}$ and $m_{e\mu}$ are related by
$\mu-\tau$   permutation symmetry
\begin{center}
$
P_{\mu\tau}=\left(
\begin{array}{cccc}
 1& 0 & 0 & 0 \\ 0& 0 &1& 0 \\ 0& 1 &0 & 0\\ 0 & 0 & 0 &1
\end{array}
\right)$,
\end{center}
in such a way that
\begin{center}$m_{e\tau}=P_{\mu\tau}^T m_{e\mu} P_{\mu\tau} . $ \end{center}
For three active neutrino case the mixing angle $\theta_{23}$ in the partner
textures linked by $\mu-\tau$ symmetry are related as
$\bar\theta_{23}=(\frac{\pi}{2} -\theta_{23})$.
However, in the 3+1 case the relation of $\theta_{23}$ between two textures related by this symmetry is not simple.
The active sterile mixing angles $\theta_{24}$ and $\theta_{34}$ are
also different in the textures
connected by $\mu-\tau$ symmetry and
are related as \cite{ggg}
\begin{equation}
\bar\theta_{12}= \theta_{12},
~~~ \bar\theta_{13} = \theta_{13},
~~~ \bar\theta_{14} = \theta_{14},
\end{equation}
\begin{equation} \label{mutau1}
\sin{\bar\theta_{24}} =  \sin\theta_{34} \cos{\theta_{24}},
\end{equation}
\begin{equation}
\sin {\bar\theta_{23}}
=\frac{\cos{\theta_{23}}\cos{\theta_{34}}-\sin{\theta_{23}}\sin{\theta_{34}}\sin{\theta_{24}}}{\sqrt{1-\cos{\theta_{24}^2}\sin{\theta_{34}^2}}},
\end{equation}
\begin{equation}
\sin {\bar\theta_{34}} \label{mutau2}
=\frac{\sin{\theta_{24}}}{\sqrt{1-\cos{\theta_{24}^2}\sin{\theta_{34}^2}}}.
\end{equation}
Due to these relations the  behaviour of $m_{e\mu}$  is different from that of
$m_{e\tau}$ unlike in three active neutrino case
where the plots of these two elements were same except for $\theta_{23}$
which differed in octant for the two cases.\\
It is found that in the limit of small $\theta_{24}$
the two active sterile mixing angles $\bar\theta_{24} \approx \theta_{34}$ from Eq (\ref{mutau1}).
The same can be seen from Eq (\ref{mutau2}) which gives
$\bar\theta_{34} \approx \theta_{24}$
for smaller values of the mixing angle $\theta_{34}$.
Thus, for these cases
the behaviour shown by $\theta_{24}$ in $m_{e\mu}$ ($m_{\mu\mu}$) is same
as shown by $\theta_{34}$ in $m_{e\tau}$ ($m_{\tau\tau}$).


In the limit of vanishing active sterile mixing angle $\theta_{34}$ this element becomes
\bea
m_{e \tau}= c_{14}(m_{e \tau})_{3 \nu}. \nonumber
\eea
Using the approximation in Eq (\ref{xnh}) for NH the above element  can be expressed as,
\bea
|m_{e\tau}|&\approx& |\sqrt{\Delta m_{23}^2}\{- s_{12}s_{23}c_{12}c_{34}\sqrt{\zeta}e^{2i\alpha}+\lambda(c_{23}c_{34}e^{i(2\beta+\delta_{13})}\chi_{13}-c_{23}c_{34}s_{12}^2
\\ \nonumber &&\chi_{13}\sqrt{\zeta}e^{i(2\alpha+\delta_{13})}+e^{i(2\gamma+\delta_{14})}\sqrt{\xi}s_{34}\chi_{14}-
e^{i(2\alpha+\delta_{14})}\sqrt{\zeta}s_{12}^2s_{34}\chi_{14}-c_{12}c_{23}\\ \nonumber &&e^{i(2\alpha+\delta_{24})}s_{12}s_{34}\chi_{24}\sqrt{\zeta})
-e^{i(\delta_{13}+\delta_{24})}(e^{2i\beta}-e^{2i\alpha}s_{12}^2\sqrt{\zeta})s_{23}s_{34}\chi_{13}\chi_{24}\lambda^2\}|.
\eea
For the case of vanishing Majorana phases and Dirac phases having the value $\pi$, this element can vanish when
\bea \label{metau:fixedphase}
&-&c_{12}c_{34}\sqrt{\zeta}s_{12}s_{23}-(1-\sqrt{\zeta}s_{12}^2)s_{23}s_{34}\lambda^2\chi_{13}\chi_{24}+\lambda(-c_{23}c_{34}\chi_{13}+\\
\nonumber && \sqrt{\zeta}c_{23}s_{12}(c_{34}s_{12}\chi_{13}+c_{12}s_{34}\chi_{24})+s_{12}^2s_{34}\chi_{14}\sqrt{\zeta}-\sqrt{\xi}s_{34}\chi_{14})
=0.
\eea
For a vanishing active sterile mixing angle $\theta_{34}$
one recovers the 3 generation case. In this limit, from Eq. (\ref{metau:fixedphase}) one observes that the leading order term and the term with $\lambda$ are of the same
order  $\sim \mathcal{O}$ (10$^{-2}$) while the $\lambda^2$ term vanishes and hence
cancellation is possible excepting for very low values of the lightest mass. We can see this in panel (a) of Fig. 6.
In panel (b) (red/light region) all the parameters are varied randomly (3+1 case) and
cancellation is seen to be possible over the whole range of $m_1$.
\begin{figure}
\begin{center}
\includegraphics[width=0.33\textwidth,angle=270]{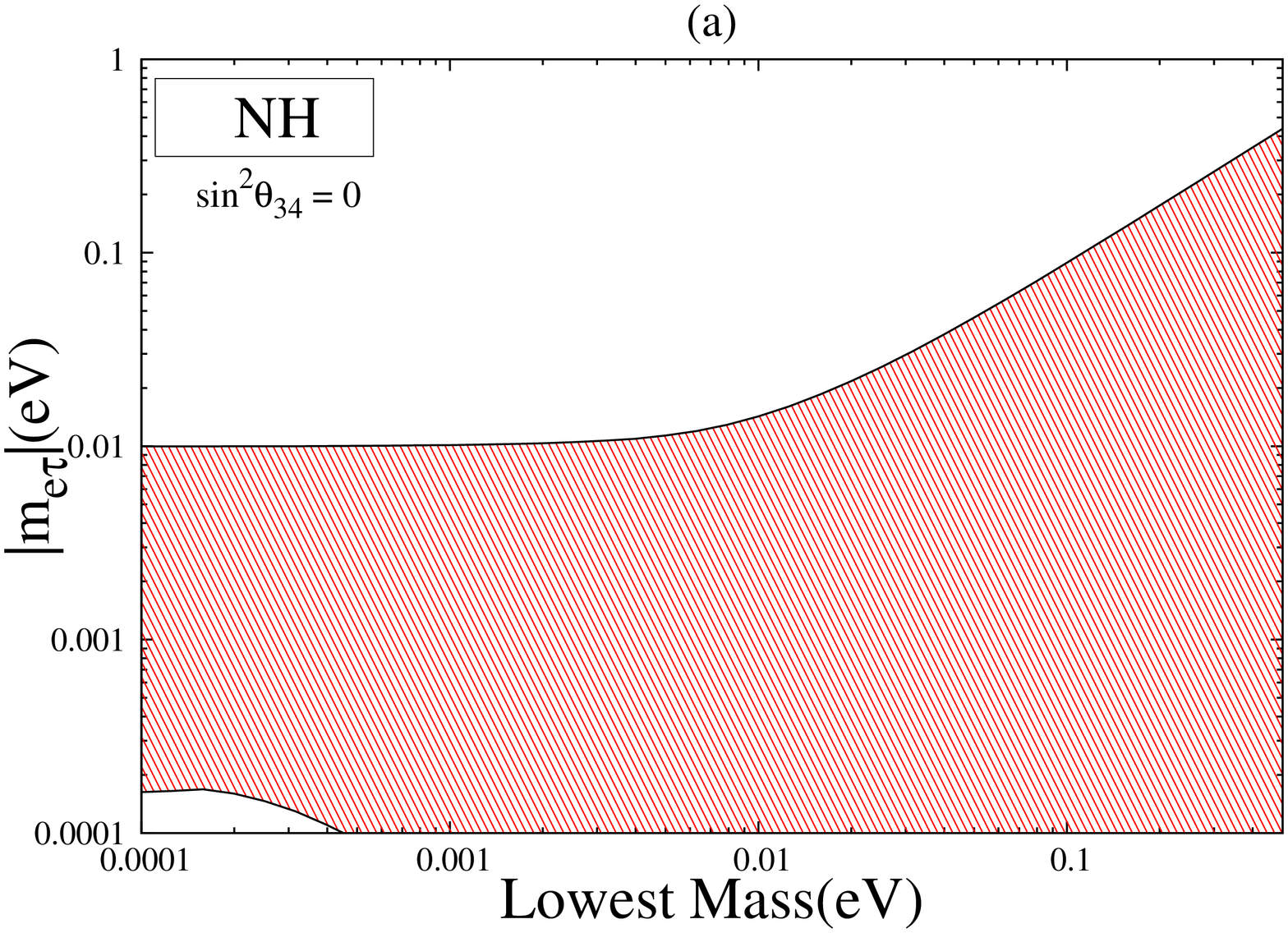}
\includegraphics[width=0.33\textwidth,angle=270]{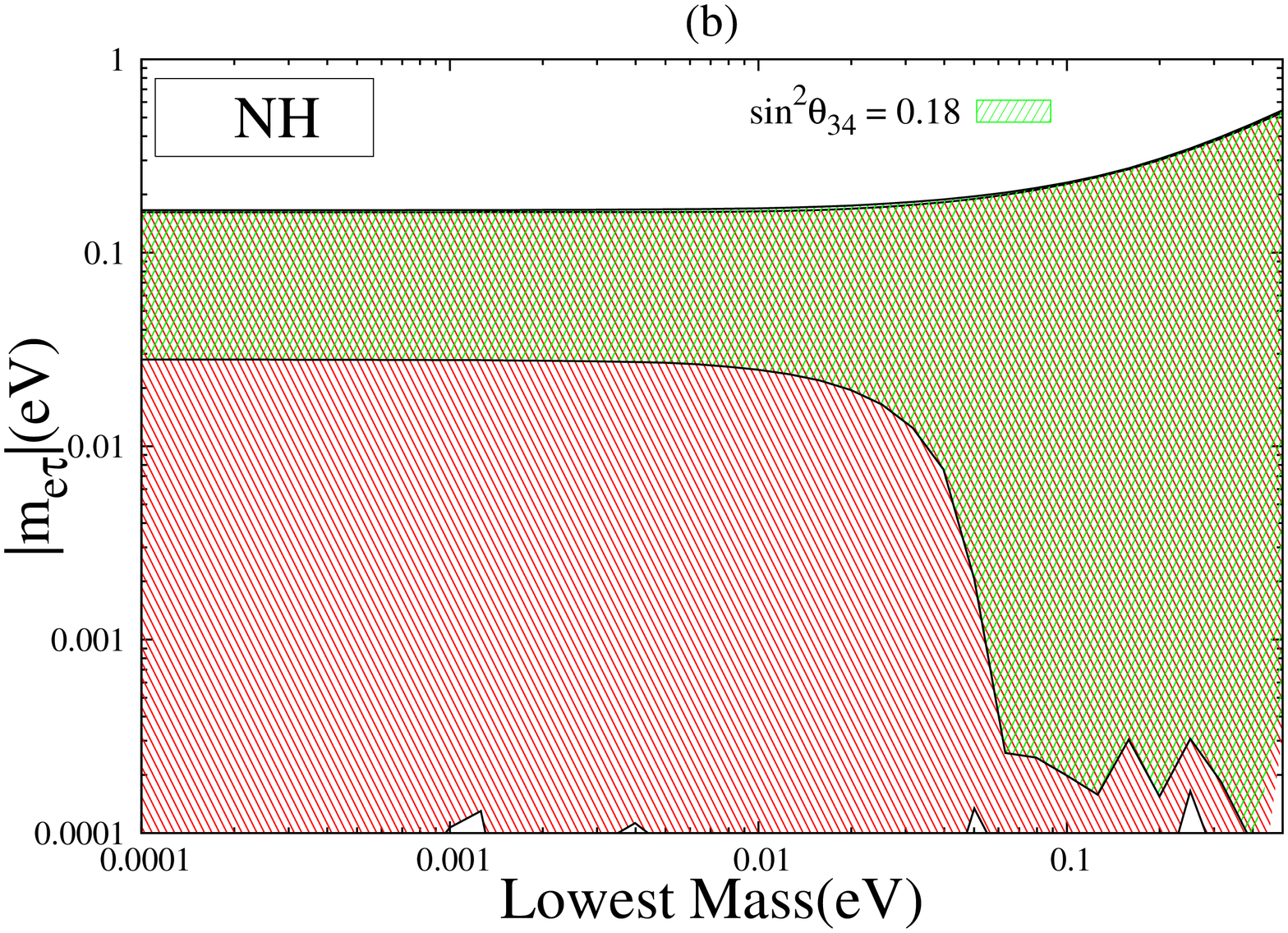}
\end{center}
Figure 6: Plots of $|m_{e\tau}|$ for normal hierarchy with lowest mass $m_1$.
 The panel (a) corresponds to three generation case. In (b) (red/light region) all the parameters are varied in their full allowed range and
 the green/dark region is for $s_{34}^2 = 0.18$ with all the other parameters covering their full range.
\label{fig6}
\end{figure}
In panel (b) (green/dark region) we also plot the element $|m_{e\tau}|$ for the upper limit of
$s_{34}^2 = 0.18$. In this case there is no cancellation for very low values of
the smallest mass. This is because
when $s_{34}^2$ is large,  the $\lambda$ term containing $\xi$ becomes large $\mathcal{O}$ (1) and there will be no cancellation.\\
For inverted hierarchy the element $m_{e\tau}$ using the approximation in Eq (\ref{xih}) becomes
\bea \nn
|m_{e\tau}|&\approx& |\sqrt{\Delta m_{13}^2}\{c_{12}c_{34}s_{12}s_{23}(-e^{2i\alpha}+1)+e^{i(\delta_{13}+\delta_{24})}
(c_{12}^2+e^{2i\alpha}s_{12}^2)s_{23}s_{34}\lambda^2\chi_{13}\chi_{24}\\ \nonumber
&-& \lambda(c_{23}c_{34}\chi_{13}e^{i\delta_{13}}(c_{12}^2+e^{2i\alpha}s_{12}^2)+e^{i\delta_{14}}s_{34}\chi_{14}(c_{12}^2+e^{i\alpha}s_{12}^2)\\
&-& e^{i(2\gamma+\delta_{14})}s_{34}\chi_{14}\sqrt{\xi}+c_{12}c_{23}s_{12}s_{34}\chi_{24}e^{i\delta_{24}}(e^{2i\alpha}-1))\}|
\eea
In the limit of vanishing Majorana phases and Dirac CP violating phases equal to $\pi$ this element becomes negligible when
\bea
\lambda(c_{23}c_{34}\chi_{13}+s_{34}\chi_{14}-s_{34}\chi_{14}\sqrt{\xi})
+s_{23}s_{34}\chi_{13}\chi_{24}\lambda^2 =0.
\eea
In panel (a) of Fig. 7 the three generation case is reproduced by putting $s_{34}^2=0$ and in (b) all the parameters are varied in their allowed range (3+1 case). In both the
figures we can see that cancellation is permissible over the whole range of $m_3$ considered.
When the CP violating phase $\alpha=0$ we see that the leading order term ($\sin2\theta_{12}s_{23}c_{34}$) vanishes and
as a result for large values of $s_{34}^2$ the cancellation is not possible
because the term with coefficient $\lambda$ becomes large ($\mathcal{O}$(10$^{-1}$)).
\begin{figure}
\begin{center}
\includegraphics[width=0.33\textwidth,angle=270]{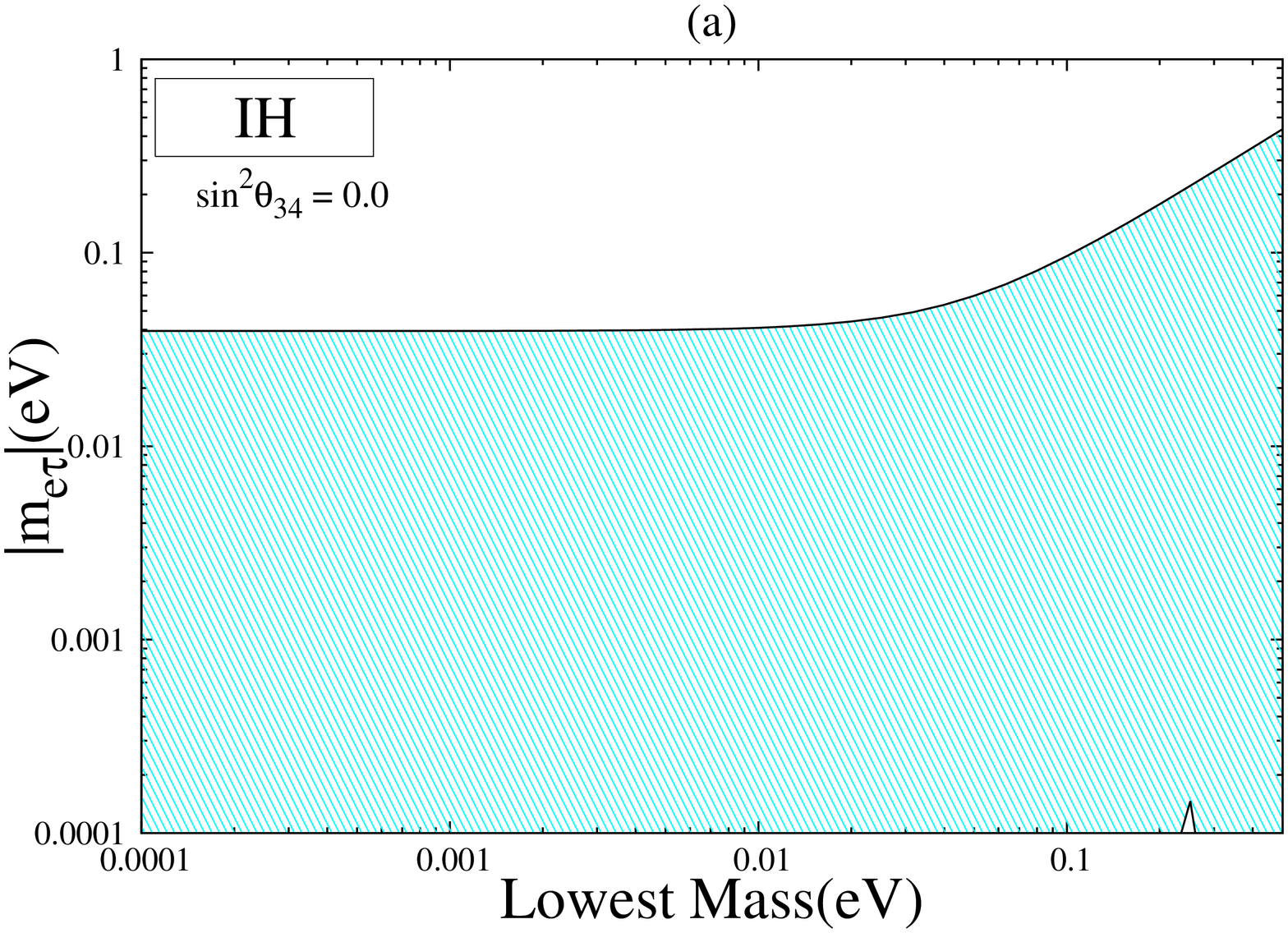}
\includegraphics[width=0.33\textwidth,angle=270]{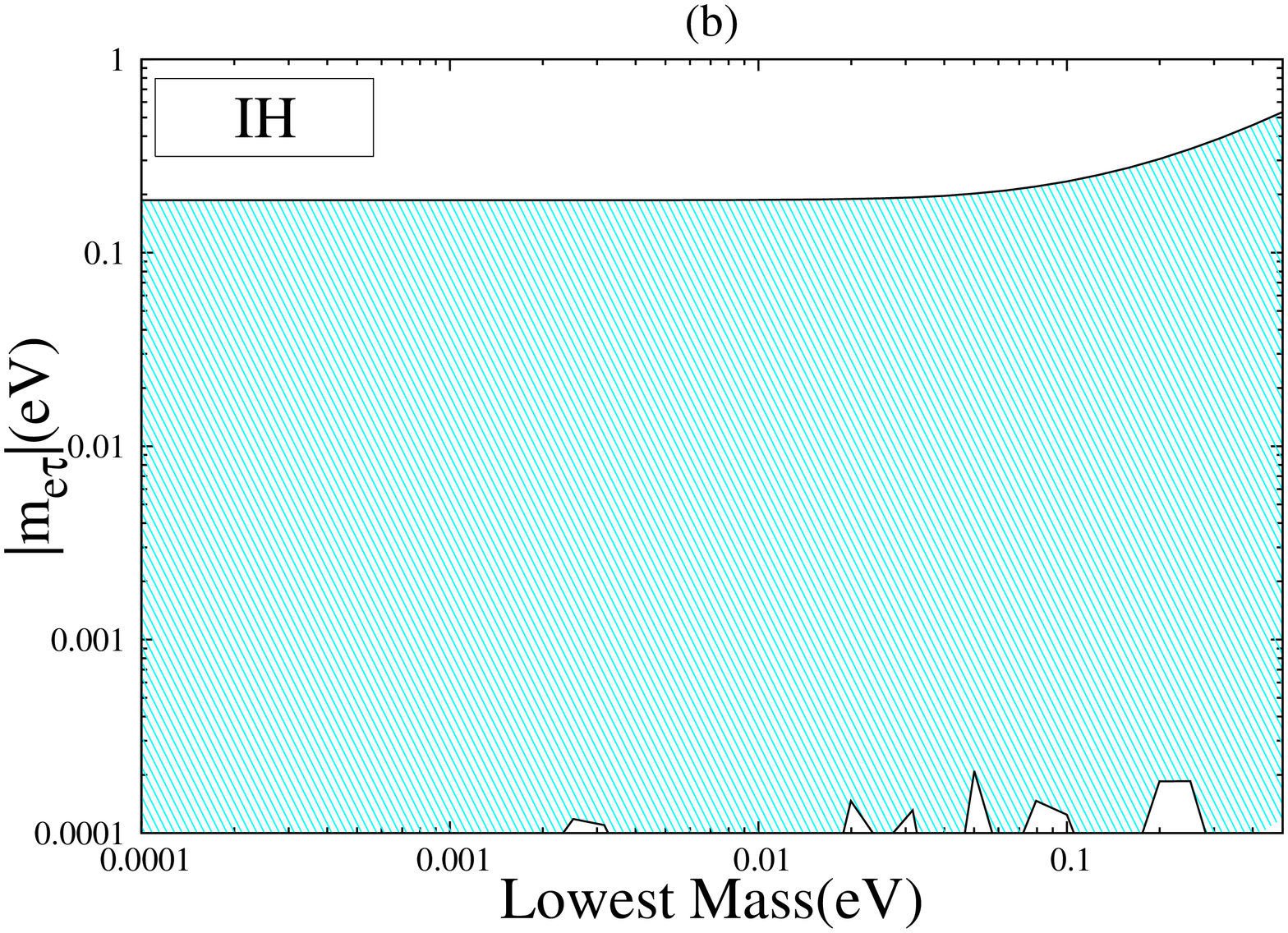}\\
\includegraphics[width=0.33\textwidth,angle=270]{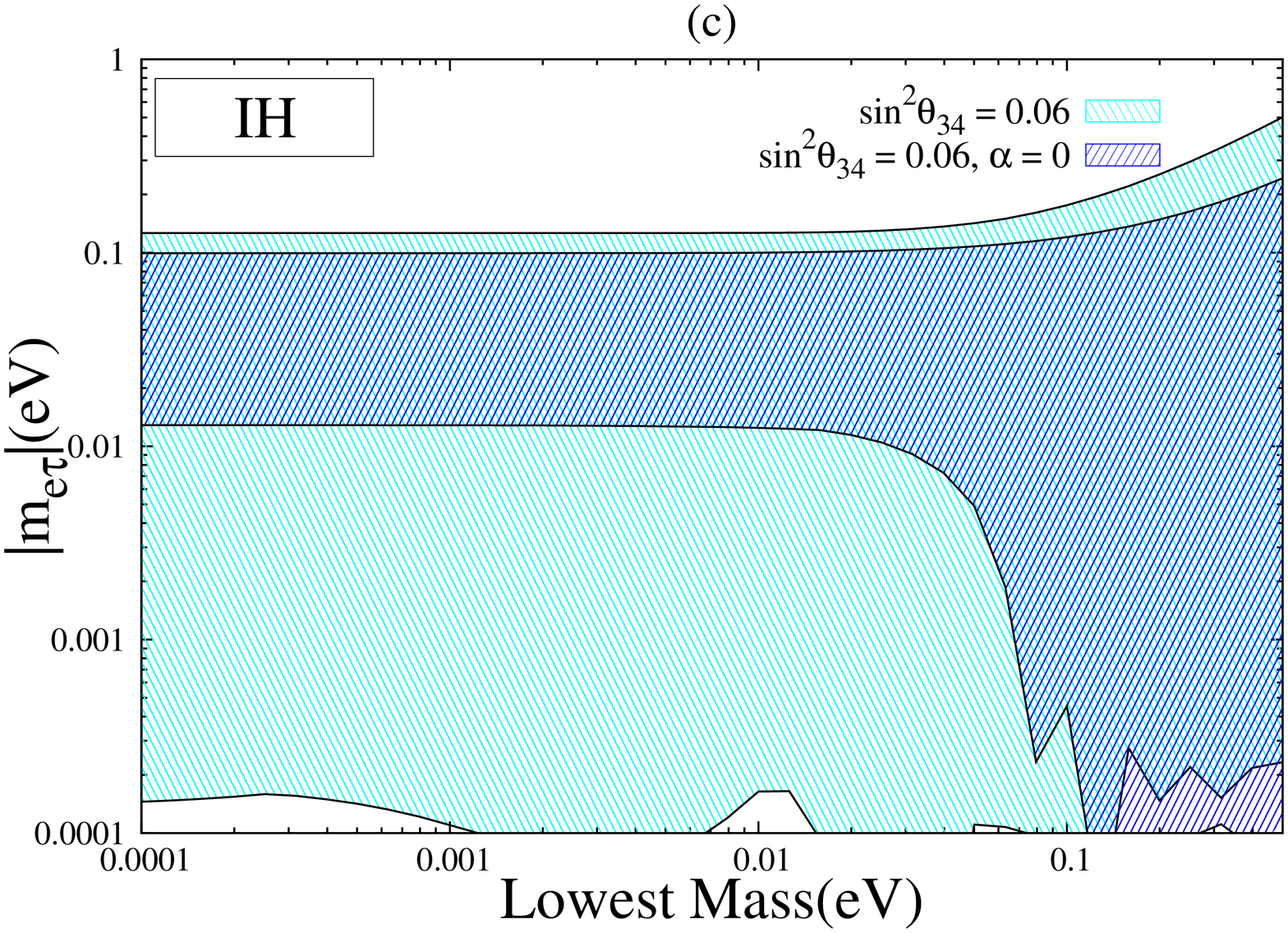}
\includegraphics[width=0.33\textwidth,angle=270]{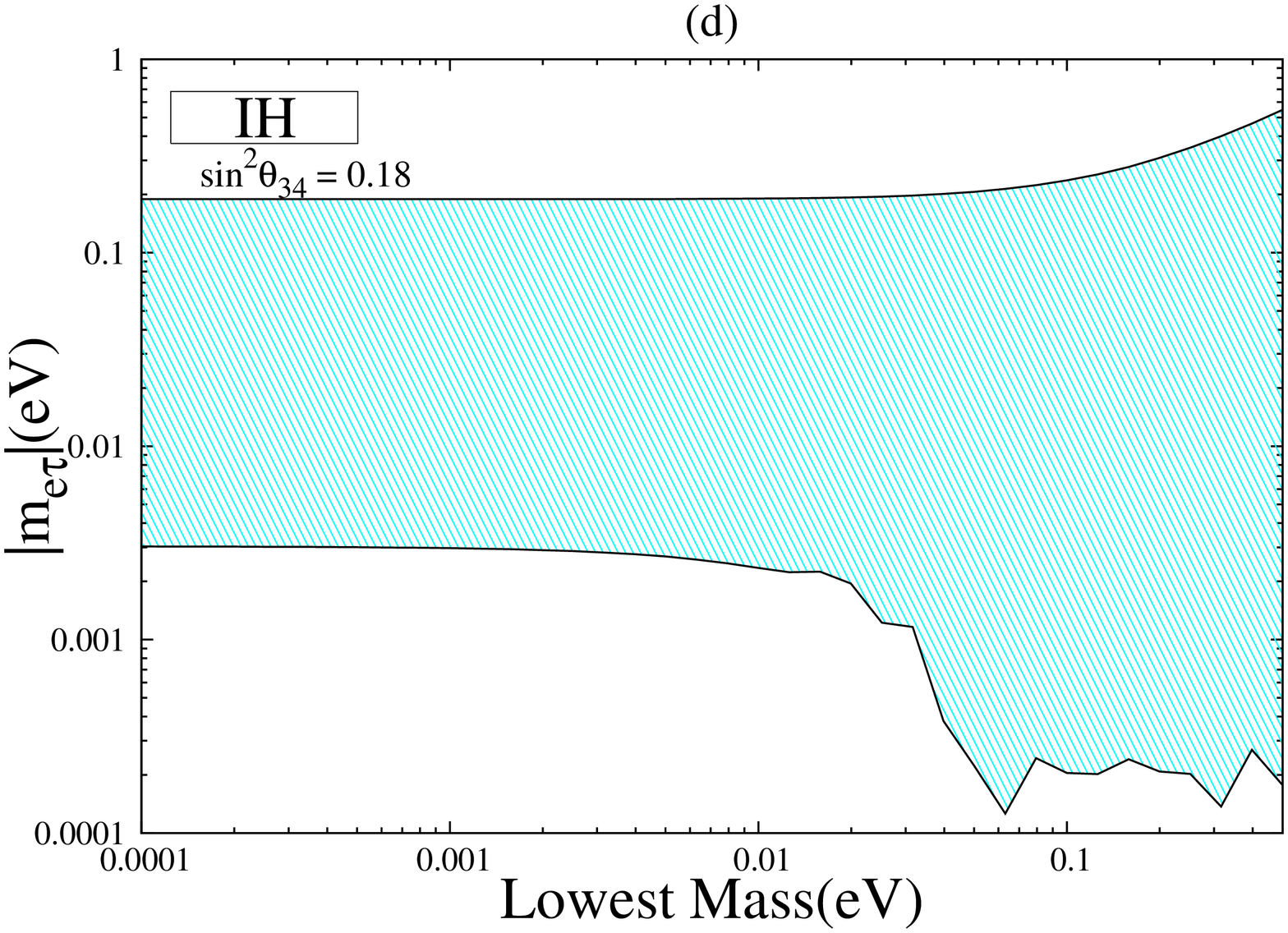}\\
\includegraphics[width=0.33\textwidth,angle=270]{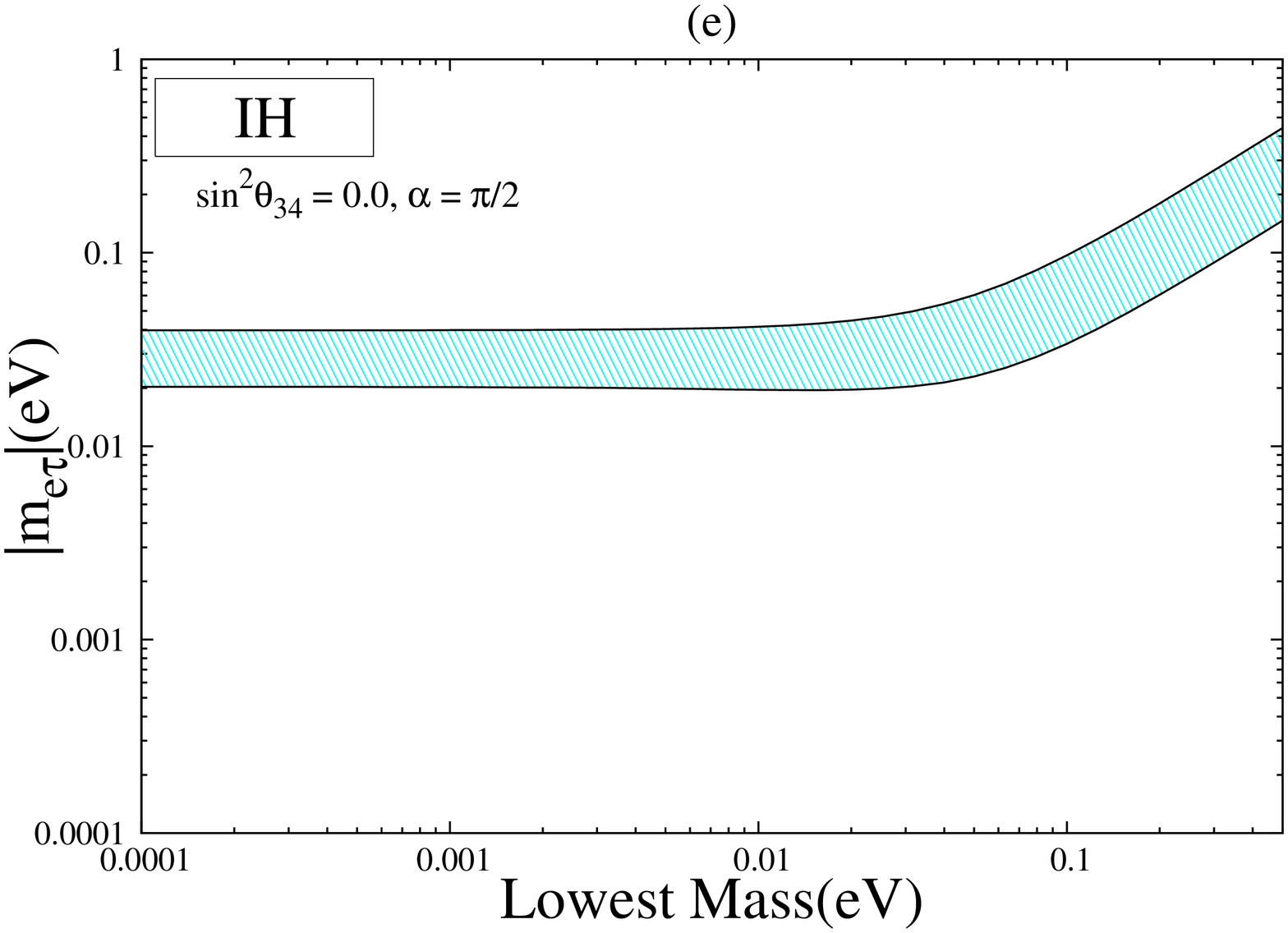}
\includegraphics[width=0.33\textwidth,angle=270]{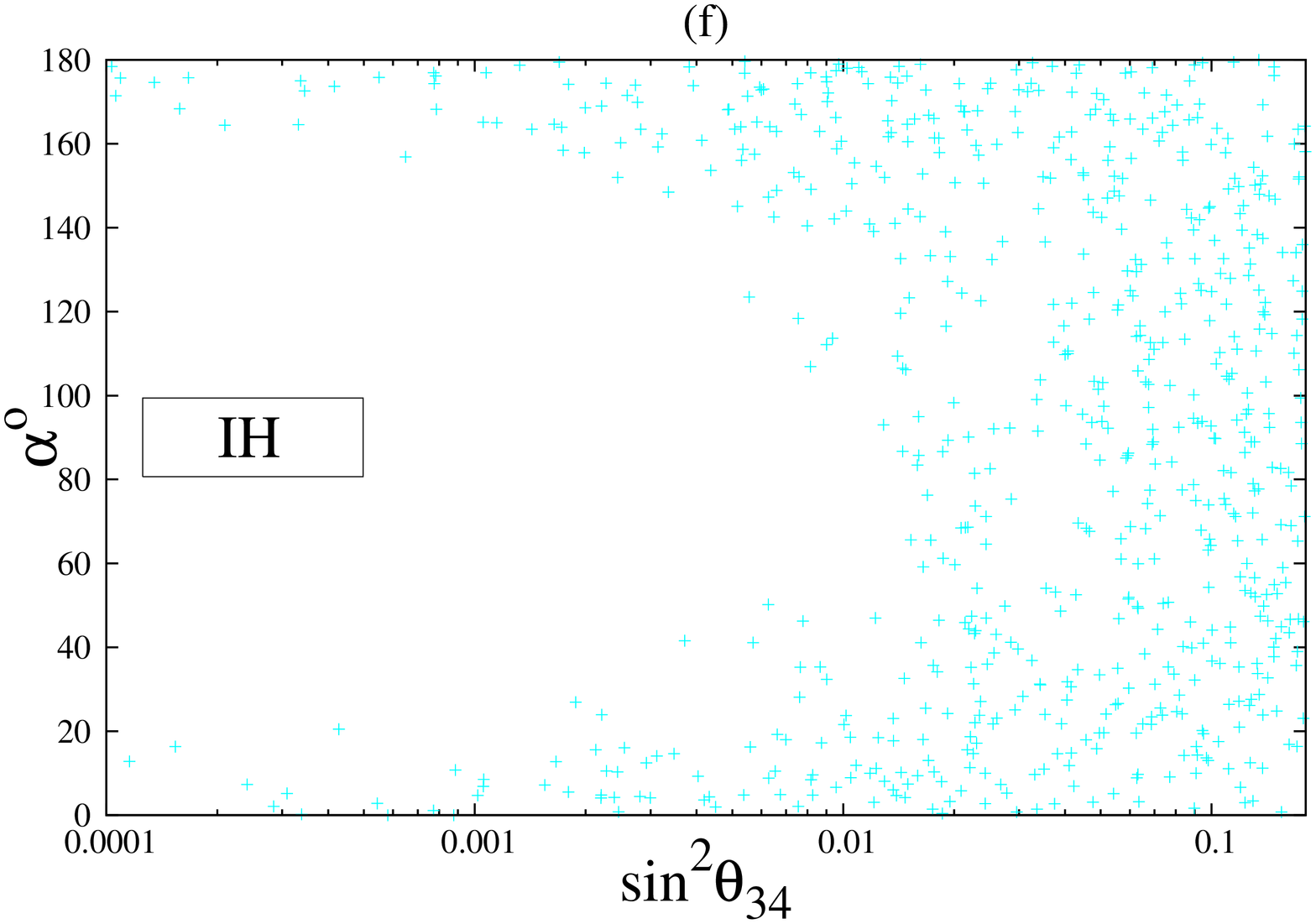}
\end{center}
Figure 7: Plots of $|m_{e\tau}|$ for inverted hierarchy with lowest mass $m_3$.
The panel (a) corresponds to three generation case. In (b) all the parameters are varied in their full allowed range (3+1).
The panel (c), (d) is for specific value of $\theta_{34}$  and $\alpha$ with all the other parameters covering their full range.
The panel (f) shows correlation between $\alpha$ and $s_{24}^2$.
\label{fig7}
\end{figure}
For non zero values of the CP violating phase $\alpha$ this leading order term is non zero
and its contribution will be significant. So in this case high values of $\theta_{34}$
are also allowed because now the leading order and the term with coefficient $\lambda$ will
be of same magnitude. When we fix $s_{34}^2 = 0.06$ and $\alpha = 0$ the region where $m_3$
is small is disallowed (panel (c) blue/dark region) but when $\alpha$ varies within its full range the
disallowed regions become allowed (panel (c) cyan/light region).
When $s_{34}^2$ approaches its upper limit, the $\lambda$ term having $\xi$ becomes very
large and cancellation is not possible even for non zero values
of  $\alpha$ which can be seen from panel (d).
However, when $\alpha=\pi/2$, very
small values of $s_{34}^2$ cannot give cancellation as the leading order term becomes large (panel (e)).
$s_{34}^2$ has to be $\geq$ 0.01 for the term to vanish which can be seen from panel (f) where we plotted the correlation between $\alpha$
and $s_{34}^2$ for $|m_{e\tau}| = 0$.

\subsection{The Mass Matrix element $m_{\mu\mu}$}
The (2,2) diagonal entry in neutrino mass matrix is given as
\bea
m_{\mu \mu} &=& e^{2 i(\delta_{14} - \delta_{24} + \gamma)} c_{14}^2 m_4 s_{24}^2 \\ \nonumber
&+& e^{2 i (\delta_{13} + \beta)} m_3(c_{13} c_{24} s_{23} - e^{i(\delta_{14} - \delta_{13} -\delta_{24})} s_{13} s_{14} s_{24})^2 \\ \nonumber
&+& m_1 \{-c_{23} c_{24} s_{12} + c_{12}(-e^{i \delta_{13}} c_{24} s_{13} s_{23} - e^{i(\delta_{14} - \delta_{24})} c_{13} s_{14} s_{24})\}^2 \\ \nonumber
&+& e^{2 i \alpha} m_2 \left\{c_{12} c_{23} c_{24} + s_{12}(-e^{i \delta_{13}} c_{24} s_{13} s_{23} - e^{i(\delta_{14} - \delta_{24})} c_{13} s_{14} s_{24})\right\}^2
\eea
This expression reduces to its three generation case if the
mixing angle $\theta_{24}$ vanishes.
Also we can see from the expression that there is no
dependence on the
mixing angle $\theta_{34}$.
Using the approximation in Eqs. (\ref{xnh})
this element can be simplified to the form
\bea
 |m_{\mu \mu}| &\approx&|\sqrt{\Delta m_{23}^2}\{ c_{12}^2 c_{23}^2 e^{2 i \alpha} \sqrt{\zeta} + e^{i(\delta_{13} + 2 \beta)} s_{23}^2 \\ \nonumber
 &-& 2 \lambda c_{12} c_{23} e^{i(\delta_{13} + 2 \alpha)} \sqrt{\zeta} s_{12} s_{23} \chi_{13} \\ \nonumber
 &+& \lambda^2\{e^{2i(\delta_{13} +\alpha)}\sqrt{\zeta}s_{12}^2s_{23}^2\chi^2_{13}+e^{i(\delta_{14}-\delta_{24})}(e^{i(2\gamma +
\delta_{14} -\delta_{24})}\sqrt{\xi}\chi_{24} \\ \nonumber
&-& 2 e^{2i\alpha}\sqrt{\zeta}c_{12}c_{23}s_{12}\chi_{14})\chi_{24}\}\}|.
\eea
For the case of Majorana CP phases having the value 0
and the Dirac phases having the value  $\pi$, this element vanishes when
\bea \label{mmmphase}
&& s_{23}^2 + c_{12}^2 c_{23}^2 \sqrt{\zeta} + c_{12} s_{12} \sin 2 \theta_{23}\sqrt{\zeta} \lambda \chi_{13} \\ \nonumber
&+&\lambda^2(s_{12}^2s_{23}^2\sqrt{\zeta}\chi_{13}^2-c_{23}\sin2\theta_{12}\sqrt{\zeta}\chi_{14}\chi_{24}+\sqrt{\xi}\chi_{24}^2)=0.
\eea
\begin{figure}
\begin{center}
\includegraphics[width=0.33\textwidth,angle=270]{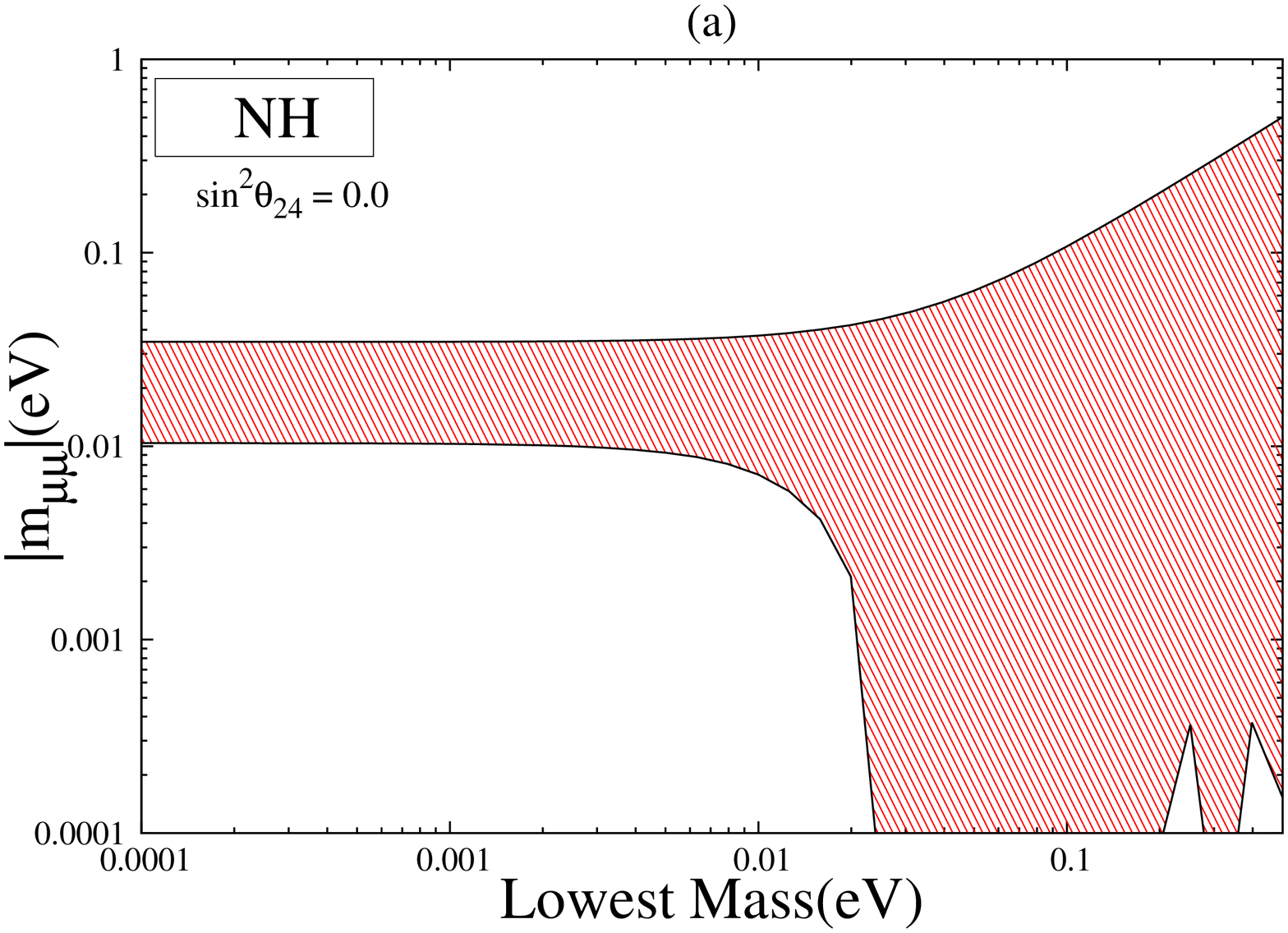}
\includegraphics[width=0.33\textwidth,angle=270]{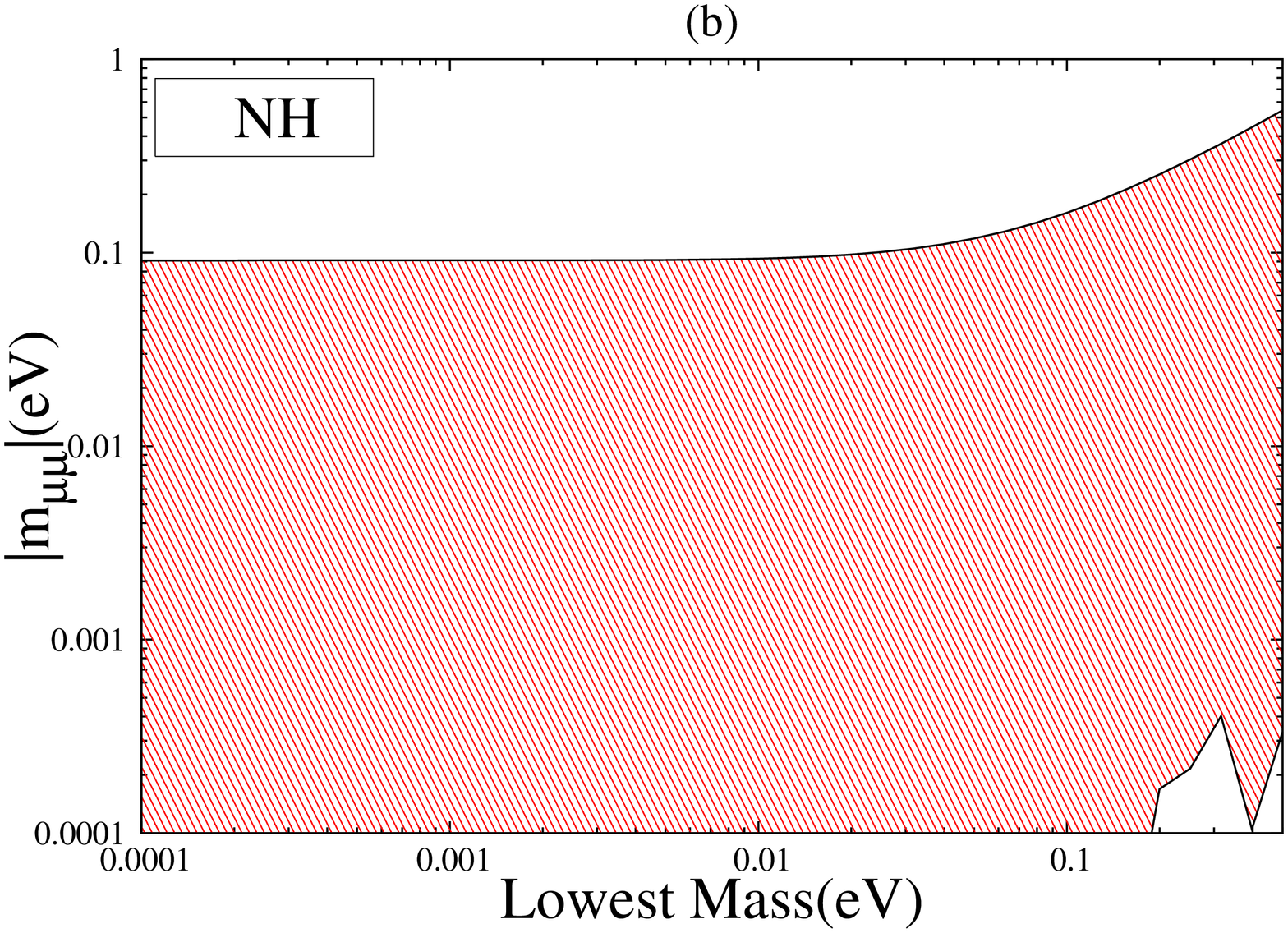}\\
\includegraphics[width=0.33\textwidth,angle=270]{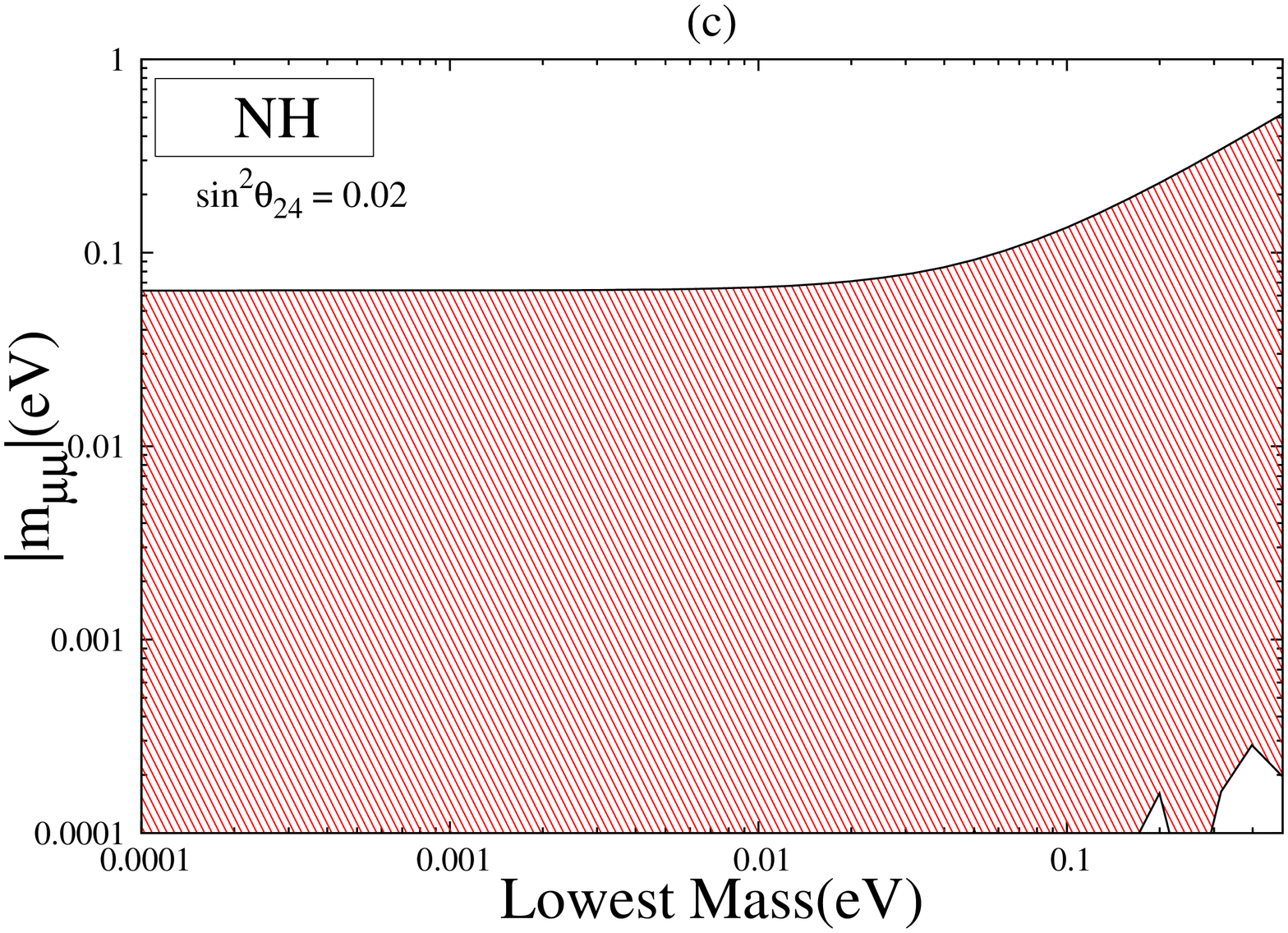}
\includegraphics[width=0.33\textwidth,angle=270]{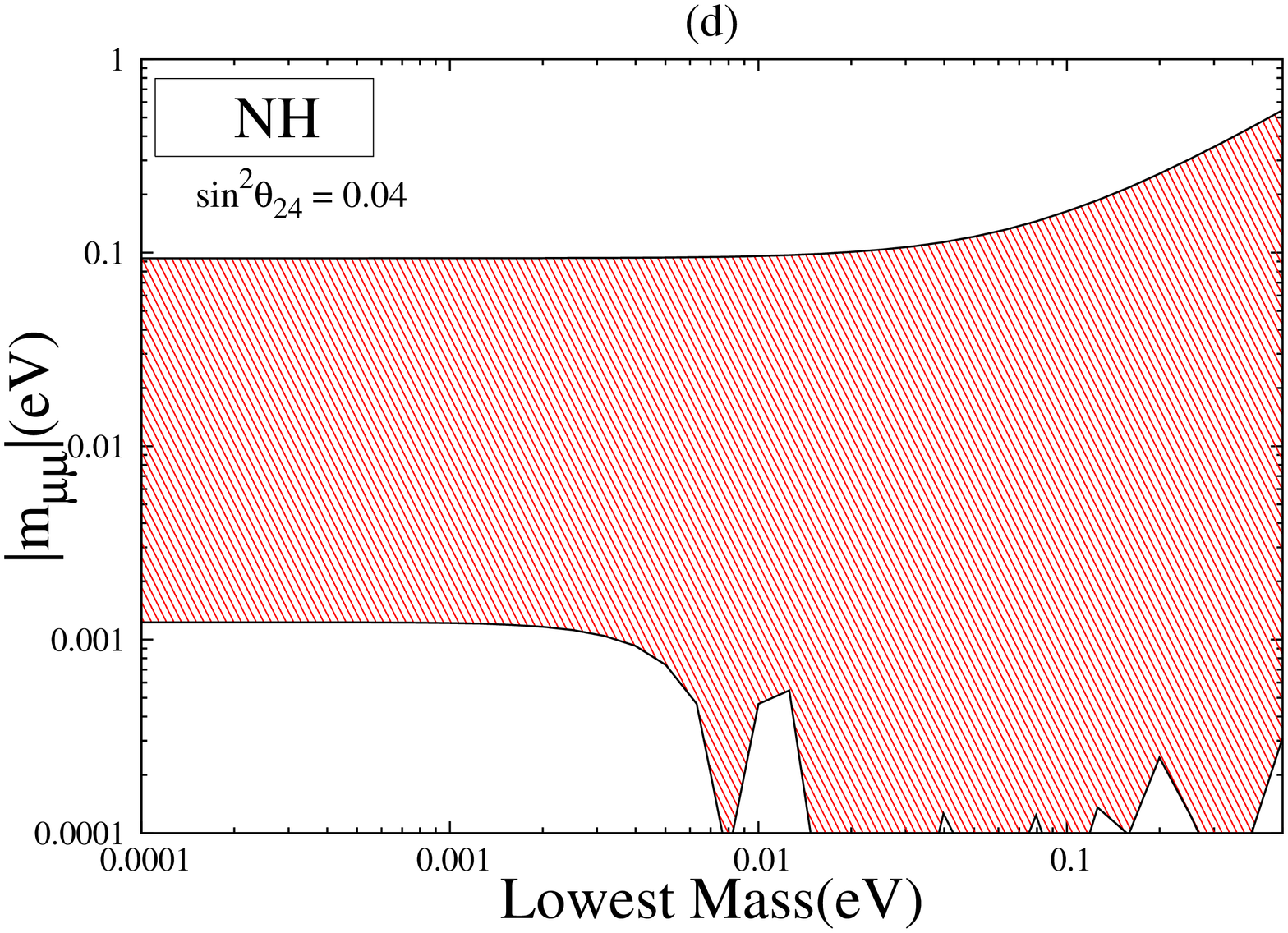}
\end{center}
Figure 8: Plots of vanishing $m_{\mu\mu}$ for normal hierarchy for different values of $\theta_{24}$
 when all other mixing angles are varied in their 3$\sigma$ ranges, Dirac CP phases are varied from 0 to $2\pi$ and Majorana phases from 0 to $\pi$.
\label{fig8}
\end{figure}
We know that for the case of 3 generations,
the elements in the $\mu -\tau$ block are quite large and cannot vanish for
normal hierarchy. In panel (a) of Fig. 8 we can see that $|m_{\mu \mu}|$
cannot vanish in small $m_1$ region for $s_{24}^2 = 0$ which is indeed the 3 generation case.
This is because the magnitude of the
first two terms in Eq. (\ref{mmmphase}) is quite large in this case,
$\sim \mathcal{O}$ (10$^{-1}$)
and for cancellation to occur the term with coefficient $\lambda^2$
has to be of the same order. This is not possible when $s_{24}^2$ is small.
However when $s_{24}^2 $ is varied in its full allowed range the contribution of the sterile
part is enhanced and this can cancel the active part as can be seen from panel (b).
Now to understand the dependence of $m_{\mu \mu}$ with $\theta_{24}$ we note that if we increase $s_{24}^2$ from its lower bound
then the two terms become of the same order.
So there will be regions in the limit of small $m_1$ for which this element
vanishes (panel (c)). We see in panel (d) of Fig. 8 that when
$\theta_{24}$ acquires very large values, the magnitude of the $\lambda^2$ ($\sqrt{\xi}\chi_{24}^2$) term becomes large, thus leading to non cancellation
of the terms with the leading order first two terms. Hence, the region with very small $m_1$ is not allowed.
Using the approximation for inverted hierarchy the element $m_{\mu\mu}$ becomes
\bea
|m_{\mu \mu}| &\approx& |\sqrt{\Delta m_{13}^2}\{c_{23}^2(s_{12}^2 + c_{12}^2 e^{2 i \alpha})+ \frac{1}{2}
 \lambda \sin2\theta_{12} \sin2\theta_{23} e^{i \delta_{13}} ( 1 - e^{2 i \alpha}) \chi_{13} \\ \nonumber
 &+& \lambda^2[\sin2\theta_{12} c_{23} e^{i(\delta_{14} - \delta_{24})}(1 - e^{2 i \alpha}) \chi_{14} \chi_{24}
 + s_{23}^2 e^{2 i \delta_{13}}(c_{12}^2 + e^{ 2 i \alpha} s_{12}^2) \chi_{13}^2 \\ \nonumber
 &+& e^{2i(\gamma + \delta_{14} - \delta_{24})} \sqrt{\xi} \chi_{24}^2]\}|.
\eea
Assuming Majorana phases to be zero
 and Dirac phases having value $\pi$, this element can vanish when
\bea
&& c_{23}^2+\lambda^2(s_{23}^2\chi_{13}^2+\sqrt{\xi}\chi_{24}^2)=0.
\eea
In panel (a) of Fig. 9 we plotted $|m_{\mu \mu}|$ for $s_{24}^2 = 0$ to reproduce 3 generation case whereas  in panel (b) all the parameters are varied in their allowed range in 3+1 scenario.
In both cases we can see that cancellation is possible for full range of $m_3$.
It can be noticed that unlike normal hierarchy, here cancellation is possible for small values of $s_{24}^2$
because in this case all the terms are of same order and there can always be cancellations. However,
  if we put $\alpha = 0$ then the term $\lambda$ ($\sin 2\theta_{12}s_{23}c_{23}\chi_{13}$) drops out from the equation and
the leading order term can not be canceled for small values of $s_{24}^2$. It can be seen from panel (c) that for $s_{24}^2 = 0.002$ and $\alpha = 0$ the regions where
$m_3$ is small is not allowed.
As the value of $\theta_{24}$ increases there is the possibility of cancellation of terms
for all the values of $\alpha$ as can be seen from panel (d) where we plot $|m_{\mu\mu}|$ with the lowest mass for $s_{24}^2=0.02$ when all the other mixing angles are
varied in 3$\sigma$ range and CP violating phases are varied in full range.
Now if we keep increasing $s_{24}^2$ then $\lambda^2$ term will become large and the chance of cancellation will be less.
\begin{figure}
\begin{center}
\includegraphics[width=0.33\textwidth,angle=270]{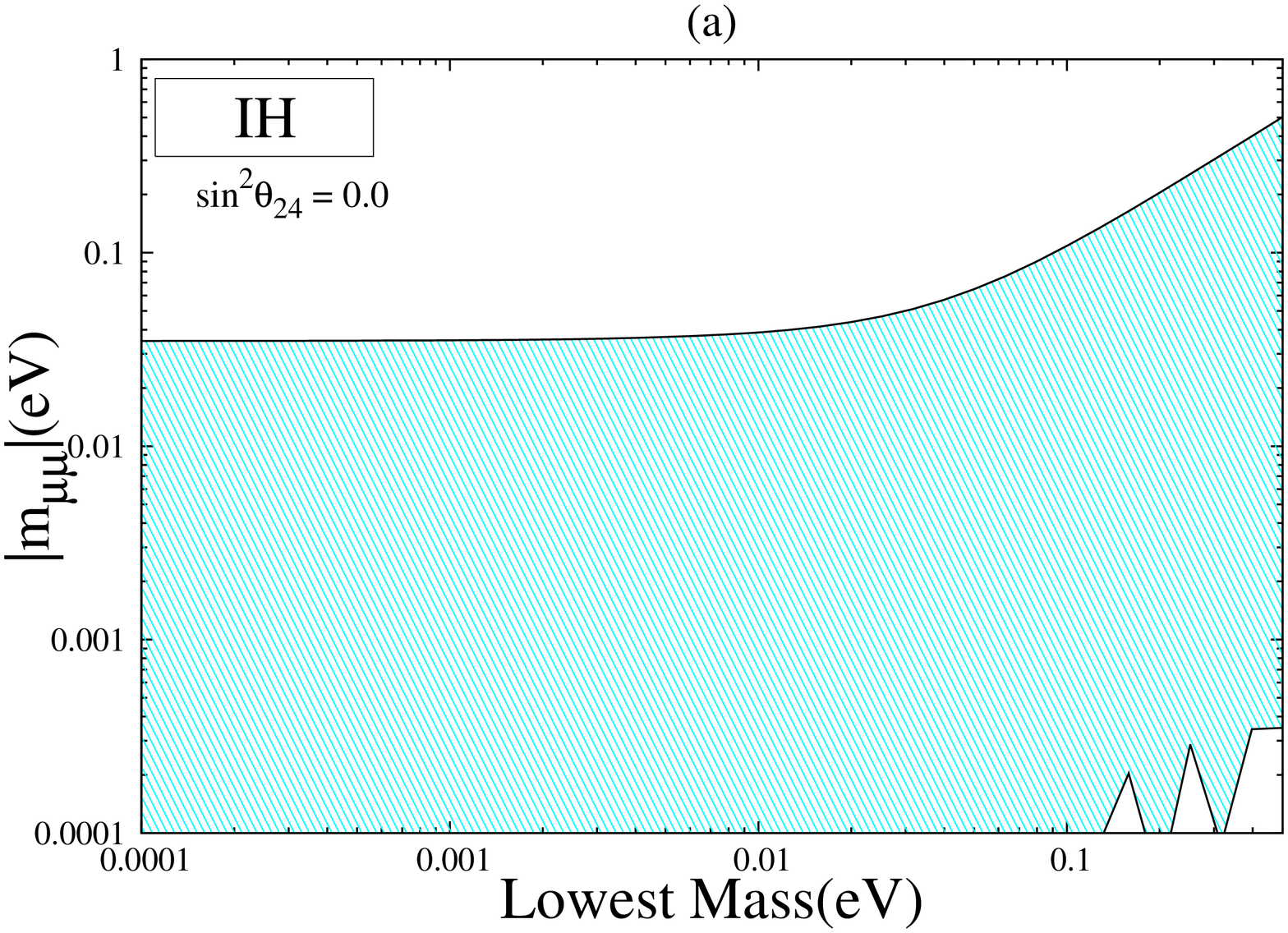}
\includegraphics[width=0.33\textwidth,angle=270]{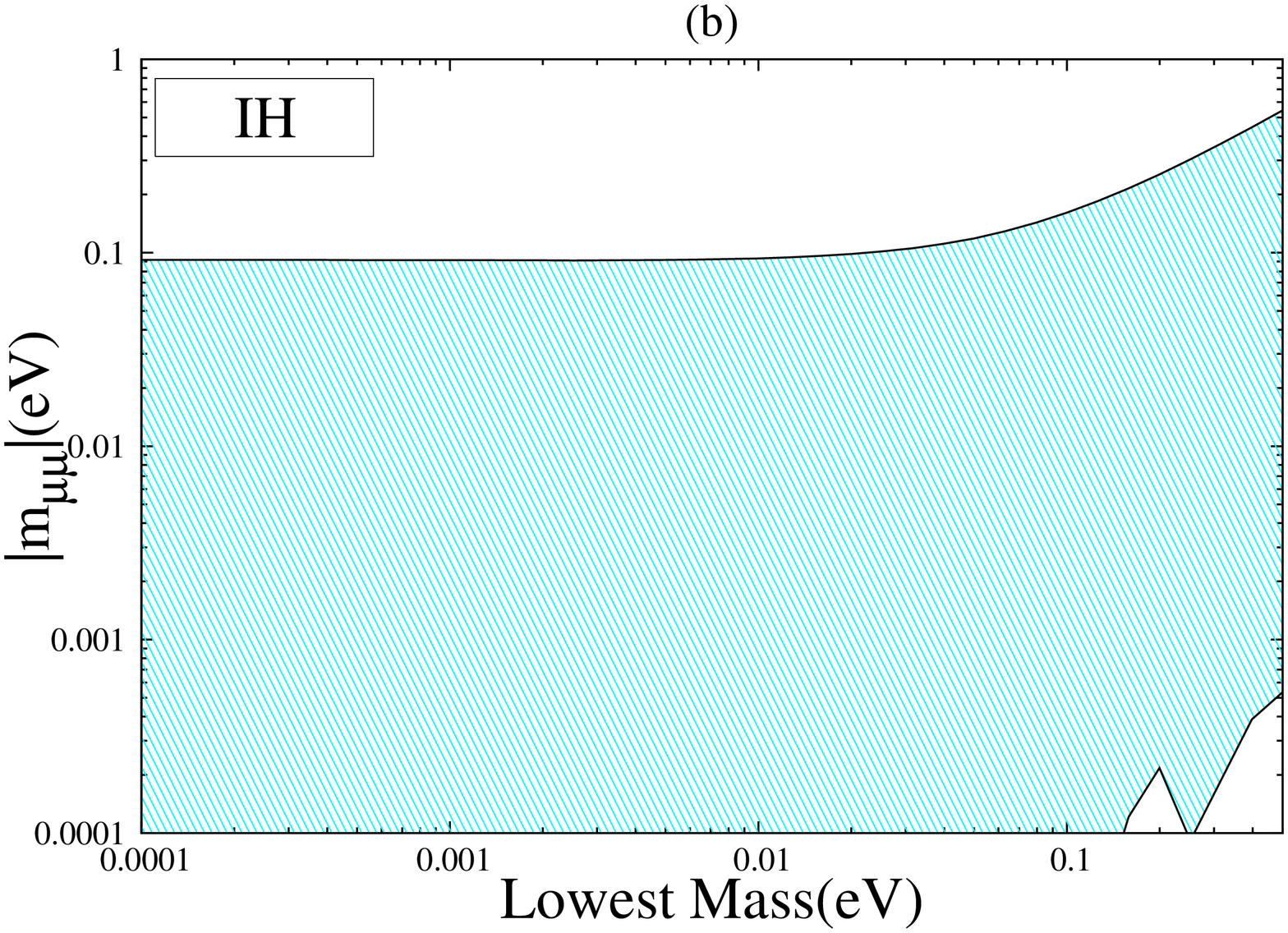}\\
\includegraphics[width=0.33\textwidth,angle=270]{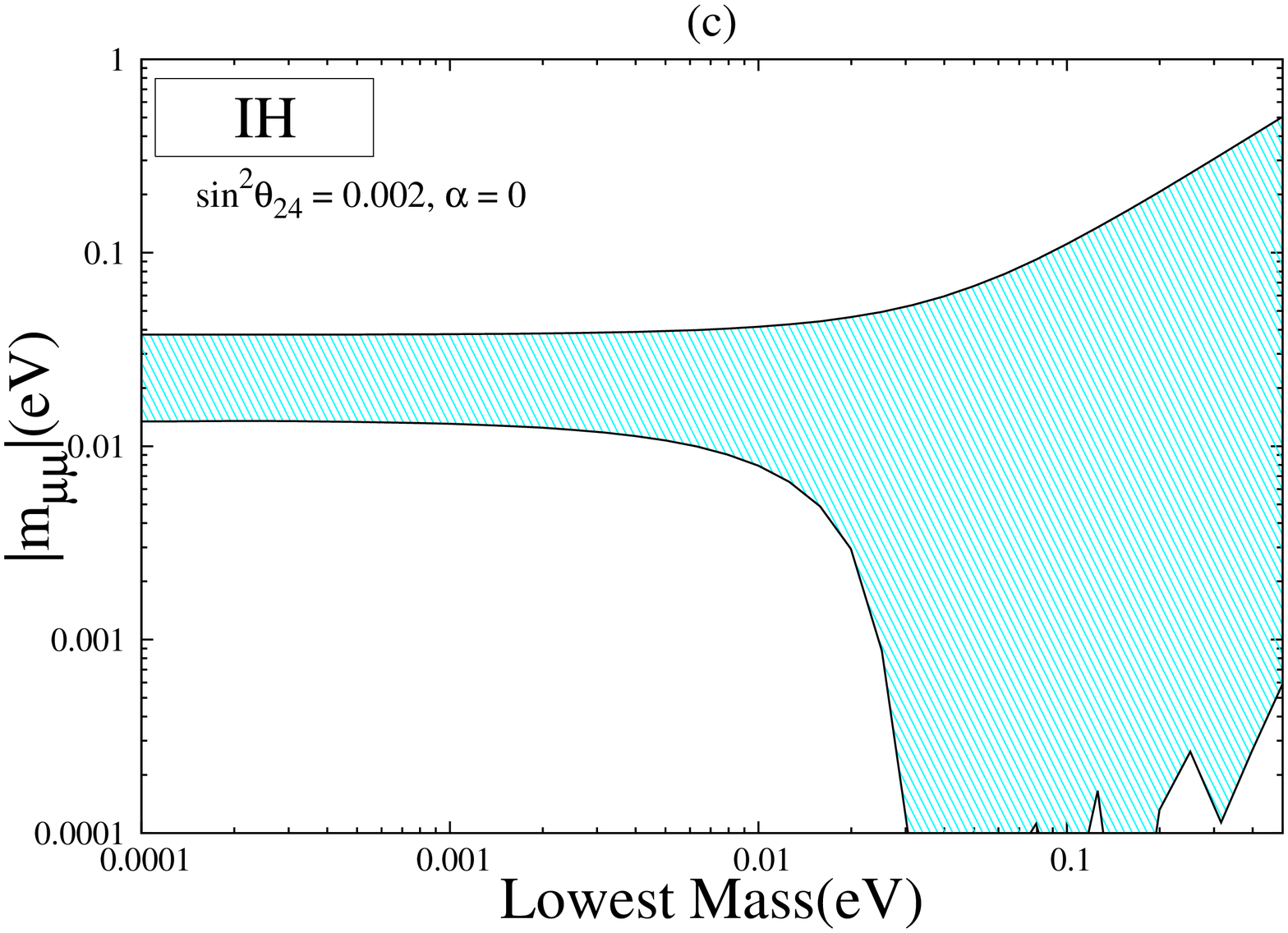}
\includegraphics[width=0.33\textwidth,angle=270]{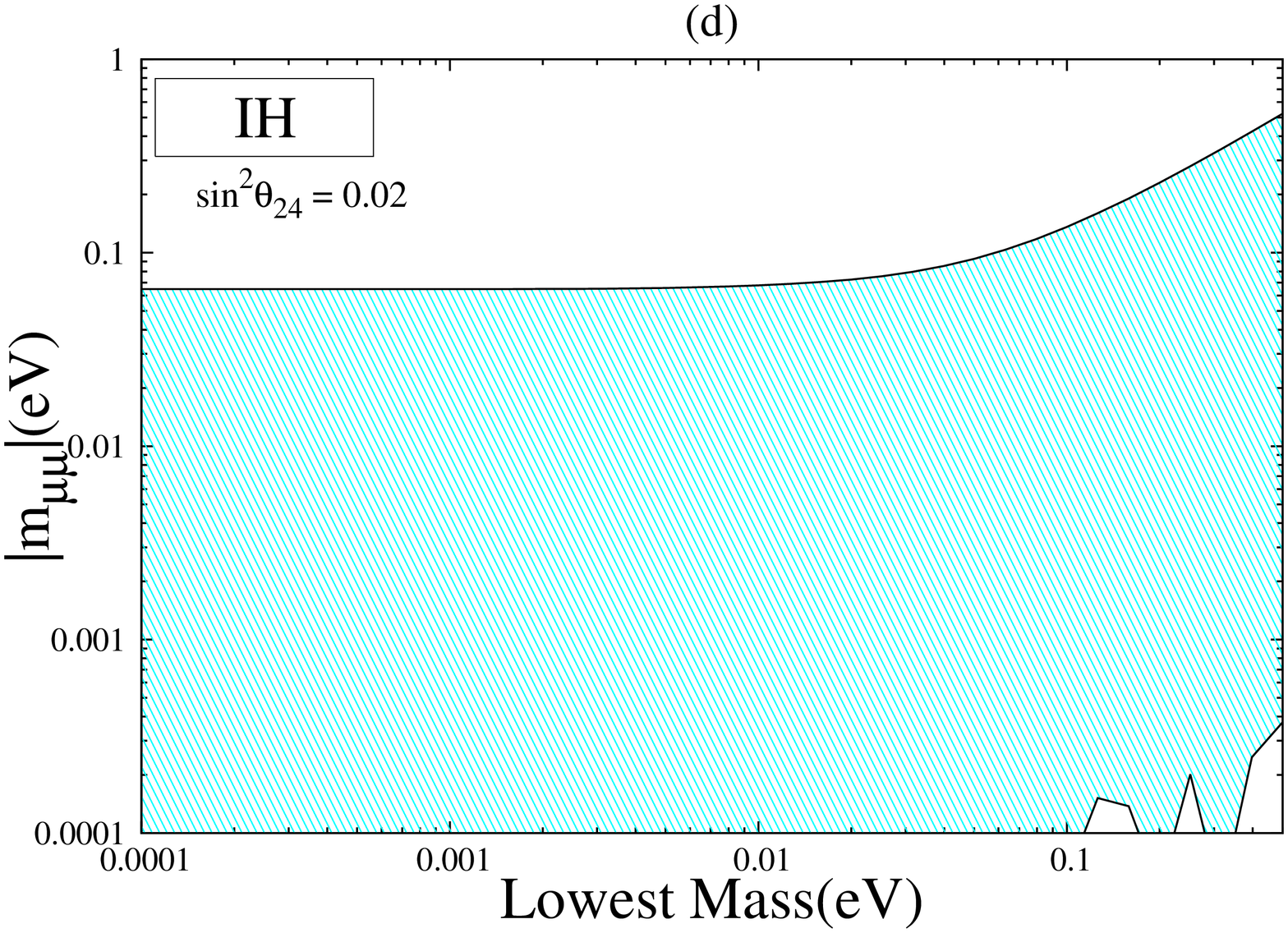}
\end{center}
Figure 9: Plots of vanishing $m_{\mu\mu}$ for inverted hierarchy with lowest mass $m_3$.
Panel (a) for 3 generation case (b) all the parameters are varied in full allowed range (3+1).  Panel (c) and (d) are for
specific values of $\alpha$ and $s_{24}^2$ are taken with all other parameters covering their full range.
\end{figure}

\subsection{The Mass Matrix element $m_{\mu\tau}$}
The (2,3) element of $M_{\nu}$ in the flavor basis becomes quite complicated in the presence of an extra sterile neutrino. The expression is
\bea
 m_{\mu \tau} &=& e^{i(2 \delta_{14} - \delta_{24} + 2 \gamma)} c_{14}^2 c_{24} m_4 s_{24} s_{34} \\ \nonumber
 &+& e^{2 i (\delta_{13} + \beta)} m_3 ( c_{13} c_{24} s_{23} - e^{i(\delta_{14} - \delta_{24} - \delta_{13})} s_{13} s_{14} s_{24}) \\ \nonumber
 && \{-e^{i(\delta_{14} - \delta_{13})} c_{24} s_{13} s_{14} s_{34}+ c_{13} (c_{23} c_{34} - e^{i \delta_{24}} s_{23} s_{24} s_{34})\} \\ \nonumber
 &+& m_1\{- c_{23} c_{24} s_{12} + c_{12}(-e^{i \delta_{13}} c_{24} s_{13} s_{23} - e^{i(\delta_{14} - \delta_{24})} c_{13} s_{14} s_{24})\} \\ \nonumber
 && [-s_{12} (-c_{34} s_{23} - e^{i \delta_{24}} c_{23} s_{24} s_{34}) \\ \nonumber
 &+& c_{12} \{ -e^{i \delta_{14}} c_{13} c_{24} s_{14} s_{34} - e^{i \delta_{13}} s_{13}(c_{23} c_{34} - e^{i\delta_{24}} s_{23} s_{24} s_{34})\}] \\ \nonumber
 &+& e^{2 i \alpha} m_2\{c_{12} c_{23} c_{24} + s_{12}(-e^{i \delta_{13}} c_{24} s_{13} s_{23} - e^{i(\delta_{14} - \delta_{24})} c_{13} s_{14} s_{24})\} \\ \nonumber
 && [c_{12}(-c_{34} s_{23} - e^{i \delta_{24}} c_{23} s_{24} s_{34}) \\ \nonumber
 &+& s_{12}\{-e^{i \delta_{14}} c_{13} c_{24} s_{14} s_{34} - e^{i \delta_{13}} s_{13}(c_{23} c_{34} - e^{i \delta_{24}} s_{23} s_{24} s_{34})\}].
\eea
It reduces to the 3 generation case when $\theta_{24} = \theta_{34} = 0$.
In the normal hierarchical region where $m_1$ can assume very small values and can be neglected, using approximations in
Eqs. (\ref{xnh}, \ref{chi1}, \ref{chi2}) we get
\bea
 |m_{\mu \tau}| &\approx & |\sqrt{\Delta m_{23}^2}\{c_{23} c_{34}(e^{2 i (\beta + \delta_{13})} - e^{2 i \alpha} \sqrt{\zeta} c_{12}^2)s_{23} \\ \nonumber
 &-& \lambda[c_{12} c_{34} e^{i(2\alpha+ \delta_{13})} \sqrt{\zeta} s_{12} \cos2\theta_{23} \chi_{13} + e^{i \delta_{24}}(e^{2i \alpha}
 c_{12}^2 c_{23}^2 \sqrt{\zeta} + e^{2i(\beta + \delta_{13})} s_{23}^2) \chi_{24}s_{34} \\ \nonumber
 &-& e^{2 i \delta_{14}}(e^{i(2 \gamma - \delta_{24})} \sqrt{\xi} \chi_{24} - c_{12} c_{23} e^{2 i \alpha} \sqrt{\zeta} s_{12} \chi_{14}) s_{34}] \\ \nonumber
 &+& \lambda^2 [\sqrt{\zeta} e^{i(2 \alpha + \delta_{13})}(e^{i \delta_{14}} s_{12} \chi_{14} + 2 c_{12} c_{23} e^{i \delta_{24}} \chi_{24})s_{12} s_{23} s_{34} \chi_{13}\\ \nonumber
 &+& e^{i \delta_{14}}(e^{i(2 \alpha - \delta_{24})} c_{12} c_{34} \sqrt{\zeta} s_{12} \chi_{24} - e^{i(2 \beta + \delta_{13})} \chi_{13} s_{34}) \chi_{14} s_{23} \\ \nonumber
 &+& c_{23} c_{34} e^{2i(\alpha + \delta_{13})} s_{12}^2 \chi_{13}^2 \sqrt{\zeta} s_{23}]\}|.
\eea
To see the order of the terms we consider the case where Majorana CP phases vanish and Dirac phases have the value $\pi$. In this limit the element becomes negligible when
\bea
&& c_{23} c_{34}s_{23}(1-c_{12}^2\sqrt{\zeta})+\lambda\{(c_{12}c_{34}s_{12}\sqrt{\zeta}\chi_{13})\cos2\theta_{23} \\ \nonumber
 &+& \chi_{24}s_{34}(s_{23}^2+c_{12}^2c_{23}^2\sqrt{\zeta})+s_{34}(\sqrt{\xi}\chi_{24}+c_{12}c_{23}s_{12}\sqrt{\zeta}\chi_{14})\} \\ \nonumber
 &+& \lambda^2\{s_{12}\chi_{13}s_{23}s_{34}\sqrt{\zeta}(s_{12}\chi_{14}+2c_{12}c_{23}\chi_{24})+\chi_{14}s_{23}(c_{12}c_{34}s_{12}^2s_{23}\sqrt{\zeta}\chi_{13}^2)\}=0.
\eea
Being an element of $\mu \tau$ block, $m_{\mu \tau}$ shows the same behaviour that of $m_{\mu \mu}$ in normal hierarchy. In panel (a) of Fig. 10 we plotted $|m_{\mu \tau}|$
for $s_{24}^2 = s_{34}^2 = 0$ which coincides
with the 3 generation case and we can see that cancellation is not possible in hierarchical region. However, when all
the parameters are varied in their allowed range in panel (b) it get contribution from the sterile part and cancellation is always possible.
It can also be seen from panel (c) of Fig. 10 that for $s_{34}^2 = 0$ there is no
 cancellation in the region when $m_1$ is small and the figure is quite similar to that of 3 generation case.
However, as this active sterile mixing angle becomes larger
there is always a possibility of allowed region towards the lower values of $m_1$ as is evident from panel (d).
This is because for the vanishing value of $\theta_{34}$
the terms with $\lambda$ and $\lambda^2$ become very small and cannot
cancel the leading term $\mathcal{O}$ (10$^{-1}$). It can also be seen that in this case (i.e $s_{34}^2 = 0$),
there is no $\chi_{24}$ term in Eq. (3.31) and this is why the figure is somewhat similar to the 3 generation case.
However, when $\theta_{34}$ increases
these two contributions become large and cancellation becomes possible.
\begin{figure}
\begin{center}
\includegraphics[width=0.33\textwidth,angle=270]{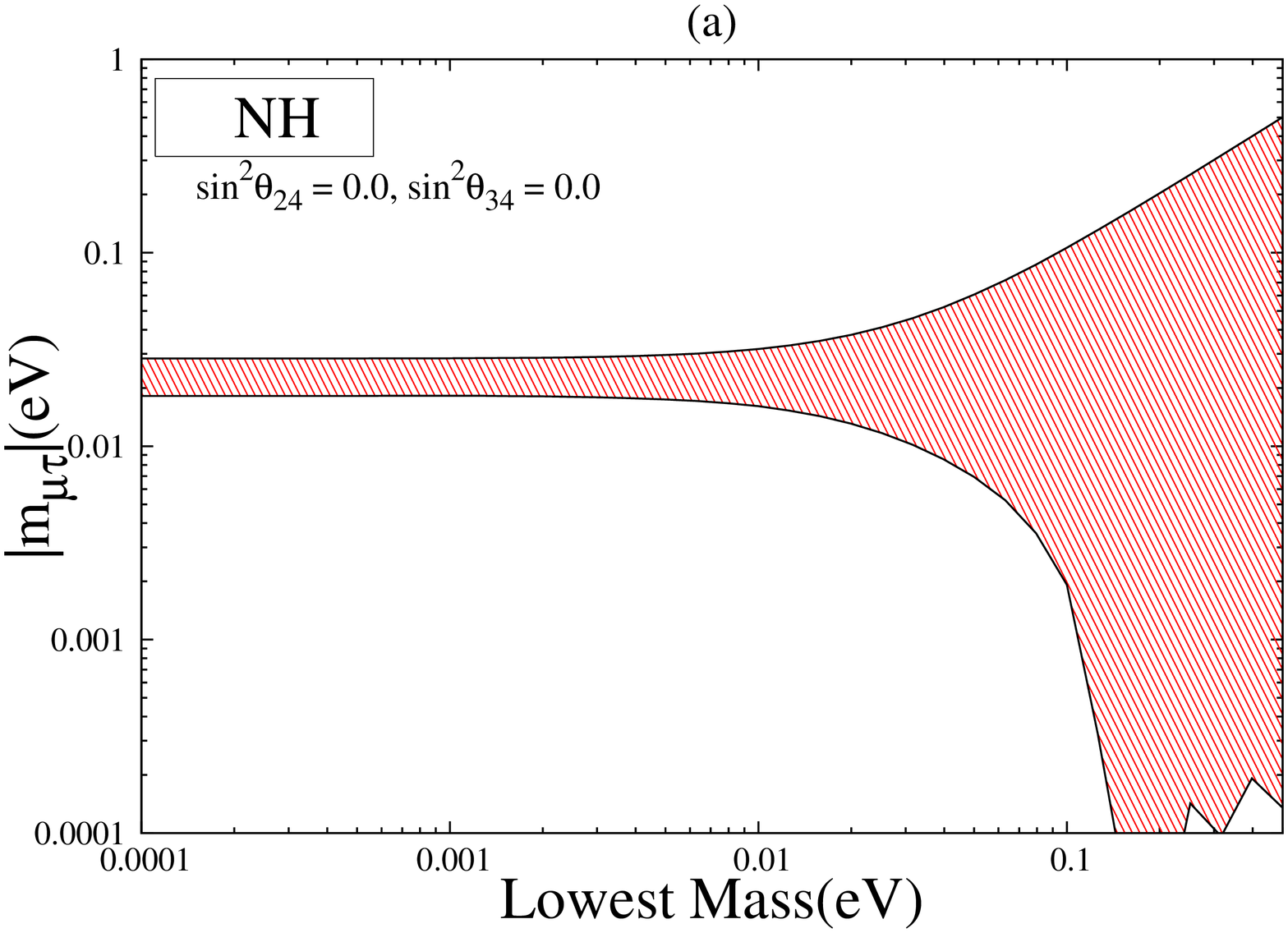}
\includegraphics[width=0.33\textwidth,angle=270]{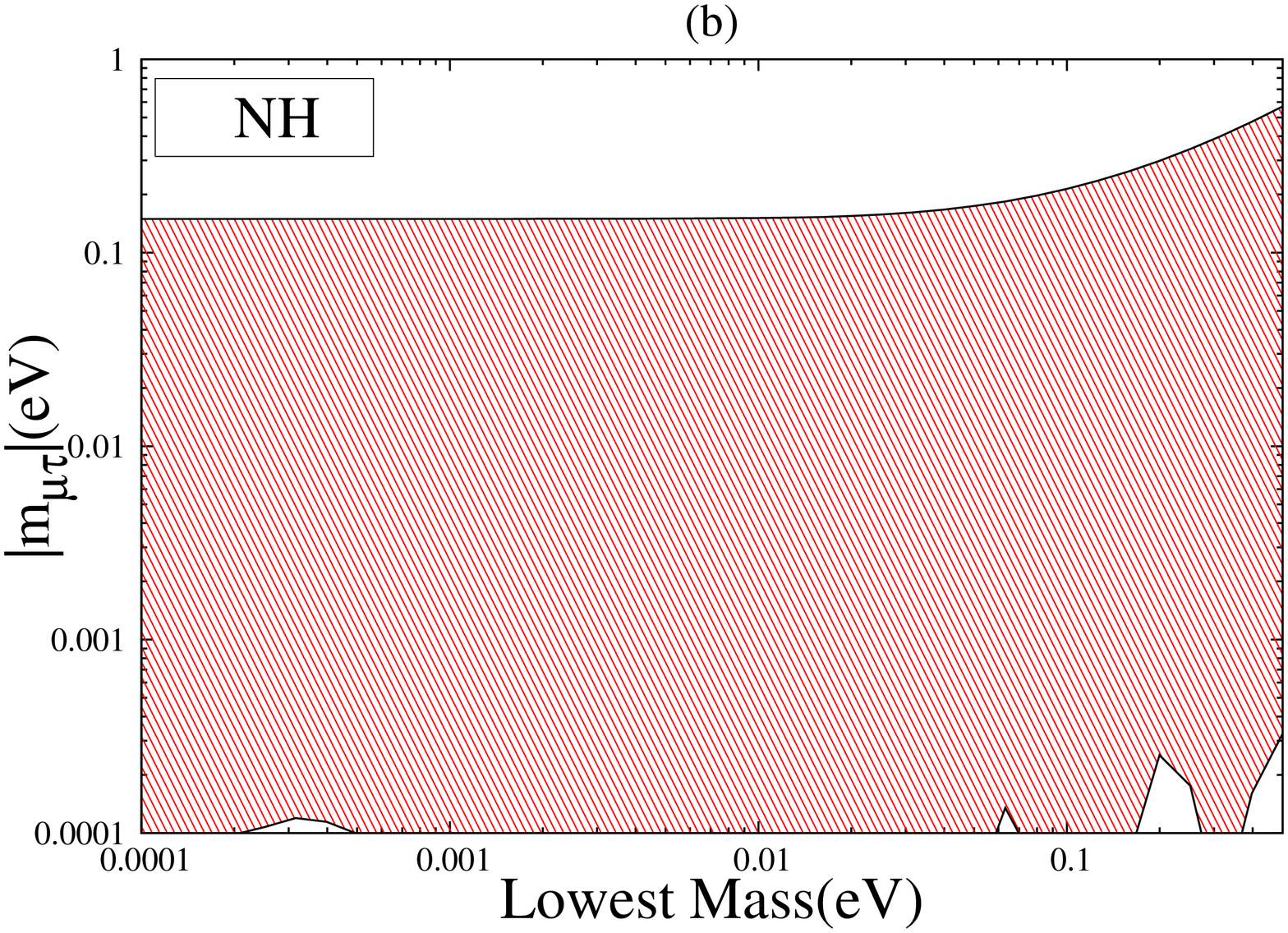}\\
\includegraphics[width=0.33\textwidth,angle=270]{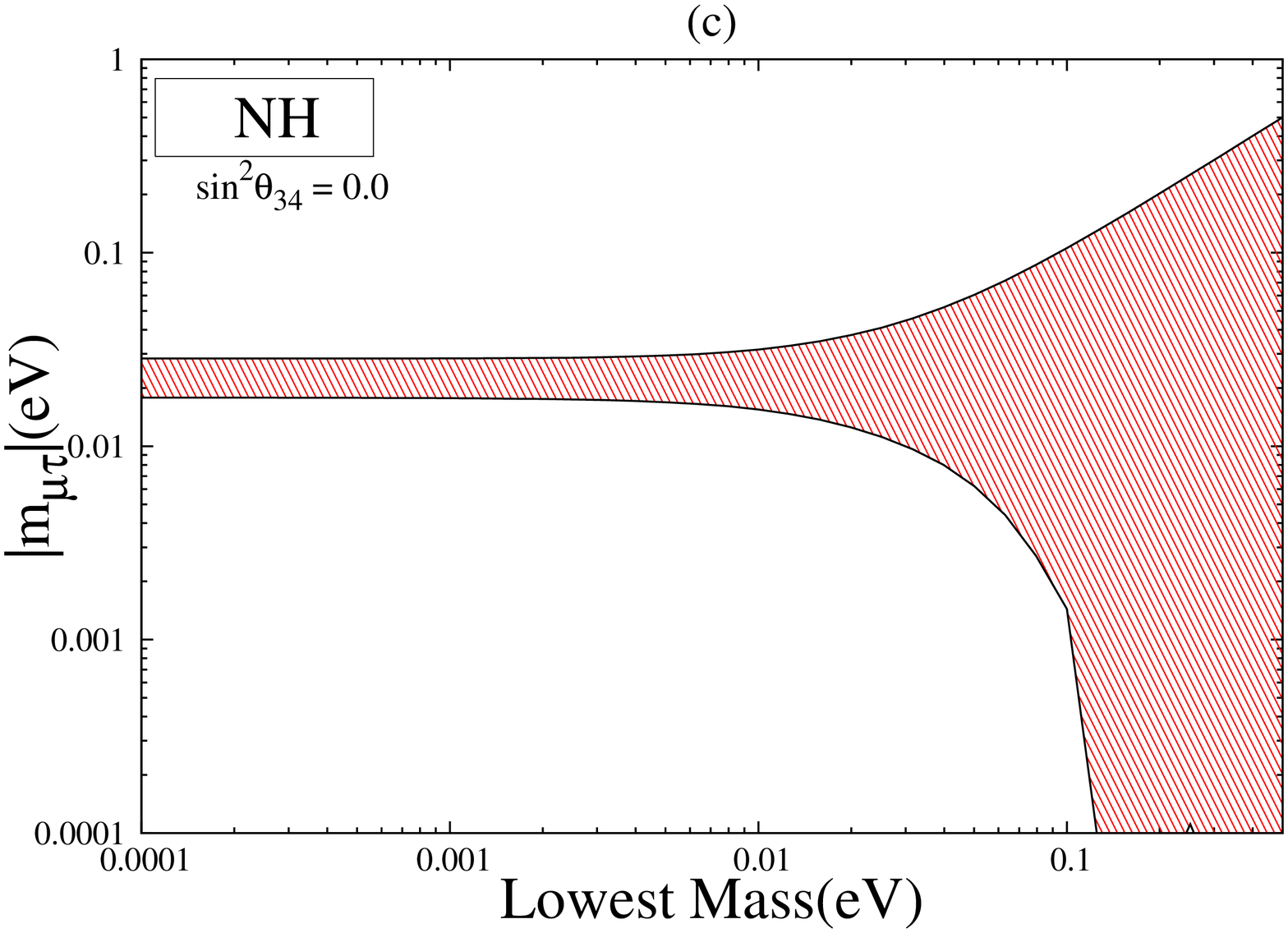}
\includegraphics[width=0.33\textwidth,angle=270]{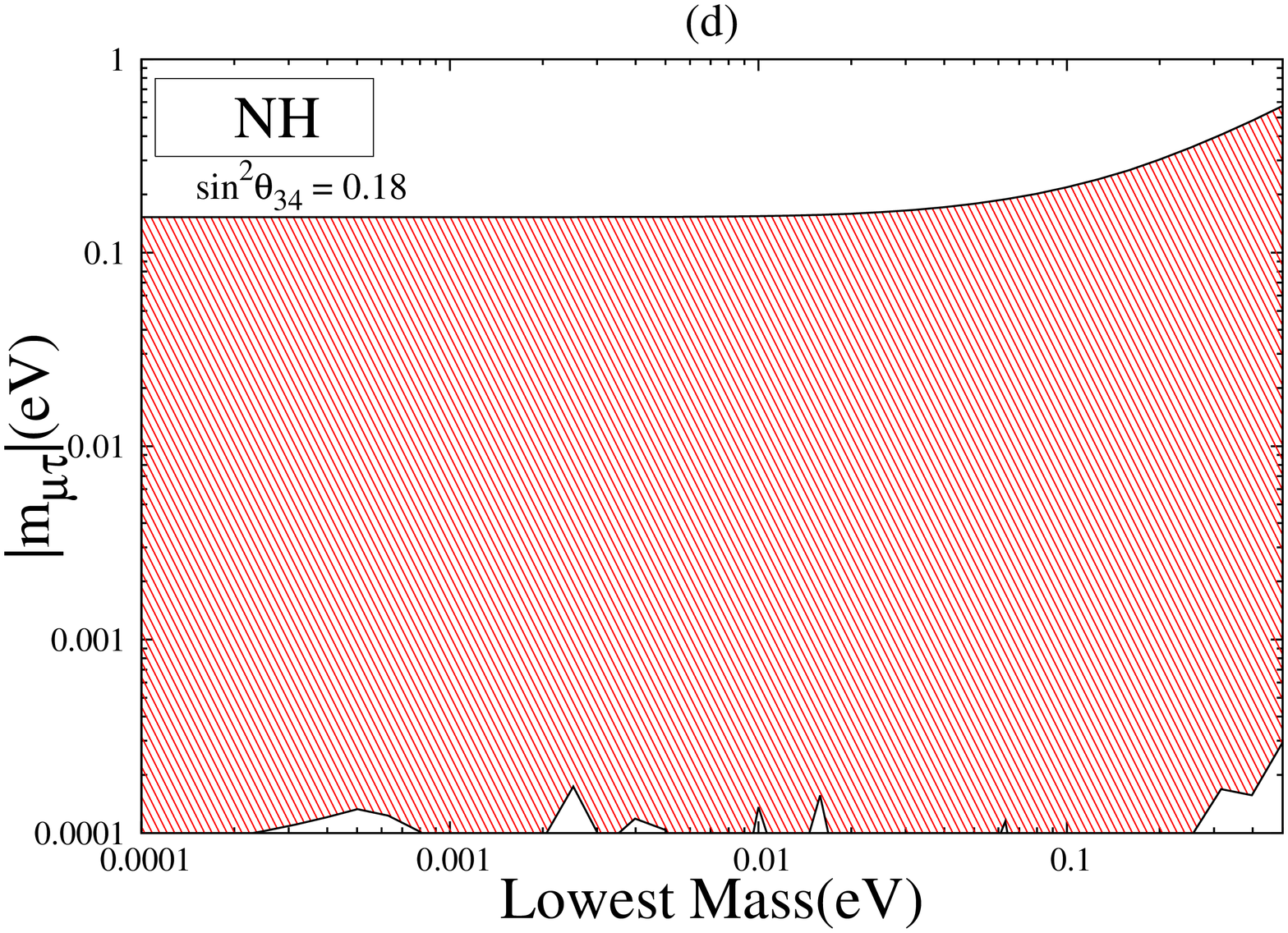}
\end{center}
Figure 10: Plots of vanishing $m_{\mu\tau}$ for normal hierarchy (a) for vanishing $\theta_{34}$ and $\theta_{24}$.
In panel (b) all parameters are varied in their full allowed range. Panel(c, d) are for specific values of $\theta_{34}$
when all other mixing angles are varied in their full range.
\label{fig10}
\end{figure}
For the case of inverted hierarchy where $m_3$ can have very small values, $m_{\mu\tau}$ becomes
\bea
 |m_{\mu \tau}| &\approx&|\sqrt{\Delta m_{13}^2}\{ -c_{23} c_{34} s_{23}(c_{12}^2 e^{2 i \alpha} + s_{12}^2) \\ \nonumber
 &+& \lambda [c_{12} s_{12} (1 - e^{2 i \alpha})(c_{34} \cos2\theta_{23} e^{i \delta_{13}} \chi_{13} + c_{23} s_{34} e^{i \delta_{14}} \chi_{14}) \\ \nonumber
 &+& s_{34}\{e^{i(2 \gamma+ 2 \delta_{14} - \delta_{24})} \sqrt{\xi} - c_{23}^2 e^{i \delta_{24}}(s_{12}^2 + c_{12}^2 e^{2 i \alpha})\}\chi_{24}] \\ \nonumber
 &+& \lambda^2 [c_{23} c_{34} s_{23} e^{ 2 i \delta_{13}}(c_{12}^2 + e^{2 i \alpha} s_{12}^2) \chi_{13}^2 \\ \nonumber
 &+& c_{12} s_{12} s_{23} (e^{2 i \alpha} - 1)(c_{34} e^{i(\delta_{14} - \delta_{24})} \chi_{14} + 2 s_{34} c_{23} e^{i(\delta_{13} + \delta_{24})} \chi_{13})\chi_{24}]\}|.
\eea
To get an idea about the magnitude of the terms we take  vanishing Majorana phases and
Dirac CP phases to be of the order $\pi$. The expression in this case for vanishing $m_{\mu \tau}$ becomes
\bea
&&-c_{23}c_{34}s_{23}+\lambda(s_{34}\chi_{24}(c_{23}^2-\sqrt{\xi}))-\lambda^2(-s_{34}\chi_{13}\chi_{14}-c_{23}c_{34}\chi_{13}^2)=0
\eea
In panel (a) of Fig. 11,
where $|m_{\mu \tau}|$ for 3 generation is plotted,
we can see that unlike $m_{\mu \mu}$ there is
no cancellation in small $m_3$ region but when plotted for the full range it gets contribution from the sterile part
and there is cancellation for the full range of $m_3$ (panel (b)).
Clearly the cancellation of the terms do not become possible for small values of $\theta_{34}$ in strict hierarchical
region. This case is similar to the three generation case in IH (cyan/light region, panel (c)). This is because for $s_{34}^2 = 0$
the contribution of $s_{24}^2$ comes from the $\lambda^2$ term. If we put the CP violating phase $\alpha$ as zero then
cancellation is not possible for whole range of $m_3$ (blue/dark region panel (c)). However,
as the value of $s_{34}^2$ increases all the terms in the above equation becomes of the same order and
cancellation for very small values of $m_3$ is possible (panel (d)).
\begin{figure}
\begin{center}
\includegraphics[width=0.33\textwidth,angle=270]{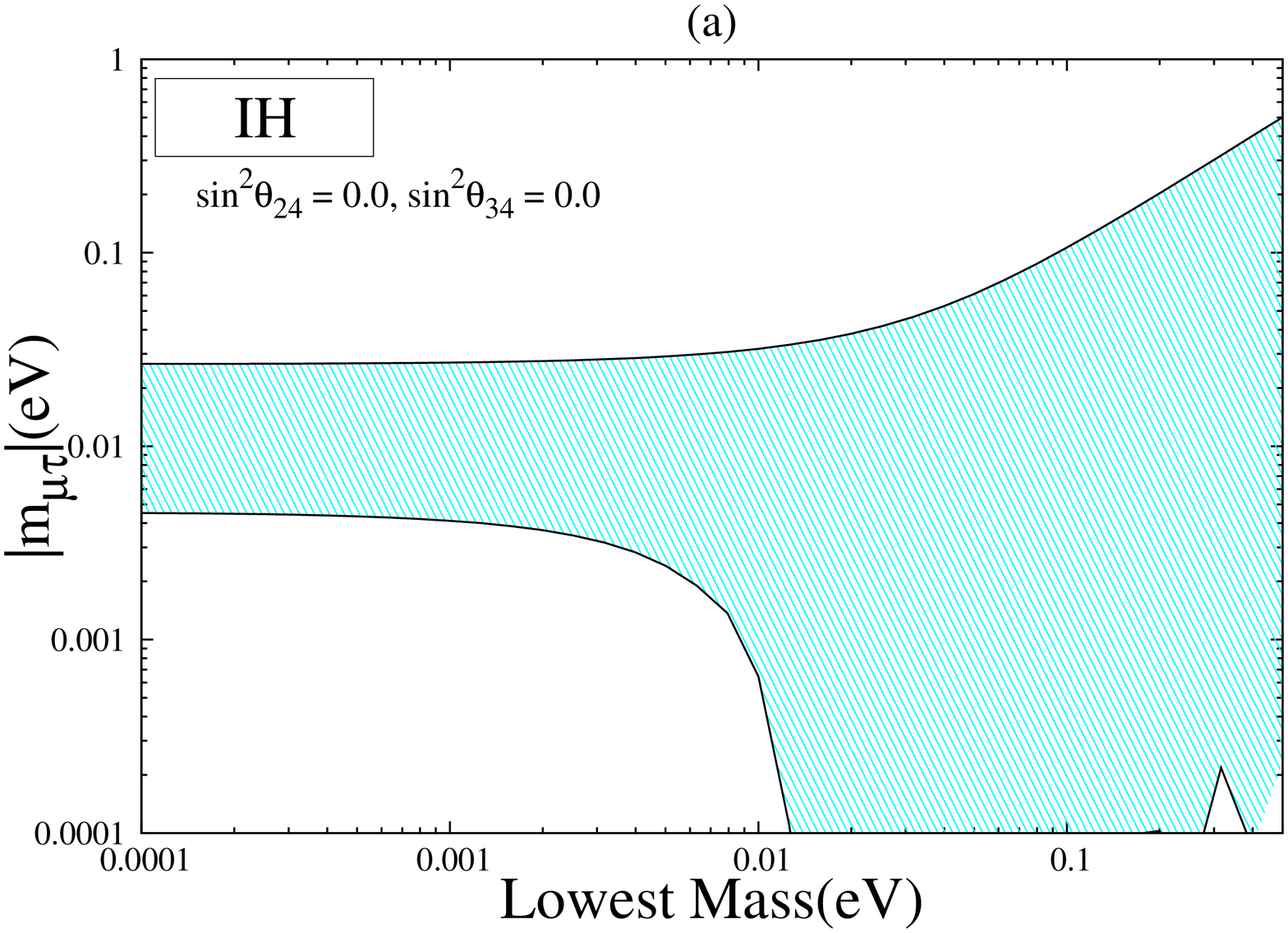}
\includegraphics[width=0.33\textwidth,angle=270]{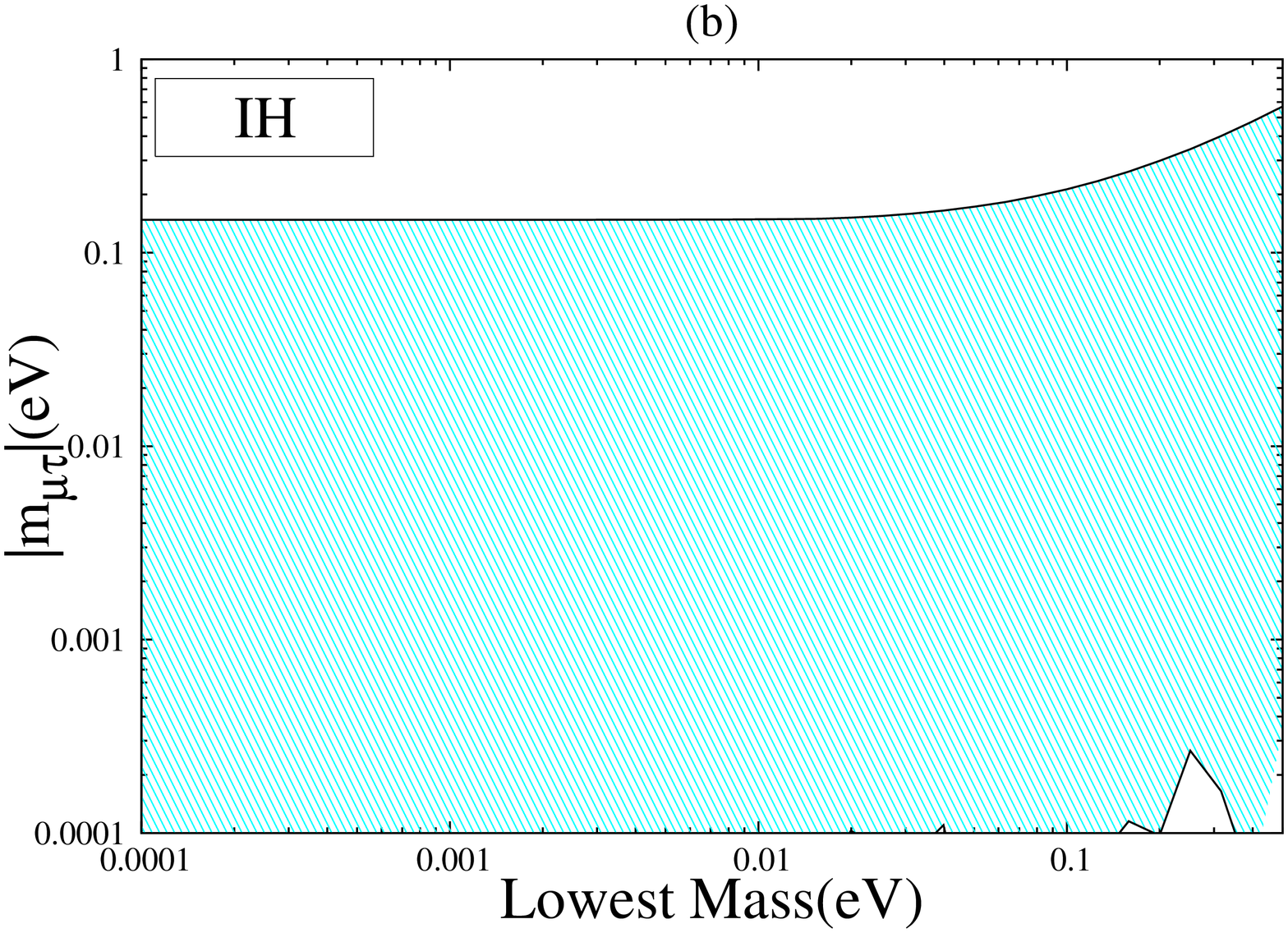}\\
\includegraphics[width=0.33\textwidth,angle=270]{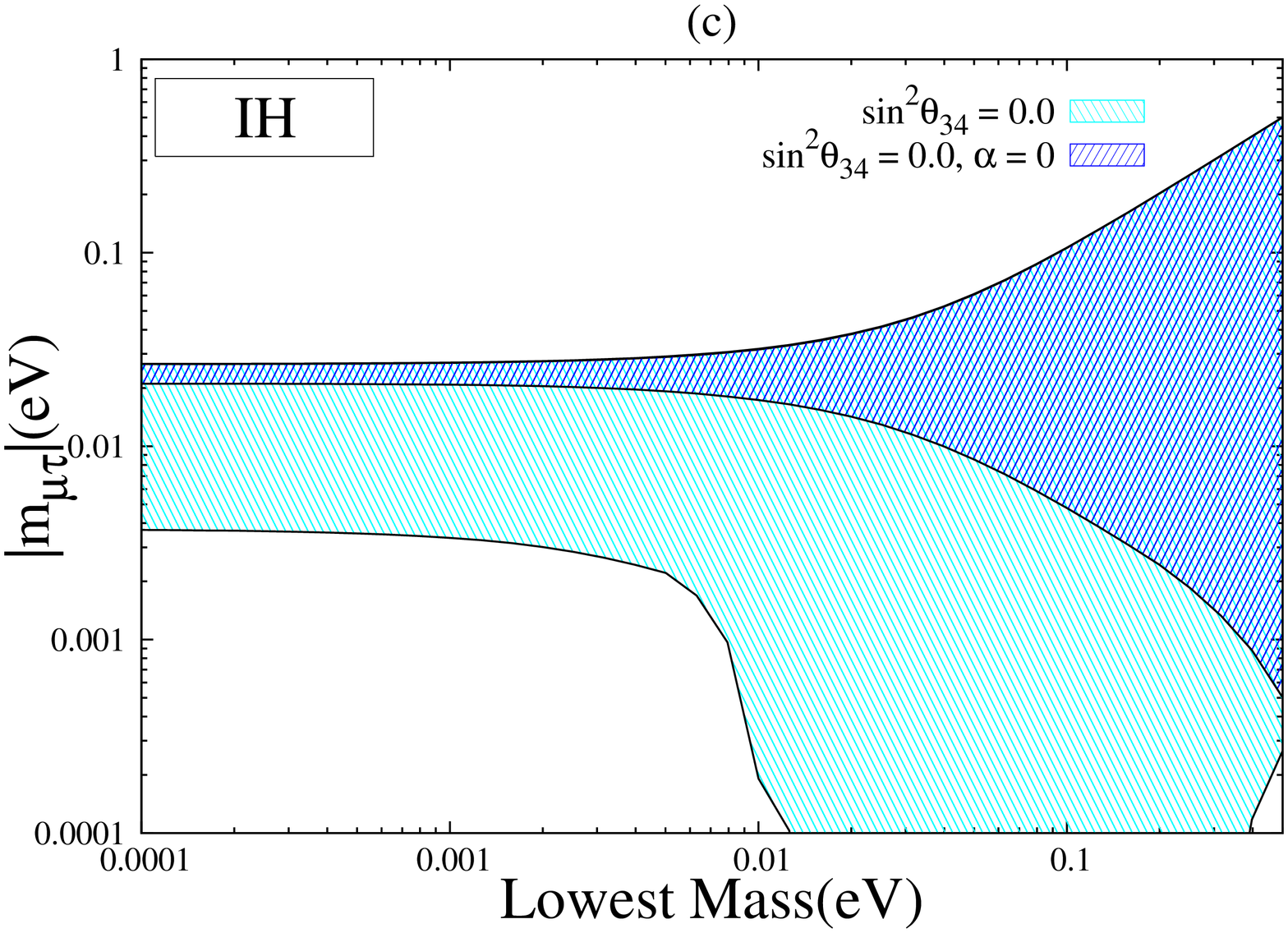}
\includegraphics[width=0.33\textwidth,angle=270]{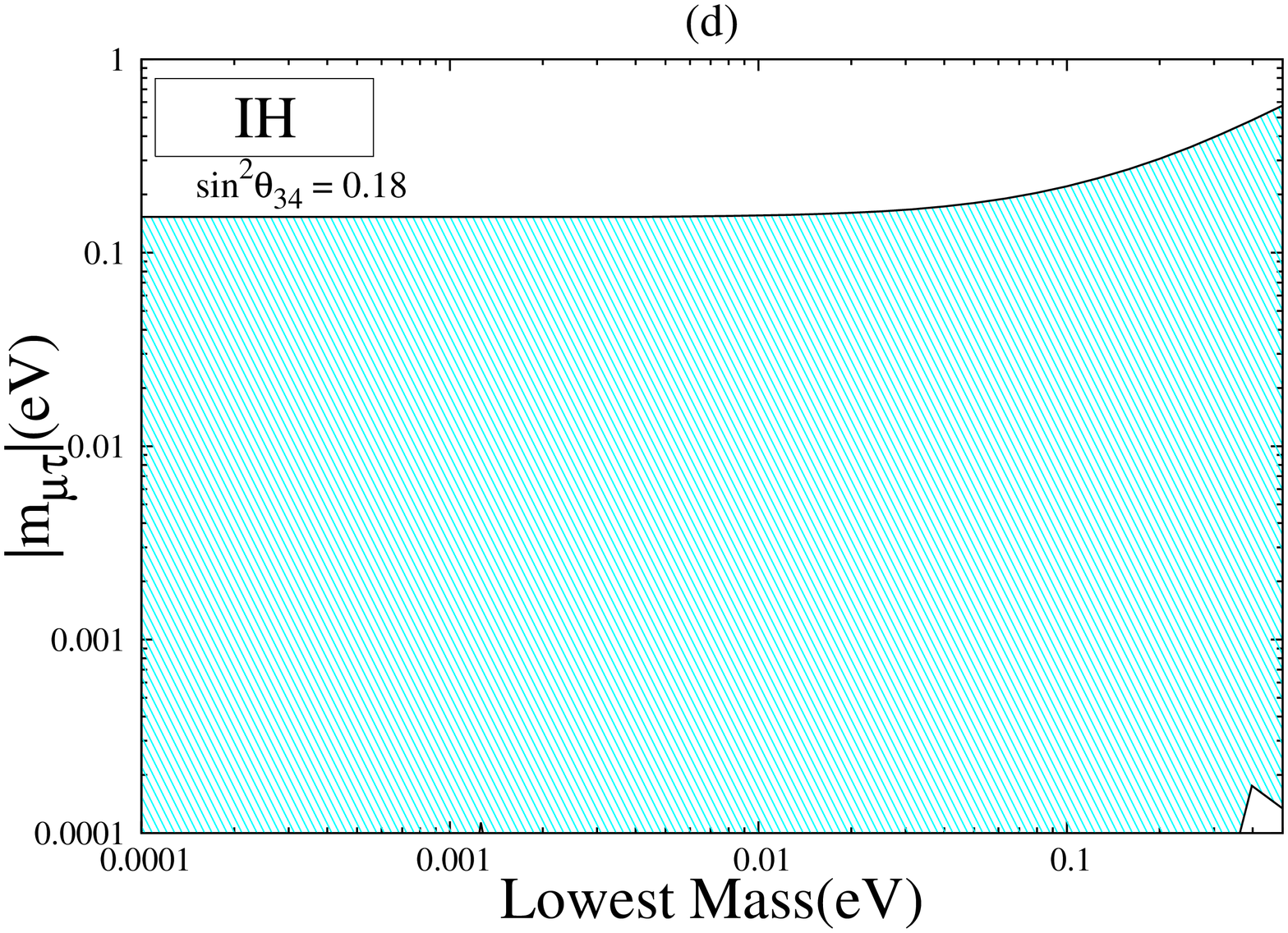}
\end{center}
Figure 11: Plots of vanishing $m_{\mu\tau}$ for inverted hierarchy (a) for vanishing $\theta_{34}$ and
$\theta_{24}$ (3 generation). In panel (b) all parameters are varied in their full allowed range (3+1).
Panel (c) and (d) are for specific values of $\theta_{34}$ and $\alpha$ when all other mixing angles are varied in their full range.
\label{fig11}
\end{figure}
\subsection{The Mass Matrix element $m_{\tau\tau}$}
This element is related to $m_{\mu\mu}$ by the $\mu-\tau$ symmetry.
As discussed earlier, in the limit when $\theta_{24}$ and $\theta_{34}$ are not very large,
the two mixing angles $\theta_{34}$ and $\theta_{24}$ will behave in the same way
in the textures related by $\mu-\tau$ symmetry.
The (3,3) element of the neutrino mass matrix in the presence of one sterile neutrino is given as
\bea
  m_{\tau \tau} &=& e^{2i(\delta_{14} + \gamma)} c_{14}^2 c_{24}^2 m_4 s_{34}^2 \\ \nonumber
  &+& e^{2i(\delta_{13} + \beta)} m_3 \{e^{i(\delta_{14} - \delta_{13})} c_{24} s_{13} s_{14} s_{34} +
  c_{13}(c_{23} c_{34} - e^{i \delta_{24}} s_{23} s_{24} s_{34})\}^2 \\ \nonumber
  &+& m_1[-s_{12}(-c_{34} s_{23} - e^{i \delta_{24}} c_{23} s_{24} s_{34}) \\ \nonumber
  &+& c_{12}\{-e^{i\delta_{14}} c_{13} c_{24} s_{14} s_{34} - e^{i \delta_{13}} s_{13}(c_{23} c_{34}
  - e^{i \delta_{24}} s_{23} s_{24} s_{34})\}]^2 \\ \nonumber
  &+& e^{2 i \alpha} m_2 [c_{12}(-c_{34} s_{23} - e^{i \delta_{24}} c_{23} s_{24} s_{34}) \\ \nonumber
  &+& s_{12}\{-e^{i \delta_{14}} c_{13} c_{24} s_{14} s_{34} - e^{i \delta_{13}} s_{13}(c_{23} c_{34}
  - e^{i \delta_{24}} s_{23} s_{24} s_{34})\}]^2.
 \eea
 It reduces to the 3 generation case for $\theta_{34} = 0$.
Using the approximation for normal hierarchy in Eqs. (\ref{xnh}, \ref{chi1}, \ref{chi2}) this becomes
\bea
|m_{\tau\tau}|&\approx& |\sqrt{\Delta m_{23}^2}\{c_{23}c_{34}s_{23}(e^{2i\beta+\delta_{13}}-c_{12}^2\sqrt{\zeta}e^{2i\alpha})
+\lambda\{-e^{i(2\alpha+\delta_{13})}\sqrt{\zeta}\\ \nonumber
&&s_{12}c_{12}c_{34}\cos2\theta_{23}\chi_{13}-\sqrt{\zeta}c_{12}c_{23}s_{34}e^{2i\alpha}(s_{12}\chi_{14}e^{i\delta_{14}} +c_{12}c_{23}\chi_{24}
e^{2i\delta_{24}})+s_{34}\chi_{24}(-s_{23}^2e^{2i(\beta + \delta_{13})+
i\delta_{24}}\\ \nonumber
&-&\sqrt{\xi}e^{2i(\gamma + \delta_{14})-i\delta_{24}})\}+\lambda^2\{\sqrt{\zeta} s_{12}^2s_{23}\chi_{13}e^{i(2\alpha +
\delta_{13})}(c_{23}c_{34}\chi_{13}e^{i\delta_{13}}+s_{34}\chi_{14}e^{i\delta_{14}})\\ \nonumber
&+&\sqrt{\zeta}c_{12}s_{12}s_{23}\chi_{24}e^{2i\alpha}(2c_{23}s_{34}\chi_{13}e^{i(\delta_{13}+\delta_{24})}+c_{34}\chi_{14}e^{i(\delta_{14}-\delta_{24})})\\ \nonumber
&-&s_{23}s_{34}\chi_{13}\chi_{14}e^{i(2\beta+\delta_{13}+\delta_{14})}\}\}|.
\eea
To get an idea of the order of the terms we consider the vanishing Majorana phases and the Dirac phases having the value equal to $\pi$. This element vanishes when
\bea
 &&c_{23}c_{34}s_{23}(1-c_{12}^2\sqrt{\zeta})+\lambda\{\sqrt{\zeta}s_{12}c_{12}c_{34}\cos2\theta_{23}\chi_{13}
+\sqrt{\zeta}c_{12}c_{23}s_{34}(s_{12}\chi_{14} \\ \nonumber
&+& c_{12}c_{23}\chi_{24})-s_{34}\chi_{24}(s_{23}^2-\sqrt{\xi})\}+\lambda^2\{\sqrt{\zeta} s_{12}^2s_{23}\chi_{13}(c_{23}c_{34}\chi_{13}+s_{34}\chi_{14})\\ \nonumber
&+&\sqrt{\zeta}c_{12}s_{12}s_{23}\chi_{24}(2c_{23}s_{34}\chi_{13}+c_{34}\chi_{14})+s_{23}s_{34}\chi_{13}\chi_{14}\}=0.
\eea
\begin{figure}
\begin{center}
\includegraphics[width=0.33\textwidth,angle=270]{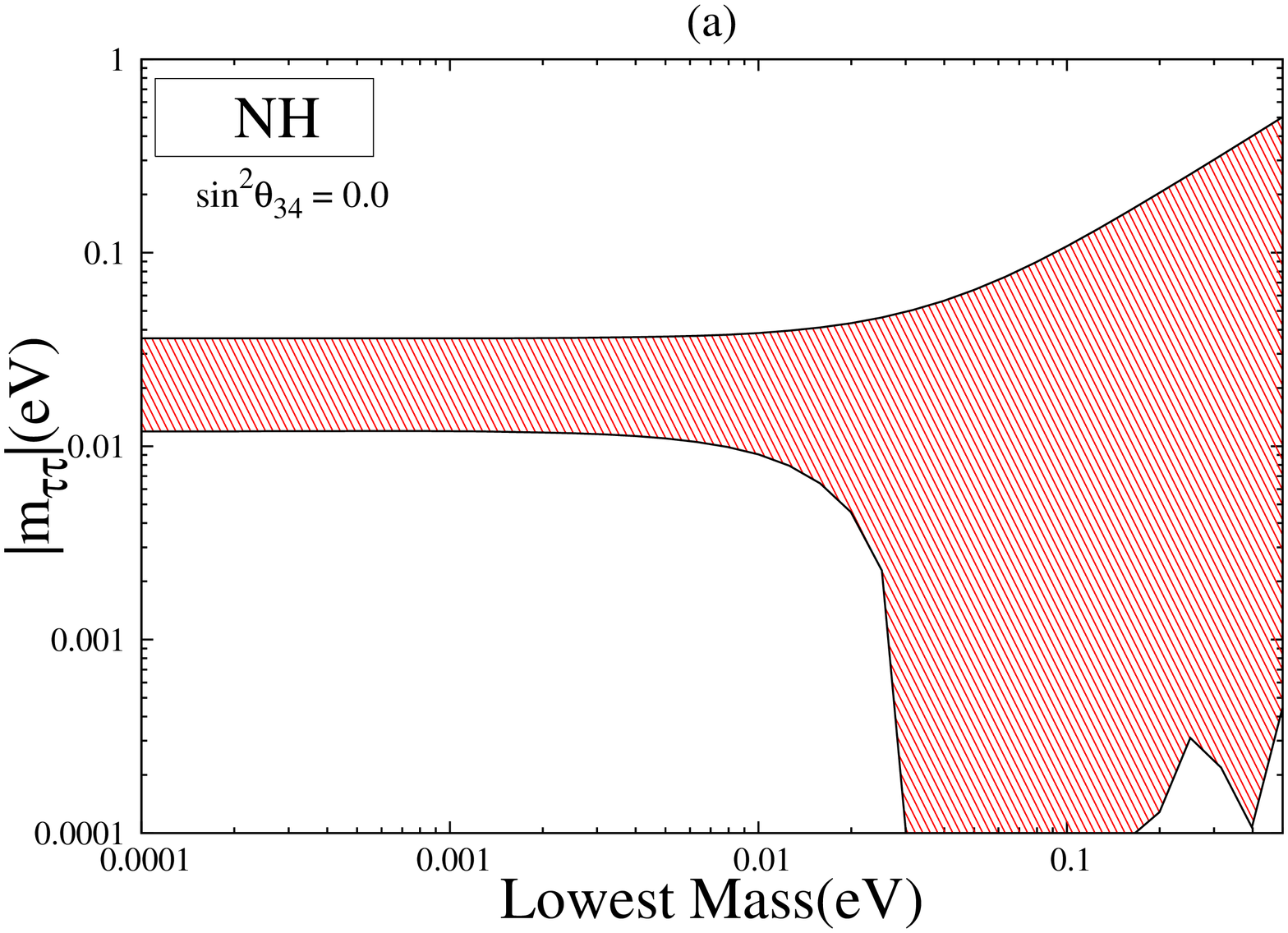}
\includegraphics[width=0.33\textwidth,angle=270]{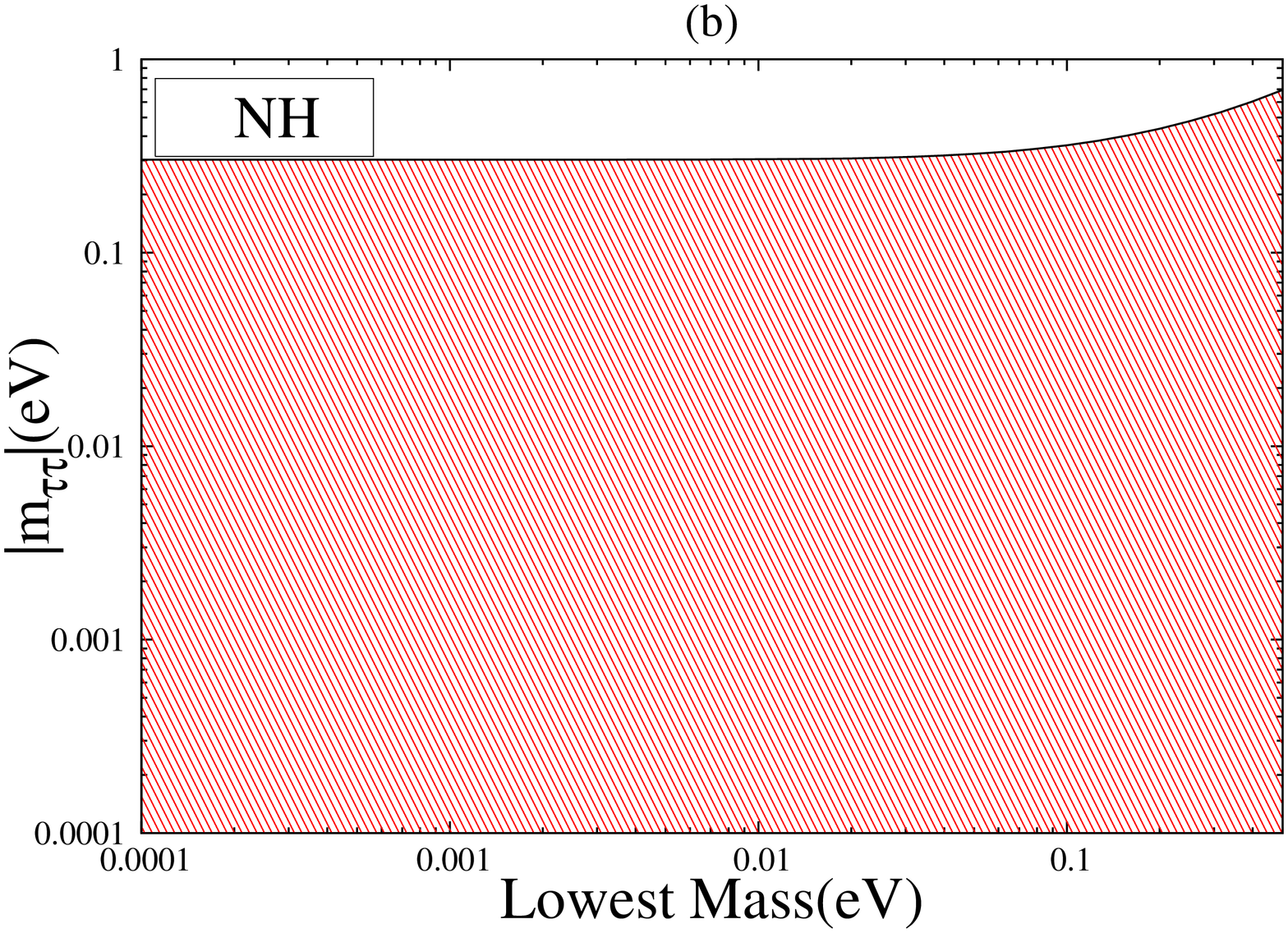}\\
\includegraphics[width=0.33\textwidth,angle=270]{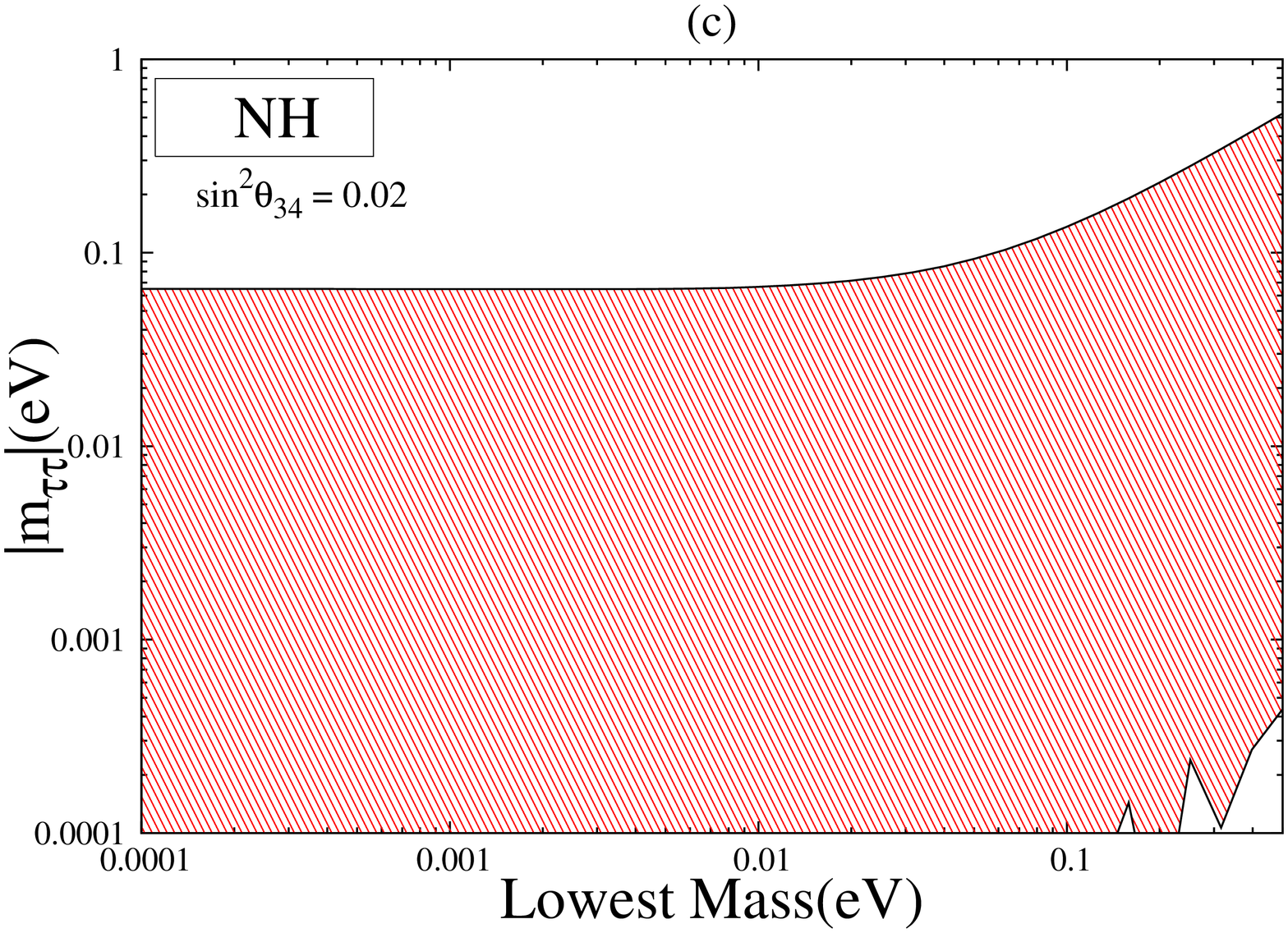}
\includegraphics[width=0.33\textwidth,angle=270]{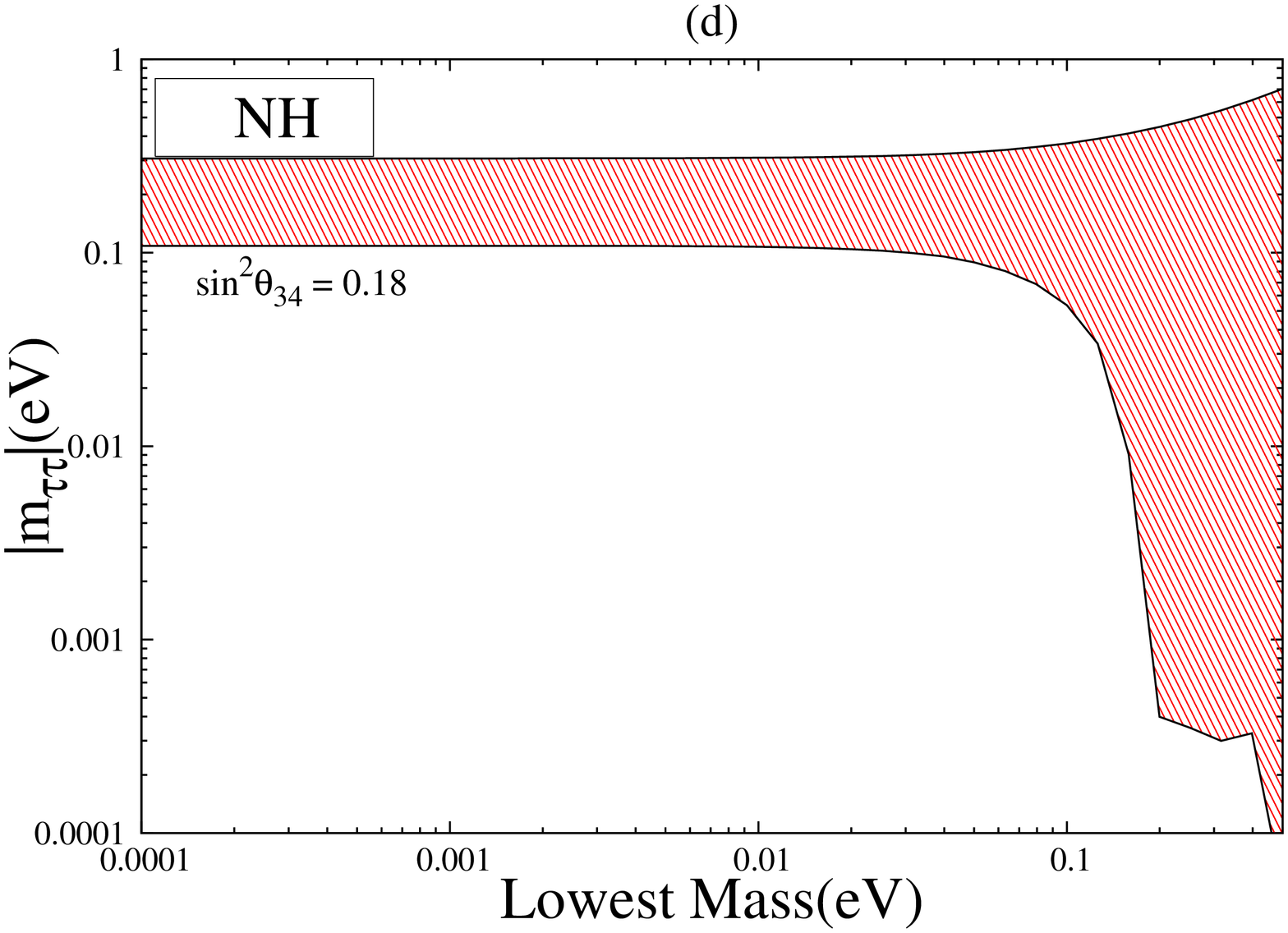}
\end{center}
Figure 12: Plots of vanishing $m_{\tau\tau}$ for normal hierarchy with lowest mass $m_1$.
Panel (a) correspond to three generation case. In panel (b) all the mixing angles are
varied in their full allowed range of parameters (3+1). Panel (c) and (d) are for some specific values of $\theta_{34}$.
\label{fig12}
\end{figure}
For vanishing $\theta_{34}$, which is the case for 3 generation, $m_{\tau \tau} = 0$ is disallowed for small $m_1$ as can be seen from panel (a) of Fig. 12.
This is the generic behaviour of a element belonging to the $\mu -\tau$ block in normal hierarchy which
we mentioned previously. This is because for $\theta_{34}$ equal to zero the leading
order term is large ($\mathcal{O}$ (10$^{-1}$)). Here the term with $\lambda^2$ is quite small
(10$^{-3}$-10$^{-4}$) and hence will not have very significant role to play.
Thus, only terms with coefficient $\lambda$ can cancel the leading order term.
However, for vanishing $\theta_{34}$ this term is small
$\mathcal{O}$ (10$^{-3}$), and cannot cancel the leading order term.
In panel (b) when all the parameters are varied in their $3 \sigma$ range we can see that cancellation is possible over the whole range of $m_1$ (3+1 case).
Now, when $\theta_{34}$ starts increasing from its lowest value there exist a region for intermediate values where both the terms become approximately
of the same order and hence there can be cancellations (panel (c)).
Towards very large values of $\theta_{34}$ the term with coefficient $\lambda$ becomes larger than
the leading order term due to which this element cannot vanish.
For the cancellation very large values of $m_1$ is required as can be seen from panel (d) of Fig 12.
\begin{figure}
\begin{center}
\includegraphics[width=0.33\textwidth,angle=270]{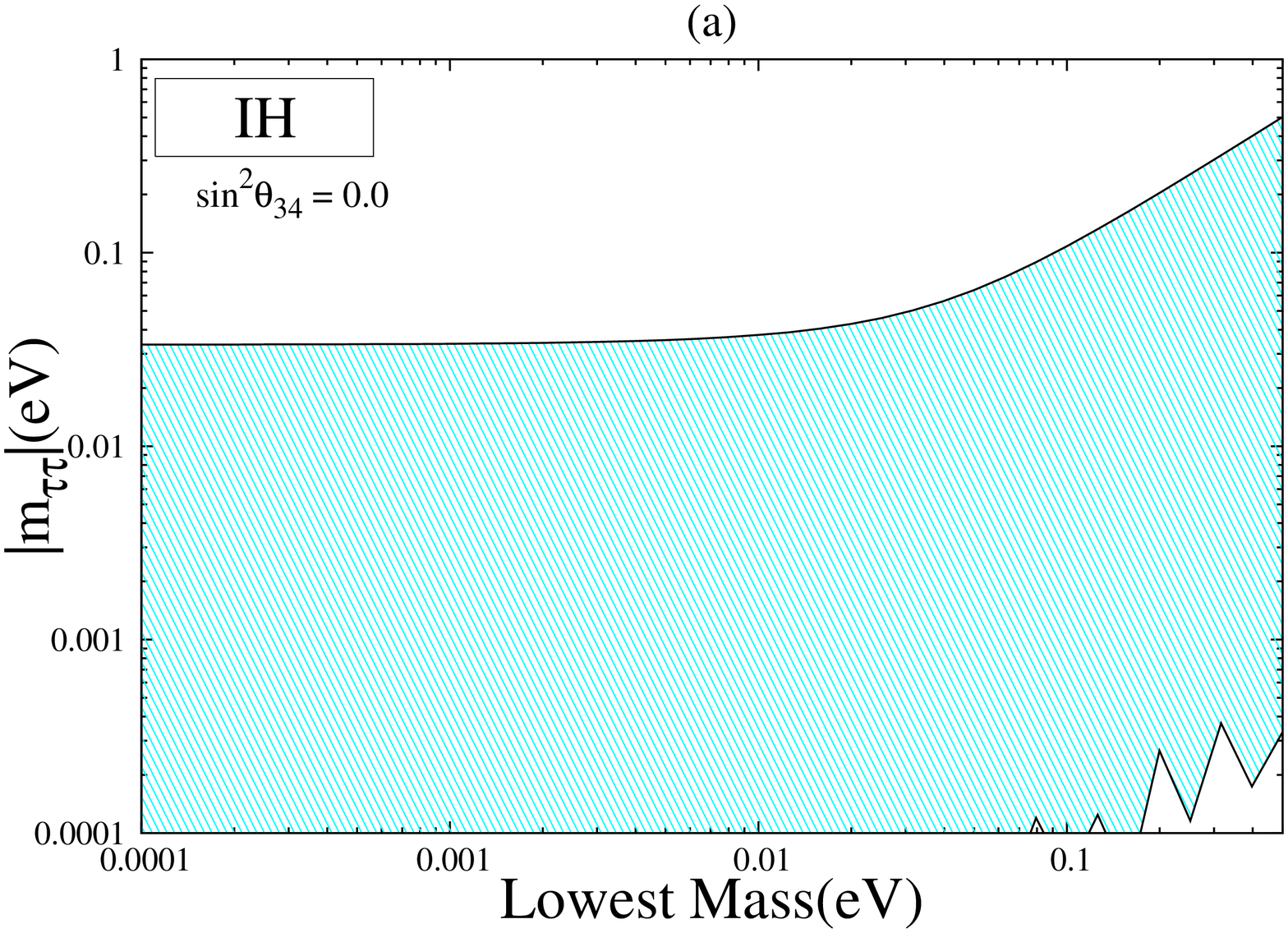}
\includegraphics[width=0.33\textwidth,angle=270]{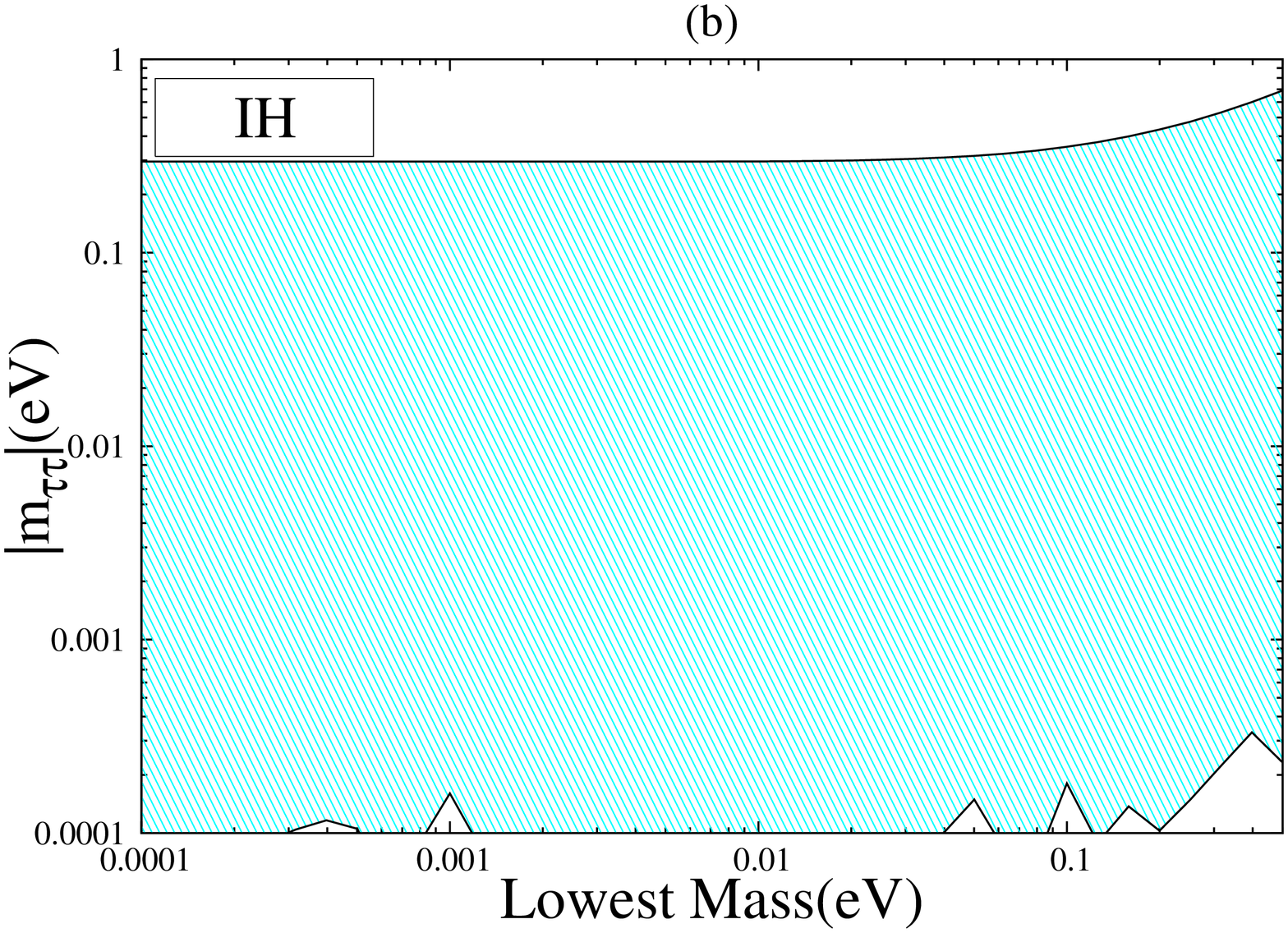}\\
\includegraphics[width=0.33\textwidth,angle=270]{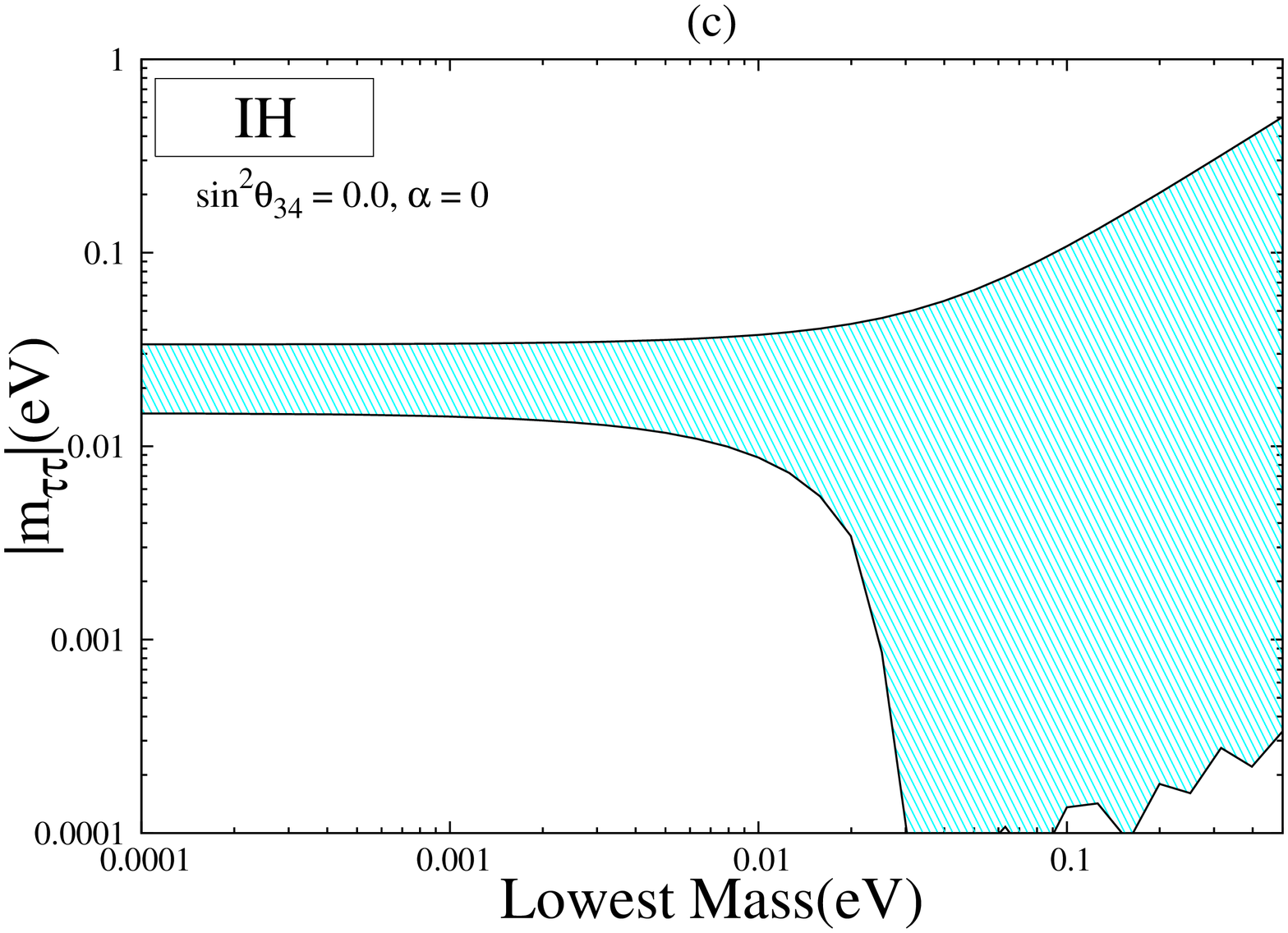}
\includegraphics[width=0.33\textwidth,angle=270]{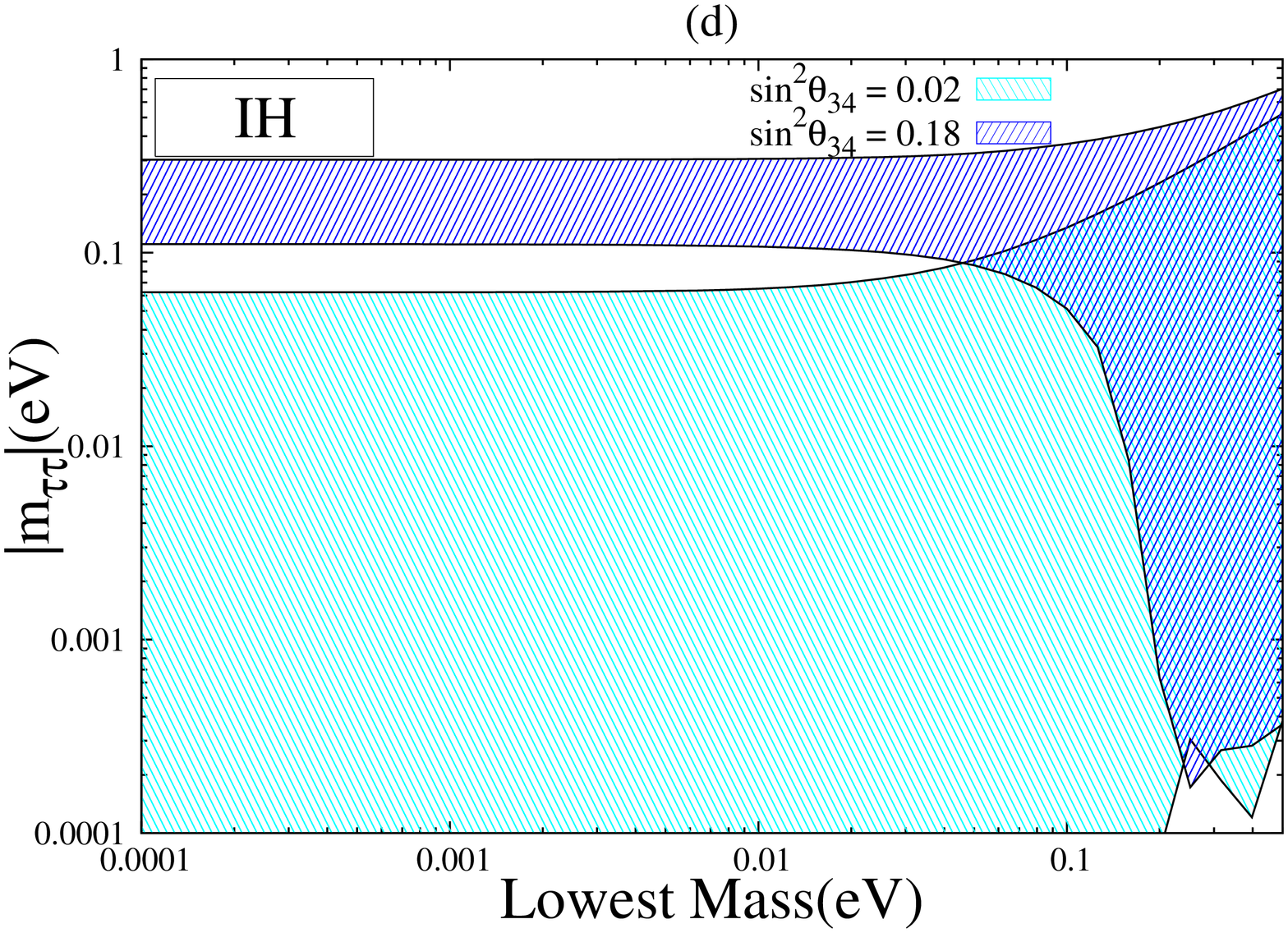}
\end{center}
Figure 13:  Plots of vanishing $m_{\tau\tau}$ for inverted hierarchy with lowest mass $m_3$. Panel (a) correspond
to three generation case. In panel (b) all the mixing angles are varied in their full allowed range of parameters
(3+1). Panel (c) and (d) are for some specific values of $\theta_{34}$ and $\alpha$.
\label{fig13}
\end{figure}
For the case of inverted hierarchy where $m_3$ approaches small values we get the expression
\pagebreak
\bea
 m_{\tau \tau} &\approx& c_{34}^2 s_{23}^2(c_{12}^2 e^{2 i \alpha} + s_{12}^2) + e^{2 i (\delta_{14} + \gamma)} \sqrt{\xi} s_{34}^2 \\ \nonumber
 &+& 2 \lambda[(e^{2 i \alpha} - 1)c_{12} c_{34} s_{12} s_{23}(c_{23} c_{34} s_{12} s_{23}(c_{23} c_{34} e^{ 2 i \delta_{13}} \chi_{13} + e^{ 2 i \delta_{14}} s_{34} \chi_{14}) \\ \nonumber
 &+& 2 c_{23} c_{34} e^{ 2 i \delta_{24}} s_{23} s_{34}(c_{12}^2 e^{ 2 i \alpha} + s_{12}^2) \chi_{24}] \\ \nonumber
 &+& \lambda^2[(c_{12}^2 + e^{ 2 i \alpha} s_{12}^2)\{c_{23} c_{34} \chi_{13} e^{ i \delta_{13}}(c_{23} c_{34} \chi_{13} e^{ i \delta_{13}} + 2 \chi_{14} s_{34}
 e^{ i \delta_{14}}) + e^{ 2 i \delta_{14}} \chi_{14}^2 s_{34}^2\} \\ \nonumber
 &+& (c_{12}^2 e^{ 2 i \alpha} + s_{12}^2)c_{23}^2 e^{ 2 i \delta_{24}} \chi_{24}^2 s_{34}^2 \\ \nonumber
 &+& 2 s_{12}(e^{2 i \alpha} - 1) e^{ i \delta_{24}}(c_{34} \chi_{13} \cos2\theta_{23} e^{ i \delta_{13}} + c_{12} c_{23} \chi_{14} s_{34}) s_{34} \chi_{24}].
\eea
For vanishing Majorana CP phases and Dirac phases having the value equal to $\pi$ this expression becomes
\bea
m_{\tau \tau} &\approx& -c_{23}c_{34}s_{23}+\lambda s_{34}\chi_{24}(c_{23}^2-\sqrt{\xi})+\lambda^2s_{23}\chi_{13}(c_{23}c_{34}\chi_{13}+s_{34}\chi_{14})
\eea
In panel (a) of Fig. 13 we reproduced the 3 generation behaviour by plotting
$|m_{\tau \tau}|$ for $s_{34}^2 = 0$ and in panel (b) all the parameters are varied randomly (3+1).
In both the cases we can see that cancellations are possible for the whole range of $m_3$.
For $s_{34}^2 = 0$ all the terms are of same order and cancellations are always possible. But if we put $\alpha = 0$ then one term with coefficient $\lambda$
 and another term with coefficient $\lambda^2$ drops out from the equation and then small values of $s_{34}^2$ can not cancel the leading order term any more.
 This can be seen from panel (c) where cancellation is not possible for lower $m_3$ region.
However when $s_{34}^2$ increases to a value of about 0.02 this element can vanish (panel (d) the cyan region).
We see that when $\theta_{34}$ increase towards its upper bound the $\lambda$ term becomes large $\mathcal{O} (1)$. Hence, the other terms are
not able to cancel this term and we do not get small $m_3$ region allowed (Panel (d), blue region).

\subsection{The Mass Matrix elements $m_{es}$, $m_{\mu s}$, $m_{\tau s}$ and $m_{ss}$ }
The elements $m_{es}$, $m_{\mu s}$, $m_{\tau s}$ and $m_{ss}$ are present in the fourth row and fourth column in the neutrino mass matrix.
They are the new elements that arises in 3+1 scenario due to the addition of one light sterile neutrino.
The expressions for $m_{es}$ and $m_{\mu s}$ are given by
\bea
  m_{es} &=& e^{i(2 \gamma + \delta_{14})} c_{14} c_{24} c_{34} m_4 s_{14} \\ \nonumber
  &+& e^{i(2 \beta + \delta_{13})} c_{14} m_3 s_{13}\{-e^{i(\delta_{14} - \delta_{13})} c_{24} c_{34} s_{13} s_{14} + c_{13}(-e^{ i \delta_{14}} c_{34} s_{23} s_{24} - c_{23} s_{34})\} \\ \nonumber
  &+& c_{12} c_{13} c_{14} m_1[-s_{12}(-e^{i \delta_{24}} c_{23} c_{34} s_{24} + s_{23} s_{34}) \\ \nonumber
  &+& c_{12}\{-e^{i \delta_{14}} c_{13} c_{24} c_{34} s_{14} - e^{i \delta_{13}} s_{13}(-e^{i \delta_{24}} c_{34} s_{23} s_{24} - c_{23} s_{34})\}] \\ \nonumber
  &+& e^{ 2 i \alpha} c_{13} c_{14} m_2 s_{12}[c_{12}(-e^{ i \delta_{24}} c_{23} c_{34} s_{24}+s_{23}s_{34})  \\ \nonumber
  &+& s_{12} \{- e^{i \delta_{14}} c_{13} c_{24} c_{34} s_{14} - e^{i \delta_{13}} s_{13}(-e^{i \delta_{24}} c_{34} s_{23} s_{24} - c_{23} s_{34})\}].
 \eea
\begin{figure}
\begin{center}
\includegraphics[width=0.33\textwidth,angle=270]{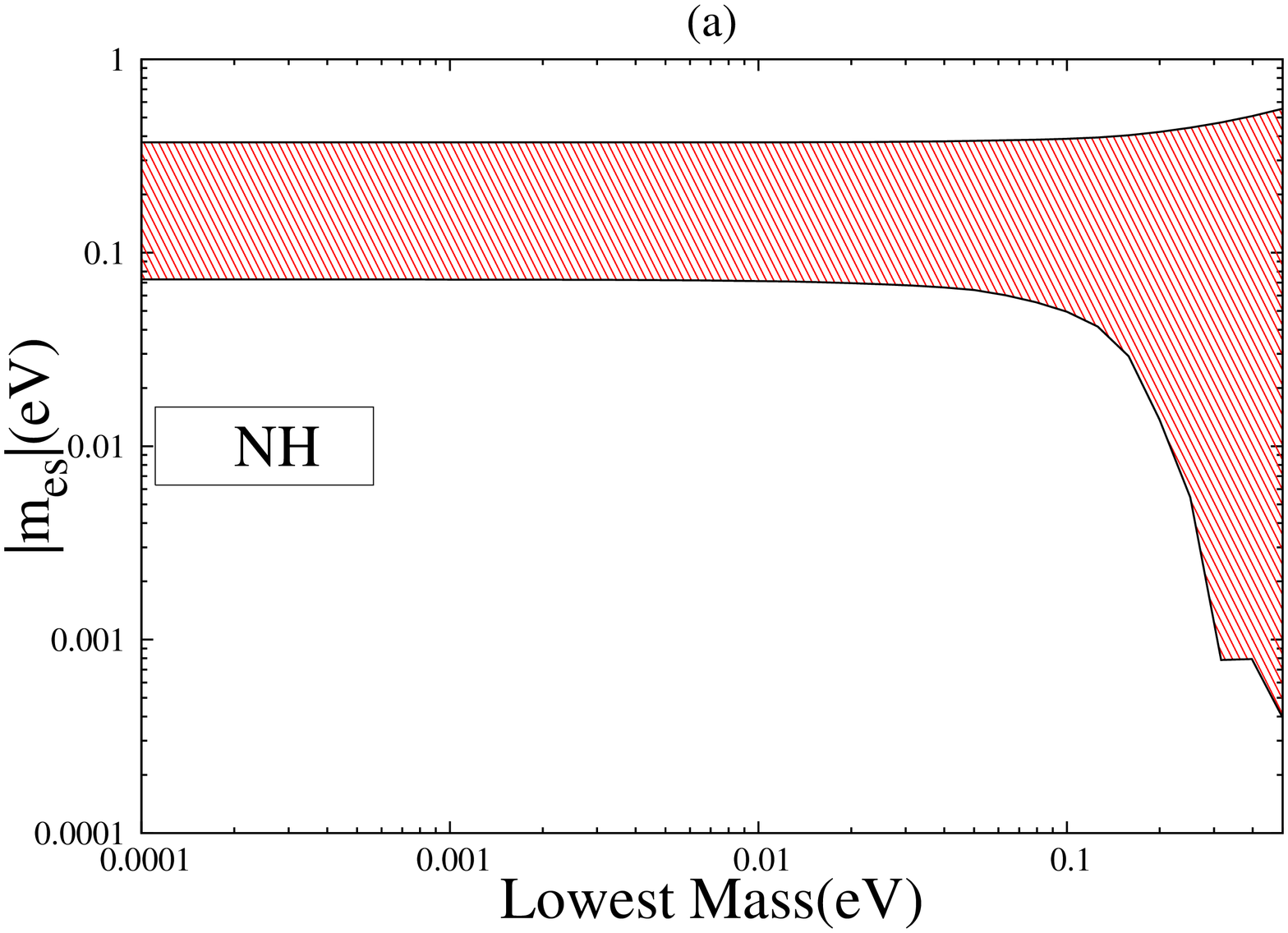}
\includegraphics[width=0.33\textwidth,angle=270]{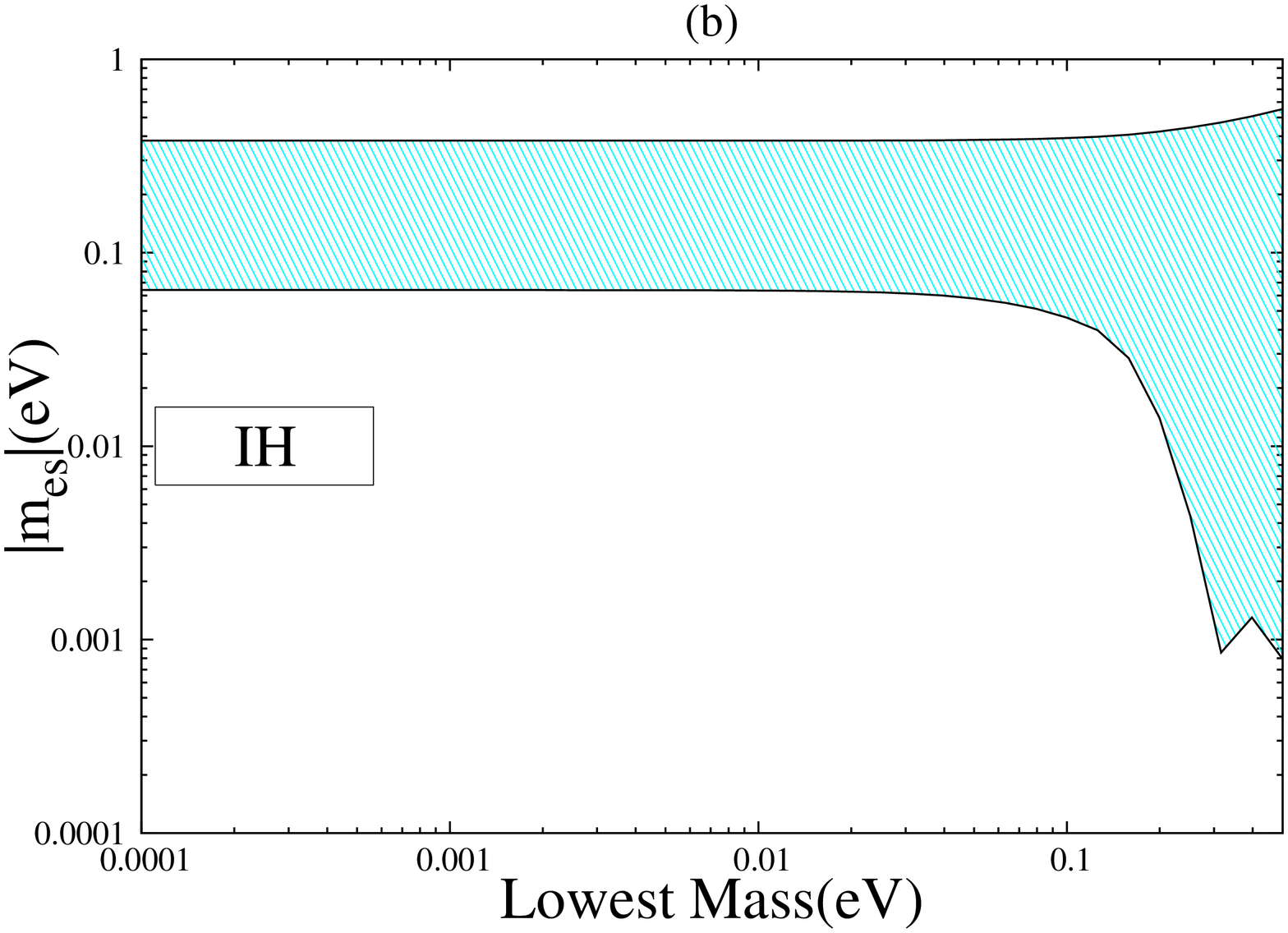} \\
\includegraphics[width=0.33\textwidth,angle=270]{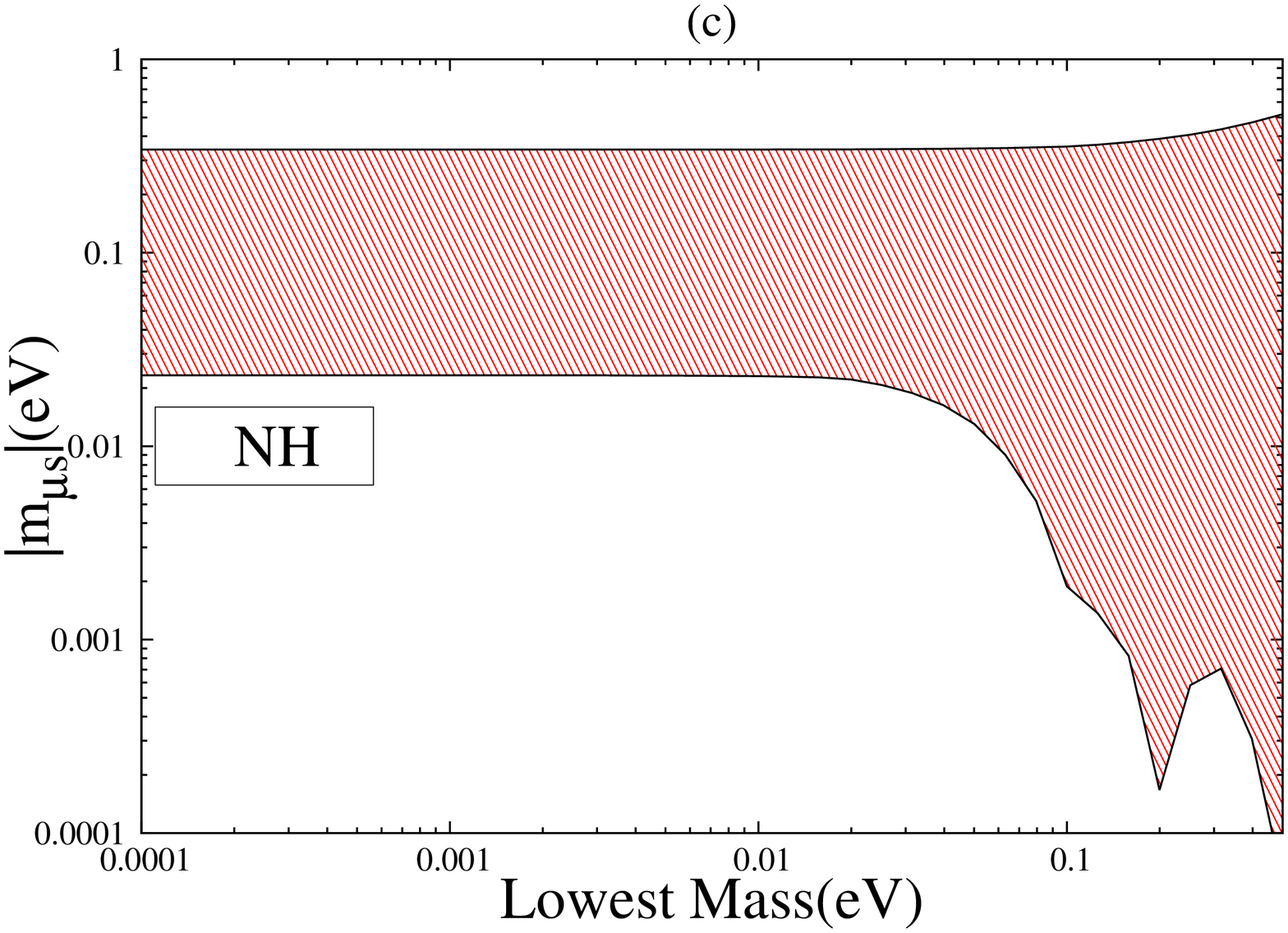}
\includegraphics[width=0.33\textwidth,angle=270]{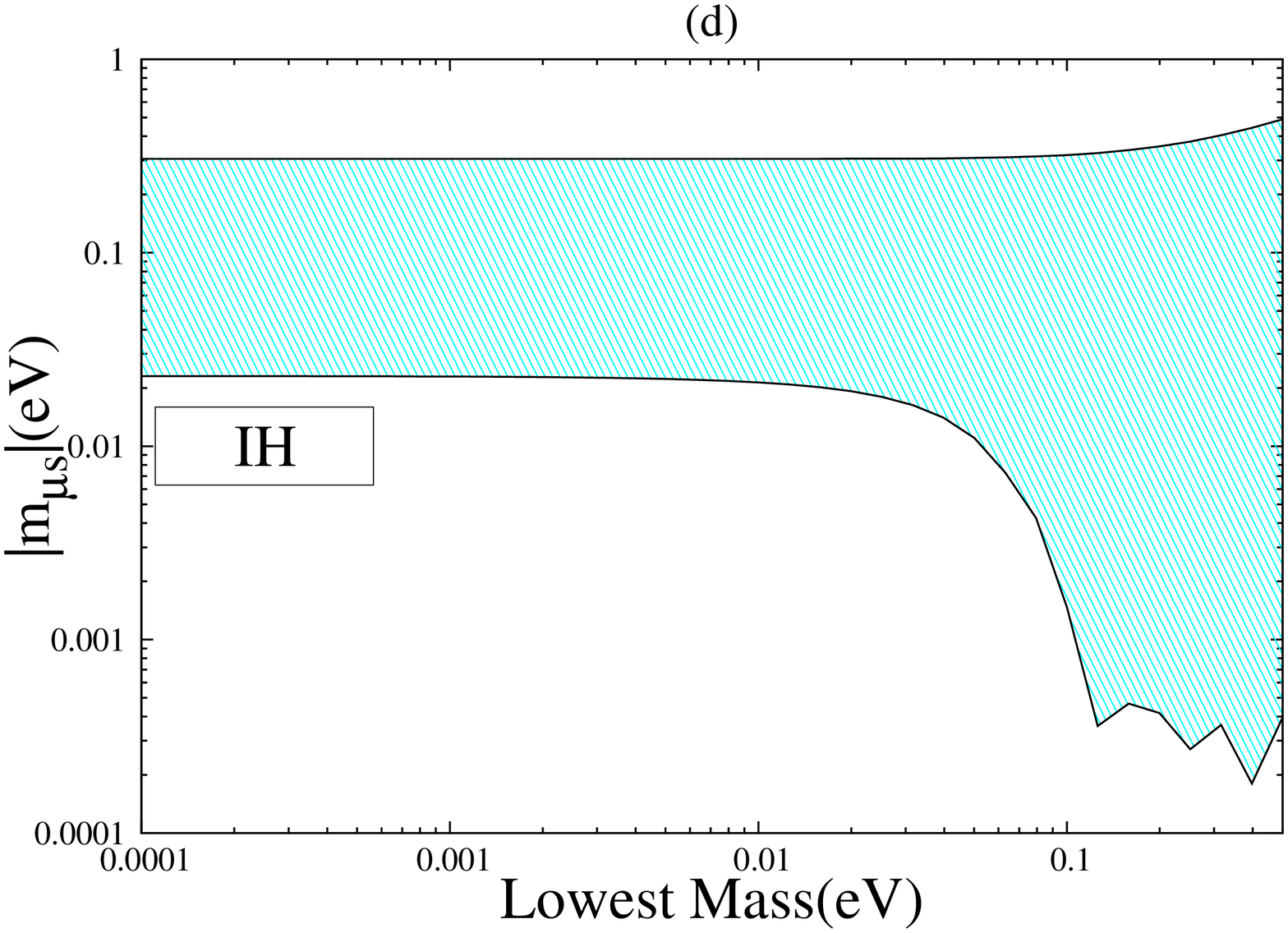} \\
\includegraphics[width=0.33\textwidth,angle=270]{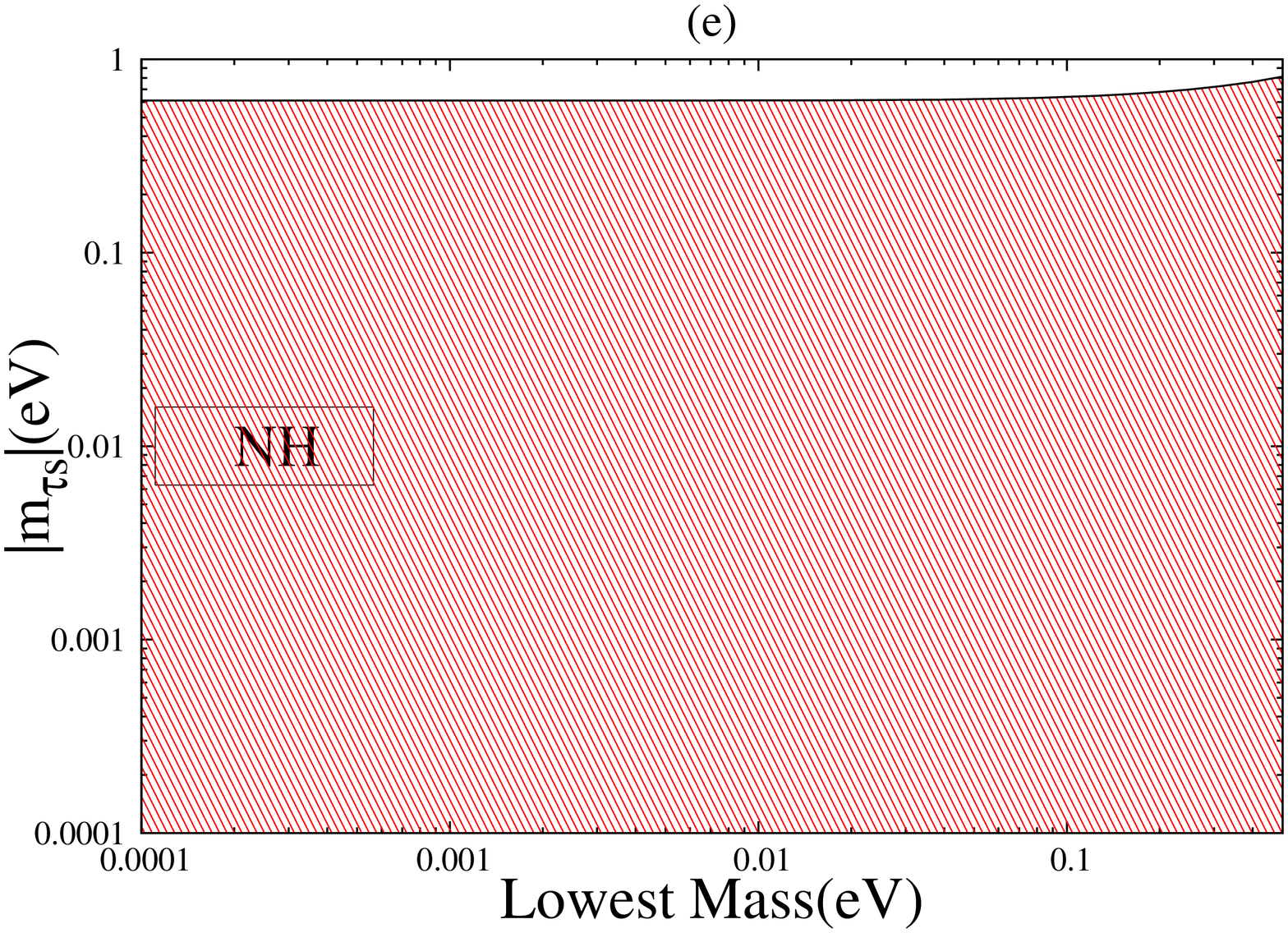}
\includegraphics[width=0.33\textwidth,angle=270]{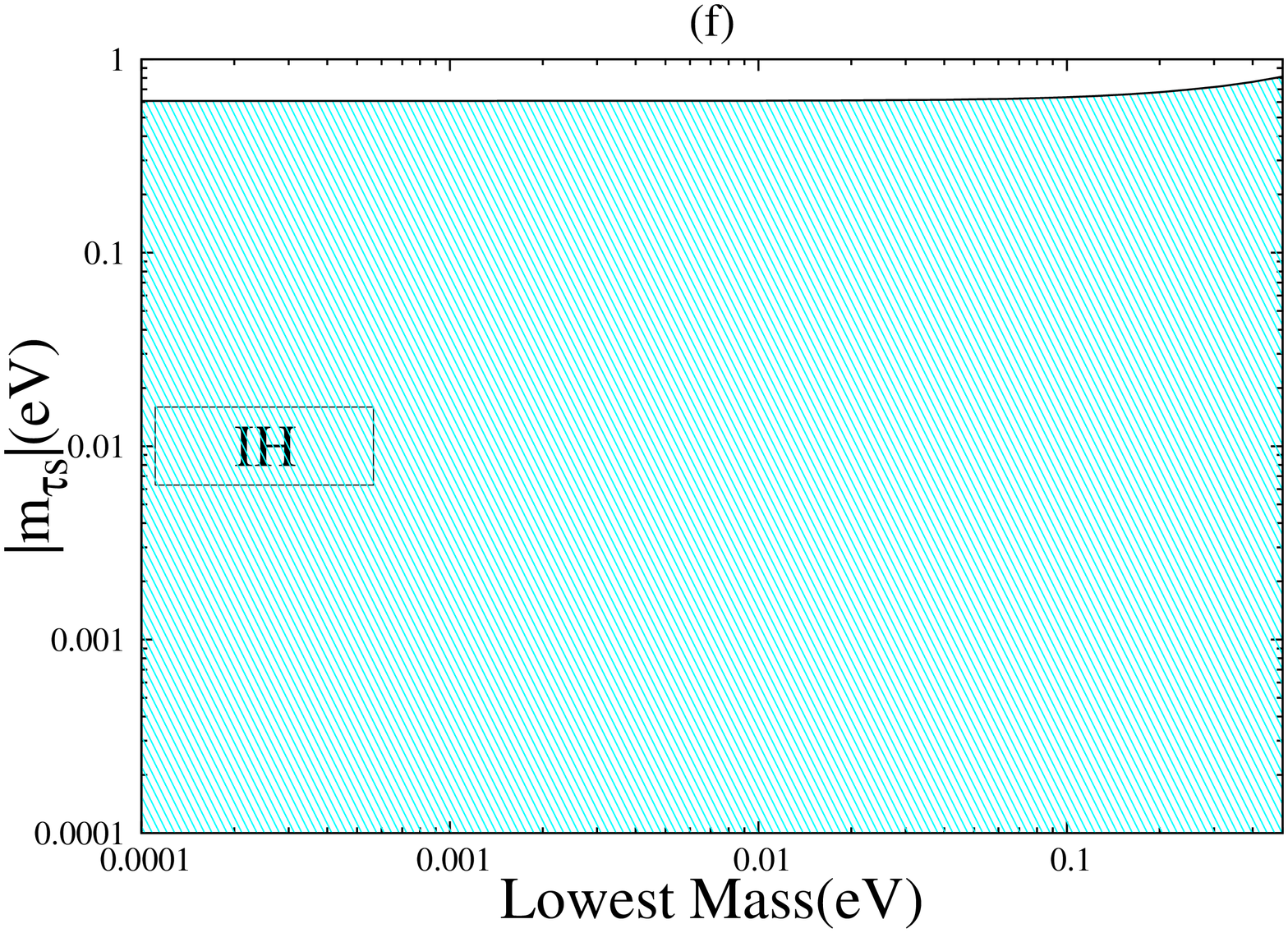} \\
\includegraphics[width=0.33\textwidth,angle=270]{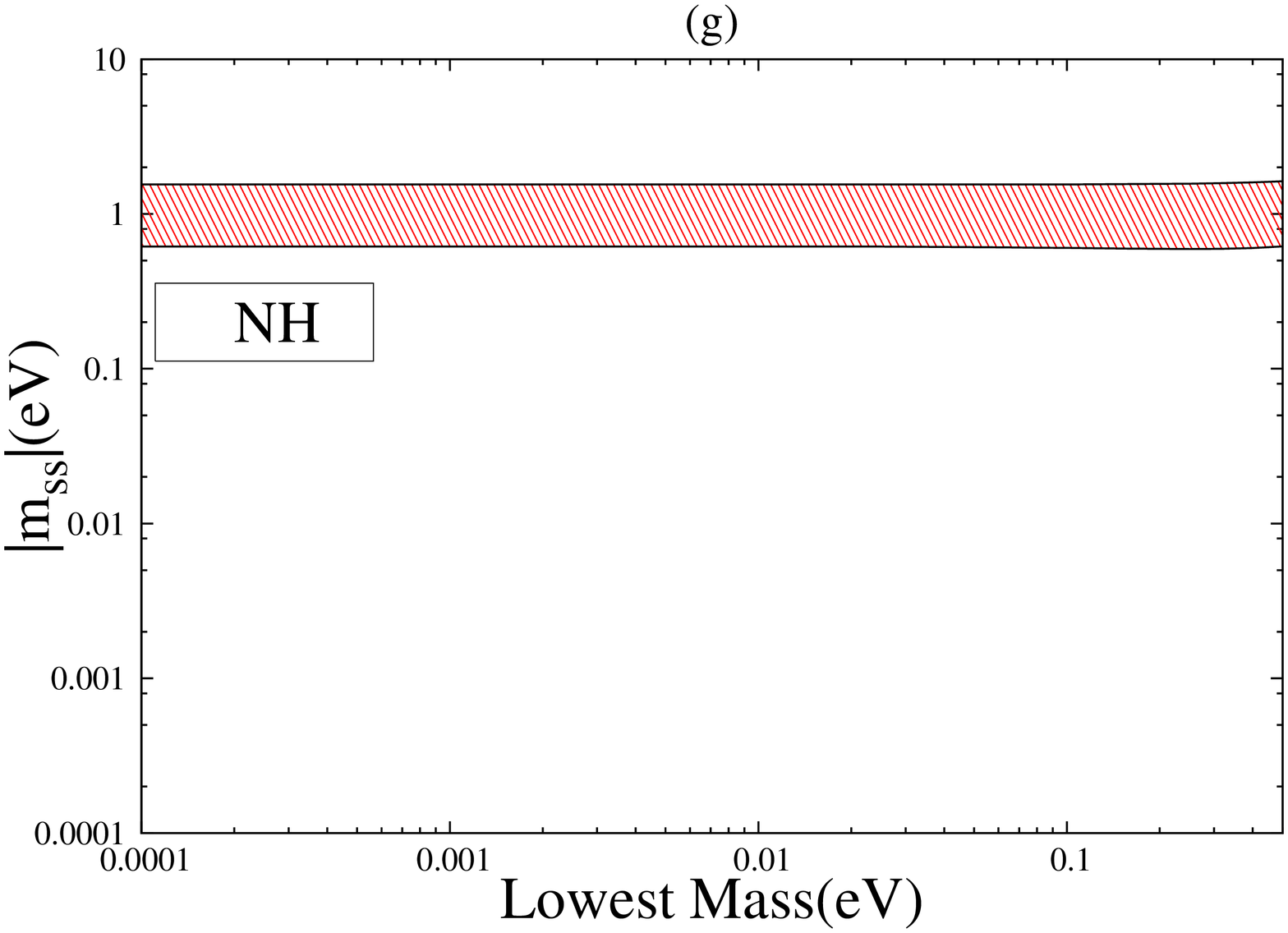}
\includegraphics[width=0.33\textwidth,angle=270]{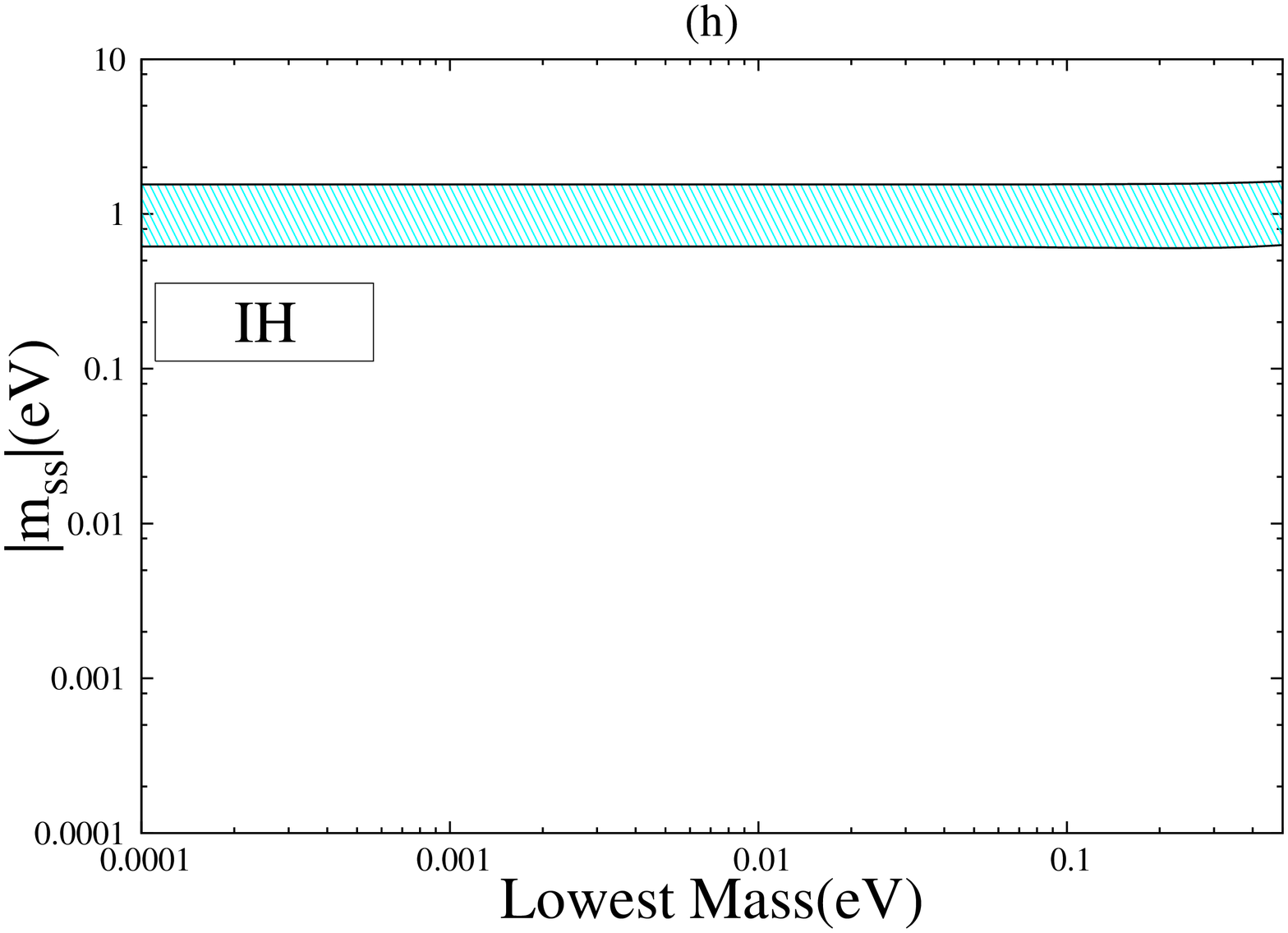}
\end{center}
Figure 14: Correlation plots for vanishing $|m_{ks}|$ for both normal and inverted hierarchy.
In these plots all the mixing angles are varied in their 3 $\sigma$ allowed range, Dirac CP phases are varied from 0 to $2\pi$ and Majorana phases from 0 to $\pi$.
\label{fig14}
\end{figure}
\begin{eqnarray}
  m_{\mu s} &=& e^{i(2 \gamma + \delta_{14})} c_{14}^2 c_{24} c_{34} m_4 s_{24} \\ \nonumber
  &+&e^{i(2 \beta + \delta_{13})} m_3(c_{13} c_{24} s_{23} - e^{i(\delta_{14} - \delta_{13} - \delta_{24})} s_{13} s_{14} s_{24}) \\ \nonumber
 && \{-e^{i(\delta_{14} - \delta_{13})} c_{24} c_{34} s_{13} s_{14} + c_{13}(-e^{ i \delta_{24}} c_{34} s_{23} s_{24} - c_{23} s_{34})\} \\ \nonumber
 &+& m_1\{-c_{23} c_{24} s_{12} + c_{12}(-e^{i \delta_{13}} c_{24} s_{13} s_{23} - e^{i(\delta_{14} - \delta_{24})} c_{13} s_{14} s_{24})\} \\ \nonumber
 &&[-s_{12}(-e^{i \delta_{24}} c_{23} c_{34} s_{24} + s_{23} s_{34}) \\ \nonumber
 &+& c_{12}\{-e^{ i \delta_{14}} c_{13} c_{24} c_{34} s_{14} - e^{i \delta_{13}} s_{13}(-e^{ i \delta_{24}} c_{34} s_{23} s_{24} - c_{23} s_{34})\}] \\ \nonumber
 &+& e^{2 i \alpha} m_2 \{c_{12} c_{23} c_{24} + s_{12}(-e^{i \delta_{13}} c_{24} s_{13} s_{23} - e^{i(\delta_{14} - \delta_{24}} c_{13} s_{14} s_{24})\} \\ \nonumber
 && [c_{12}(-e^{i \delta_{24}} c_{23} c_{34} s_{24} + s_{23} s_{34}) \\ \nonumber
 &+& s_{12}\{-e^{i \delta_{14}} c_{13} c_{24} c_{34} s_{14} - e^{i \delta_{13}} s_{13}(-e^{ i \delta_{24}} c_{34} s_{23} s_{24} - c_{23} s_{34})\}].
 \end{eqnarray}
Though the equations seem very complex, one can easily understand the properties of these elements by just looking at the $m_4$ terms.
The $m_4$ term in  $m_{es}$ is proportional to $s_{14}$. So in general it is quite large ($\mathcal{O}(1)$). For this
element to become negligible very small values of $s_{14}^2$ is required. But as this angle is bounded by the SBL experiments,
complete cancellations never occurs for both normal and inverted hierarchy (Panel (a), (b) of Fig. 14). Similar predictions are obtained for $m_{\mu s}$ element which
cannot vanish since $s_{24}^2$ has to be negligible which is not allowed by the data. This can be seen from Panel (c), (d) of Fig. 14.

For the element $m_{\tau s}$ the scenario is quite different.
\bea
  m_{\tau s} &=& c_{14}^2 c_{24}^2 c_{34} e^{ 2 i(\delta_{14} + \gamma)} m_4 s_{34} \\ \nonumber
  &+& e^{2 i (\beta + \delta_{13})} m_3\{-c_{24} c_{34} e^{i(\delta_{14} - \delta_{13})} s_{13} s_{14} + c_{13}(-c_{23} s_{34} - c_{34} e^{ i \delta_{24}} s_{23} s_{24})\} \\ \nonumber
  &&\{-c_{24} e^{i(\delta_{14} - \delta_{13})} s_{13} s_{14} s_{34} + c_{13}(c_{23} c_{34} - e^{i \delta_{24}} s_{23} s_{34}s_{24})\} \\ \nonumber
  &+& m_1[-s_{12}(-c_{23} c_{34} e^{i \delta_{24}} s_{24} + s_{23} s_{34}) \\ \nonumber
  &+& c_{12}\{-c_{13} c_{24} c_{34}  e^{ i \delta_{14}} s_{14} - e^{ i \delta_{13}} s_{13}(-c_{23} s_{34} - c_{34} e^{i \delta_{24}} s_{23} s_{34})\}] \\ \nonumber
  &&[-s_{12}(-c_{34} s_{23} - c_{23} e^{i \delta_{24}} s_{34}s_{24}) \\ \nonumber
  &+& c_{12}\{-c_{13} c_{24} e^{ i \delta_{14}} s_{14} s_{34} - e^{i \delta_{13}} s_{13}(c_{23} c_{34} - e^{ i \delta_{24}} s_{23} s_{34}s_{24})\}] \\ \nonumber
  &+& e^{ 2 i \alpha} m_2[c_{12}(-c_{23} c_{34} e^{i \delta_{24}} s_{34} + s_{23} s_{34}) \\ \nonumber
  &+& s_{12}\{-c_{13} c_{24} c_{34} e^{ i \delta_{14}} s_{14} - e^{ i \delta_{13}} s_{13}(-c_{23} s_{34} - c_{34} e^{ i \delta_{24}} s_{23} s_{24})\}] \\ \nonumber
  &&[c_{12}(-c_{34} s_{23} - c_{23} e^{ i \delta_{24}} s_{34}s_{24}) \\ \nonumber
  &+& s_{12}\{-c_{13} c_{24} e^{ i \delta_{14}} s_{14} s_{34} - e^{ i \delta_{13}} s_{13}(c_{23} c_{34} - e^{i \delta_{24}} s_{23} s_{34}s_{14})\}].
 \eea
In this case the $m_4$ term is proportional to $\theta_{34}$ and there is no lower bound on it from the SBL
experiments i.e. it can approach smaller values. As a result the term with
$m_4$ can be very small. Thus this matrix element can possibly vanish in both hierarchies for whole range of the lowest mass
(Panel (e), (f)).\\
The (4,4) element of the neutrino mass matrix is given as
\bea
 m_{ss} &=&  c_{14}^2 c_{24}^2 c_{34}^2 e^{2 i (\gamma + \delta_{14})} m_4 \\ \nonumber
 &+& e^{2 i(\beta+ \delta_{13})} m_3\{-c_{24} c_{34} e^{i(\delta_{14} - \delta_{13})} s_{13} s_{14} + c_{13}(-c_{23} s_{34} - c_{34} e^{ i \delta_{24}} s_{23} s_{24})\}^2 \\ \nonumber
 &+& m_1[-s_{12}(-c_{23} c_{34} e^{ i \delta_{24}} s_{24} + s_{23} s_{34}) \\ \nonumber
 &+& c_{12}\{-c_{13} c_{24} c_{34} e^{ i \delta_{14}} s_{14} - e^{ i \delta_{13}} s_{13}(-c_{23} s_{34} - c_{34} e^{ i \delta_{24}} s_{23} s_{24})\}]^2 \\ \nonumber
 &+& e^{ 2 i \alpha} m_2[c_{12}(-c_{23} c_{34} e^{ i \delta_{24}} s_{24} + s_{23} s_{34}) \\ \nonumber
 &+& s_{12}\{-c_{13} c_{24} c_{34} e^{ i \delta_{14}} s_{14} - e^{i \delta_{13}} s_{13}(-c_{23} s_{34} - c_{34} e^{i \delta_{24}} s_{23} s_{34})\}]^2
 \eea

The $m_4$ term for $m_{ss}$ is proportional to  $c_{14}^2 c_{24}^2 c_{34}^2$.
One can see that this term is of order one as a result this element can
never vanish as is evident from panel (g, h).



\section{Conclusions}

In this paper we analyze systematically the one-zero textures
of the $4\times4$ mass matrix in presence of a sterile neutrino.
Assuming  neutrinos to be Majorana particles,
this is a symmetric matrix with 10
independent entries.
We use the information on the
active sterile mixing angles from the short baseline experiments.
We analyze if the current constraints
on oscillation parameters allow each of these  entries
to assume a vanishing
value. We also study the implications and correlations among the
parameters when each matrix element is zero.
We expand the mass matrix element in terms of a parameter $\lambda$
with suitable coefficients $\chi_{13}, \chi_{14}$ and $\chi_{24}$
corresponding to the
mixing angles $\theta_{13}$, $\theta_{14}$ and $\theta_{24}$.
This is motivated by the observation that these angles
are of same order with $\lambda \equiv 0.2$.
These expressions facilitate the analytic understanding of the
numerical results presented in the different plots.
We study the vanishing condition as a function of the
lowest mass $m_1$ (NH) or $m_3$ (IH) by varying
the lightest mass in
the range 0.0001 - 0.5 eV.

We find that $|m_{ee}| =0$ is possible for NH only for higher values
of the  smallest mass $m_1$ while for IH it is possible even for
lower values. This is in sharp contrast with the 3 generation case
where complete cancellation can never take place for IH.
The current and upcoming $0\nu\beta\beta$ experiments like
GERDA, CUORE, MAJORANA, EXO, SuperNEMO, KamLAND-ZEN, SNO+ \cite{Gando:2012zm,Auger:2012ar,meeexp}
can lower the present sensitivity
by one order of magnitude ($\sim$0.012 - 0.06 eV)
and hence can probe the IH region for the
three neutrino scheme \cite{werner-jg}.
However for 3+1 scenario,
$m_{ee}$ can be in the range of the expected sensitivity of the future
$0\nu\beta\beta$ experiments, even for NH.
Thus if the existence of sterile neutrinos is confirmed by future experiments
\cite{sterile-future-smirnov} then
it may be difficult to probe the hierarchy from $0\nu\beta\beta$ alone.

$|m_{e\mu}|$ can vanish over the whole range of the smallest mass
for both 3 and 3+1 neutrino scenarios. However for larger values of
the mixing angle $s_{24}^2$, cancellation is not achieved for smaller
$m_1$ for NH.
For IH
the cancellation condition depend on the Majorana phase
$\alpha$ and the mixing angle $\theta_{24}$.
We obtain the correlations
between these two parameters required for making this
element vanishingly small.

Cancellation is achieved for the element $m_{e\tau}$ for
the full range of the lowest mass in the 3+1 scenario.
The element $m_{e\mu}$ is related to the
element $m_{e\tau}$ by $\mu-\tau$ symmetry. However unlike three
generation case $\theta_{23}$ in these textures are not related
simply by $\overline\theta_{23} = (\pi/2 - \theta_{23})$.
The mixing angles $\theta_{24}$ and $\theta_{34}$ are also
different in these two textures in general. However for small
values of $\theta_{24}$ we get $\overline\theta_{24} = \theta_{34}$
in these textures. Consequently the role played by $\theta_{24}$
for $m_{e\mu}$ is played by $\theta_{34}$ in $m_{e \tau}$ in this limit.
Thus in this case cancellation is not achieved for larger values
of $s_{34}^2$ in the hierarchical regime for NH.
For IH we obtain correlations between
$\alpha$ and $\sin^2\theta_{34}$ for fulfilling the condition
for cancellations.

The elements $m_{\mu \mu}$ and $m_{\tau \tau}$ are related by
$\mu-\tau$ symmetry.
For these cases, cancellation is not possible in the
hierarchical zone for IH, in the 3 generation case.
However the extra contribution coming from the sterile part
helps in achieving cancellation in this region.
For IH one can obtain correlations between the Majorana phase
$\alpha$ and the mixing angle $\theta_{24}$($\theta_{34}$) for
$|m_{\mu \mu}|=0$($|m_{\tau \tau}|=0$).

For $m_{\mu \tau}$ element cancellation was possible for three
generation case only for higher values of the lightest mass.
However if one includes the sterile neutrino then this element
can vanish over the whole range of the lightest neutrino mass
considered.

With the current constraints on sterile parameters it is not possible
to obtain $m_{ss}=0$ while $m_{es}$ and $m_{\mu s}$ can only vanish
in the QD regime of the active neutrinos.
However, the element $m_{\tau s}$ can  be vanishingly small
in the whole mass range. This is because the angle
$\theta_{34}$
can take very small values  and hence cancellation is possible
even for smaller values of the lowest mass.

The above results can be useful for building models for light
sterile neutrinos and shed light on the underlying new physics
if future experiments and analyses reconfirm the explanation of
the present anomalies in terms of  sterile neutrinos.

\section{Acknowledgements}
The work of C. S. K and S. G. is supported by the National Research Foundation of Korea (NRF) grant funded by
Korea government of the Ministry of Education, Science and Technology (MEST) (Grant No. 2011-0017430) and (Grant No. 2011-0020333).

\end{document}